\documentclass[fleqn,usenatbib]{mnras}

\usepackage{newtxtext,newtxmath}

\usepackage[T1]{fontenc}

\DeclareRobustCommand{\VAN}[3]{#2}
\let\VANthebibliography\thebibliography
\def\thebibliography{\DeclareRobustCommand{\VAN}[3]{##3}\VANthebibliography}

\usepackage{graphicx}	
\usepackage{amsmath}	
\usepackage{adjustbox}  
\usepackage{wasysym}    
\usepackage{bm}         
\usepackage{xspace}
\usepackage{enumitem}

\makeatletter 
  \patchcmd{\NAT@citex}
    {\@citea\NAT@hyper@{%
      \NAT@nmfmt{\NAT@nm}%
      \hyper@natlinkbreak{\NAT@aysep\NAT@spacechar}{\@citeb\@extra@b@citeb}%
      \NAT@date}}
    {\@citea\NAT@nmfmt{\NAT@nm}%
    \NAT@aysep\NAT@spacechar\NAT@hyper@{\NAT@date}}{}{}

  \patchcmd{\NAT@citex}
    {\@citea\NAT@hyper@{%
      \NAT@nmfmt{\NAT@nm}%
      \hyper@natlinkbreak{\NAT@spacechar\NAT@@open\if*#1*\else#1\NAT@spacechar\fi}%
        {\@citeb\@extra@b@citeb}%
      \NAT@date}}
    {\@citea\NAT@nmfmt{\NAT@nm}%
    \NAT@spacechar\NAT@@open\if*#1*\else#1\NAT@spacechar\fi\NAT@hyper@{\NAT@date}}
    {}{}
\makeatother

\newcommand{\thz}{\mbox{\textsc{thesan-zoom}}\xspace}
\newcommand\orcid[1]{\href{http://orcid.org/#1}{\adjustbox{trim={-.15\width} {0\height} {-.15\width} {0\height},clip}{\includegraphics[height=12pt]{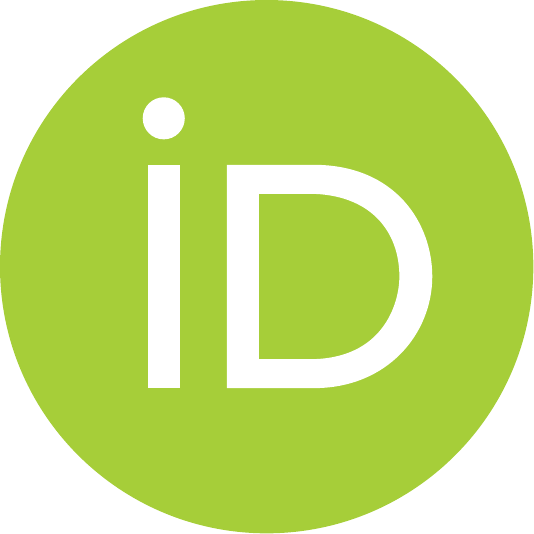}}}}
\newcommand\Msun{\text{M}_{\astrosun}} 
\newcommand\Zsun{\text{Z}_{\astrosun}} 

\defcitealias{Sodini2024}{S24}

\title[Interpreting Pop~III-enriched gas candidates with \thz]{The THESAN-ZOOM project: interpreting candidate Population~III-enriched absorption line systems during reionization}

\author[G. Pruto et al.]{%
Giulia~Pruto\orcid{0000-0003-0828-7192},$^{1}$\thanks{E-mail: \href{mailto:giulia.pruto@ed.ac.uk}{giulia.pruto@ed.ac.uk}}
Laura~Keating\orcid{0000-0001-5211-1958},$^{1}$ 
Rahul~Kannan\orcid{0000-0001-6092-2187},$^{2}$
Ewald~Puchwein\orcid{0000-0001-8778-7587},$^{3}$
Aaron~Smith\orcid{0000-0002-2838-9033},$^{4}$
\newauthor
Enrico~Garaldi\orcid{0000-0002-6021-7020},$^{5}$
Mark~Vogelsberger\orcid{0000-0001-8593-7692},$^{6}$
William~McClymont\orcid{0009-0009-5565-3790},$^{7}$
Xuejian~Shen\orcid{0000-0002-6196-823X}$^{6}$
\newauthor
and Sandro~Tacchella\orcid{0000-0002-8224-4505}$^{8, 9}$
\\
\\
$^{1}$Institute for Astronomy, University of Edinburgh, Blackford Hill, Edinburgh, EH9 3HJ, UK\\%
$^{2}$ Department of Physics and Astronomy, York University, 4700 Keele Street, Toronto, ON M3J 1P3, Canada\\%
$^{3}$ Leibniz-Institut f\"ur Astrophysik Potsdam, An der Sternwarte 16, 14482 Potsdam, Germany\\%
$^{4}$ Department of Physics, The University of Texas at Dallas, Richardson, Texas 75080, USA\\%
$^{5}$ Kavli Institute for the Physics and Mathematics of the Universe, The University of Tokyo, 5-1-5 Kashiwanoha, Kashiwa, 277-8583, Chiba, Japan\\%
$^{6}$ Department of Physics, Kavli Institute for Astrophysics and Space Research, Massachusetts Institute of Technology, Cambridge, MA 02139, USA\\%
${7}$ Canadian Institute for Theoretical Astrophysics, University
of Toronto, 60 St. George Street, Toronto, ON M5S 3H8,
Canada\\%
$^{8}$ Kavli Institute for Cosmology, University of Cambridge, Madingley Road, Cambridge CB3 0HA, UK\\%
$^{9}$ Cavendish Laboratory, University of Cambridge, 19 JJ Thomson Avenue, Cambridge, CB3 0HE, UK
}

\date{Accepted XXX. Received YYY; in original form ZZZ}

\pubyear{\the\year{}}

\begin{document}
\label{firstpage}
\pagerange{\pageref{firstpage}--\pageref{lastpage}}
\maketitle

\begin{abstract}
{
With \textit{JWST}, metal absorption lines are now being detected at increasingly high redshift, where signatures of the first generation of stars (Population~III) may be observable.
Interpreting observations of possible Pop~III-enriched gas is challenging, as metals released by stars inevitably mix with pristine gas accreting onto galaxies and recycled gas from different episodes of star formation. 
Inference of Pop~III enrichment from low-ionisation absorption lines requires placing stellar nucleosynthesis yields within the broader context of galaxy evolution. 
With \thz, a set of high-resolution radiation hydrodynamical zoom-in simulations of galaxies, we investigate the metal abundance patterns observable in absorption. As our fiducial simulations do not include Pop~III enrichment channels, our primary goal is to test how galaxy formation processes can shape observed abundance ratios.
Our results indicate that a single abundance ratio plane, such as the commonly used [Si/O] vs. [C/O], cannot conclusively distinguish Pop~III enrichment from enrichment by subsequent populations of stars. Robust detections instead require characterising multiple abundance ratios.
In our simulations, a larger dispersion in abundance ratio distributions (often interpreted as possible evidence of pollution from the first stellar populations) naturally arises with increasing redshift, when the metallicity and dust content are lower. However, this alone does not reproduce all the observed results.
By running an additional simulation with an updated \thz model containing Pop~III stars, we show that this has a large impact on the simulated abundance ratios, pushing them closer to those observed in potential Pop~III-enriched gas. 
}

\end{abstract}




\section{Introduction}

The emergence of the first stars, also referred to as Population III (Pop~III), marked the beginning of radiative and chemical transformation in the early Universe.
Theoretical studies predict Pop~III stars to form from primordial gas collapsing within dark matter haloes from $z\sim30$, possibly continuing until $z\sim5$ \citep[][]{Klessen2023, Zier2025b}.
Because of the lack of metals, cooling in this primordial gas below temperatures of $10^4$\,K is less efficient, and simulations have shown that this leads to the formation of more massive protostellar clumps \citep[e.g.,][]{Abel2000, Bromm2001, Bromm2002}.
Consequently, Pop~III stars are generally expected to have higher masses than stars in subsequent stellar populations.
The presence of these massive, metal-free stars has a significant impact on the early Universe, as they produce large amounts of ionising radiation, contributing to the onset of reionization, and enrich the primordial gas with the first heavy elements, leaving distinctive chemical signatures in their surroundings \citep[e.g.,][]{HegerWoosley2010, Vanni2023}.
Despite extensive theoretical studies predicting the properties of Pop~III stars, robust direct observational detections have yet to be reported.
Improving the constraints used to identify these first stars is therefore essential.

Locally, the search for Pop~III stars has focused on the identification of extremely metal-poor stars in the Milky Way \citep[][]{Frebel2015}. Observations have revealed stars with metallicities as low as $\textrm{Z}_\star \lesssim 10^{-5}\,\Zsun$ \citep[e.g.,][]{Caffau2011, Keller2014} exhibiting significant carbon enhancement in their atmospheres, which can potentially be explained by different types of core-collapse supernovae originating from the first generations of stars. 
However, despite extensive observational efforts, no metal-free stars have yet been seen in the Milky Way, suggesting that they might be very rare or alternatively massive enough that they did not survive until the present day \citep{Rossi2021}.

With the advent of the James Webb Space Telescope (\textit{JWST}), the search for the first stars has been pushed to increasingly higher redshifts, with multiple studies attempting to identify Pop~III host galaxies.
These searches are typically focused on finding low-metallicity sources with hard ionising spectral features \citep[e.g., ][]{Fujimoto2025, Maiolino2026}.
However, the large scatter in metallicity observed at high redshift points toward complex enrichment histories, which can produce strong oxygen line emission in $z>10$ galaxies \citep[e.g.,][]{Witstok2026, Harikane2026}, while also allowing for few metal-poor candidates at $z\simeq6$ \citep[e.g.,][]{Bolamperti2026, Morishita2025}.
The detection of Pop~III stars through this method is further biased by selection effects, which generally favour the identification of more metal-rich galaxies.
Studies also showed the importance of follow-up spectroscopic observations to rule out contamination due to galaxies with exceptionally strong Balmer jumps, which could be interpreted as metal-free galaxies from photometric data \citep[][]{Fujimoto2025b}.
Moreover, the short timescale over which hard spectral features are expected to persist (only a few Myr after a starburst), and the presence of additional processes able to produce similar spectral signatures further complicate the picture \citep[][]{Venditti2026}.

A complementary approach to identify the first stars is to study the chemical abundances in the gas surrounding galaxies.
In fact, especially in low-mass haloes ($\approx 10^6\,\Msun$), where the first stars have a high probability of forming, strong stellar feedback could drive gas enriched with Pop~III signatures out into the circumgalactic medium (CGM) or even further into the intergalactic medium (IGM). 
This material will then mix with pristine gas and become further enriched by ejecta from subsequent stellar populations, or it could be reaccreted onto the halo \citep[][]{Tumlinson2017}.
Such diffuse gas is generally not visible in emission, and its metal enrichment together with its ionisation state is typically studied through absorption spectroscopy \citep[][]{Fan2023}.
This technique allows us to identify damped Lyman-$\alpha$ (Ly$\alpha$) absorbers (DLAs), defined as absorption systems with a neutral hydrogen column density ($N_{\rm HI}$) so high ($N_{\rm HI} > 2 \times 10^{20}$\,cm$^{-2}$) that their absorption profile is saturated and off-resonance absorption of photons falling in the Lorentzian damping wings becomes detectable \citep[][]{Wolfe1986}.
At these high neutral hydrogen densities, the gas is self-shielded and most of the abundances can be computed from measuring the column density of low-ionisation elements with only small ionization corrections needed \citep[][]{Viegas1995}, making DLAs ideal candidates to study the metal enrichment of the more diffuse gas in the Universe \citep[e.g.,][]{Wolfe2005}.

At intermediate redshifts ($2 < z < 4$), multiple studies have revealed very metal-poor DLAs with enhanced $\alpha$-element enrichment, similar to the composition of the most metal-poor stars in the Milky Way \citep[][]{Cooke2011, Saccardi2023, Welsh2024}.
During reionization ($z \gtrsim 5$), these metal-poor systems are expected to be more frequent, since gas would have generally had less time to be enriched by subsequent stellar populations \citep[][]{Becker2011, Becker2012, Poudel2018, Poudel2020}. 
Similar metal-poor sites, where carbon and oxygen abundances are considered too low to cool gas efficiently to form stars of solar masses or below, have been detected even at $z\approx6$ \citep[][]{Bosman2026}.

However, during reionization, at $z \gtrsim 6$, neutral hydrogen density in the IGM is so high that Gunn-Peterson troughs begin to emerge \citep[][]{GunnPeterson1965}, complicating measurements of hydrogen abundances.
Metal enrichment can still be studied thanks to the presence of metal line absorbers at wavelengths longer than Ly$\alpha$, where the quasar's continuum is unaffected by the neutral hydrogen absorption and can be accurately modelled. 
The identification of DLA analogue systems at these high redshifts mainly relies on the detection of \ion{O}{i} absorption lines, usually associated with the presence of other low-ionisation elements such as \ion{C}{ii}, \ion{Si}{ii} and \ion{Mg}{ii} \citep[][]{Becker2011, Becker2019, Sebastian2024}.
By comparing the abundances with predictions from semi-analytical models \citep[][]{Saccardi2023, Vanni2023, Rusta2025}, observations of intervening absorbers during reionization have found several candidates of Pop~III-enriched gas \citep[e.g.,][]{Christensen2023, Sodini2024, Higginson2025, Li2026}.
However, these models are typically based on stellar evolution calculations that predict the chemical yields of a stellar population as a function of their initial mass and metallicity. 
Some of them additionally account for gas dilution and enrichment from subsequent stellar populations, highlighting how an increasing contribution from Pop~III enrichment can lead to a larger scatter in abundance ratios \citep[][]{Vanni2024}.
Nevertheless, these approaches generally do not provide a self-consistent treatment of the spatial distribution and mixing of metals, nor of the impact of different star formation histories on the metallicity of the absorbing gas.

Exploring galaxies during reionization with \textit{JWST} has provided the opportunity to directly connect the properties of gas in absorption with those of galaxies in emission.
At $z \approx 6$, wide-field spectroscopic surveys in quasar fields like EIGER \citep{Kashino2023} and ASPIRE \citep{Wang2026} allow the distribution of galaxies close to quasar absorption lines to be measured. 
Similar observations has been used to show how one of the potential Pop~III-enriched absorbers identified in \citet{Sodini2024} resides close to an overdensity of galaxies, whose Lyman-Werner flux might have delayed star formation in the surrounding haloes \citep[][]{Li2026}.
{\textit{JWST}} spectroscopic observations have further extended the analysis of metal absorbers within the galaxies themselves \citep{Zhu2026, Pollock2026, Nakane2026}. 
However, independent constraints derived from the same datasets have reached different conclusions, partly reflecting the different methodologies adopted and the large uncertainties coming from medium-resolution spectroscopy.
While all studies find evidence that the key conditions for baryon cycling were already established in the first several hundred million years after the Big Bang, only one of them identifies a strong Pop~III candidate \citep[][]{Pollock2026}.
However, chemical enrichment in the early Universe is not homogeneous, and alongside detections of galaxies already enriching their surroundings at $z\approx7$, neutral sites where the metallicity is extremely low ($\log (Z/\Zsun )<-4$) have also been observed in absorption at the same redshift \citep[][]{Simcoe2012}, corroborating the idea that the first generation of stars could be detected even at later cosmic times.

To provide a more physically consistent framework to identify Pop~III from observations of gas in absorption, it is necessary to quantify how the physical processes missing in semi-analytical models may affect the interpretation of observational results. Therefore, in this study we compare the abundance patterns observed in absorption to theoretical predictions from the \thz radiation-hydrodynamic simulations \citep[]{Kannan2025}.
By combining high spatial resolution in the gas with detailed subgrid models, we can study how metal abundances are shaped by the complex interplay between star formation, stellar feedback, inflows and outflows of gas, and metal mixing.
Since Pop~III stars are not included in the fiducial \thz simulations, the main goal of this study is to assess whether the abundance patterns of observed Pop~III-enriched gas candidates can be reproduced by a hydrodynamical model in which enrichment arises exclusively from later stellar populations. 
By comparing the observed abundance patterns with those predicted by the simulations, we aim to place more stringent constraints on chemical abundance criteria to identify Pop~III-enriched gas. However, we further explore an extension to the \thz model which includes Pop~III enrichment, to quantify how this may change the abundance patterns.

The paper is organised as follows: in Section~\ref{sec:method} we present the simulations used in this study, followed by the methods to identify systems similar to the observed absorbers. 
We then present metal abundance ratios and compare our results to observations in Section~\ref{sec:results}, followed by a discussion on the nature of our metal abundance distributions in Section~\ref{sec:discussion}.
Finally, we present our conclusions in Section~\ref{sec:conclusions}. In this work, we assume the \citet[][]{PlanckCosmo2016} cosmology, with $H_0 = 100 h \, \rm{km}\,\rm{s}^{-1}\,\rm{Mpc}^{-1}$, $h = 0.6774$, $\Omega_m = 0.3089$, $\Omega_{\Lambda} = 0.6911$ and $\sigma_8 = 0.8159$.

\section{Modelling DLA-like absorbers}
\label{sec:method}

\subsection{The THESAN-ZOOM simulations}
\label{sec:thz}

In this work, we analyse a subset of the \thz simulation suite.
\thz is a set of high-resolution zoom-in radiation-hydrodynamic simulations first presented in \citet{Kannan2025}.
It consists of $14$ different zoom-in regions, each focused on a target halo of different mass selected at $z = 3$ from the \textsc{thesan-dark-1} simulation, the dark-matter (DM) only version of the large-volume radiation hydrodynamic simulation \textsc{thesan} \citep[]{Kannan2022, Garaldi2022, Smith2022, Garaldi2024}.
Each zoom-in simulation resamples, at $4\times$ higher resolution ($m_{\rm gas} \approx 9.09 \times 10^{3}\,\Msun$), the Lagrangian region of the material that ends up within $4$~virial radii of the target halo at $z=3$, and evolves it from $z=127$.
Target haloes are selected to sample a wide mass range ($10^8 - 10^{13}\,\Msun$ at $z=3$), and throughout this paper we will label each zoom-in region by the base-10 logarithm of its target halo mass at $z=3$, as adopted in \citet{Kannan2025}. For example, m13.0 denotes a zoom-in region centred on a halo with a mass of $10^{13}~\Msun$ at $z=3$.

Hydrodynamic equations are solved using a quasi-Lagrangian method implemented in \textsc{arepo-rt} \citep{Kannan2019}, a radiation hydrodynamic extension of \textsc{arepo} \citep{Springel2010, Weinberger2020} with the implementation of ``node-to-node'' communication presented in \citet[][]{Zier2024}.
Stars are formed stochastically in gas cells that are self-gravitating and Jeans unstable with densities above $10\,\text{cm}^{-3}$. In our fiducial model, we do not form Pop~III stars; however, this will be tested later on.
Star particles have masses comparable to the target gas mass and represent stellar populations with Chabrier initial mass function \citep[]{Chabrier2003} with masses ranging from $0.1\,\Msun$ to $100\,\Msun$.
The simulation tracks the evolution of $9$ chemical species (H, He, C, O, N, Ne, Mg, Si and Fe) produced by stellar populations and ejected into the gas by galactic processes. A key difference between these zoom-in simulations and the parent \textsc{thesan} simulation is the treatment of star-forming gas.
While the latter employs an effective equation of state to describe unresolved multiphase gas \citep[][]{SpringelHernquist2003}, in \thz the high resolution allows the multiphase structure of the interstellar medium (ISM) to be followed directly. This is accounted for through a custom version of the Stars and MUltiphase Gas in GaLaxiEs (SMUGGLE) model \citep[]{Marinacci2019}, as described in \citet{Kannan2020, Kannan2025}.
Stars with mass $\textrm{M}_\ast > 8\,\Msun$ explode as SN type II (SN~II), and energy, momentum and mass are injected into the closest gas cells together with dust and metals \citep{Vogelsberger2013, McKinnon2016}, according to the yields presented in \citet{Portinari1998}.  
In addition, the simulations also account for SN type Ia (SN~Ia) using metal yields from \citet{Thielemann2003} as reported in \citet{Travaglio2004} and stellar winds from OB and asymptotic giant branch stars (AGB) with metal yields from \citet{Karakas2010}.
Early stellar feedback, which has been proven essential in our cosmological zoom-in simulations to match the observed stellar masses \citep[][]{Kannan2025}, is further included in \thz. 
Although implemented as an empirical prescription, early stellar feedback serves to compensate for the effects of physical processes not explicitly modelled, such as cosmic rays, Ly$\alpha$ radiation pressure and magnetic fields, as well as to account for uncertainties in subgrid physics, for example, those associated with stellar winds from metal-poor stars.
This feedback operates during the first $5 \,\text{Myr}$ after star-particle formation and injects momentum into the surrounding gas, thereby providing a mechanism for regulating feedback before the explosion of the first SNe.

The radiation content of each cell is described across seven bins of frequency from infrared to \ion{He}{ii} ionising radiation, as specified in \citet[][]{Kannan2025}.
We account for contributions arising from inside and outside each zoom-in region; the former is computed using the BPASS stellar population synthesis model \citep{Eldridge2017}, which accounts for binary star evolution, while the latter is derived from the parent \textsc{thesan} simulation.
The radiation field and its coupling to gas are evolved using a moment-based approach \citep[]{Levermore1984}.
To reduce the computational cost, the simulations adopt a reduced speed of light, $\tilde{c} = 0.01\,c$, where $c$ is the speed of light in vacuum. 
Within each hydrodynamical step, the radiation field is sub-cycled $16$ times, allowing its evolution and interaction with the gas to be followed on shorter timescales than those set by the hydrodynamical evolution.
Radiation is coupled to gas, in particular, the simulations track different states of hydrogen (\ion{H}{i}, \ion{H}{ii}, H$_2$) and helium (\ion{He}{i}, \ion{He}{ii}, \ion{He}{iii}), used to compute primordial elements cooling.
Metal cooling is implemented based on \textsc{cloudy} tables, as described in \citet{Vogelsberger2013}.
We additionally include the evolution of dust with an updated implementation of the model presented in \citet{McKinnon2016,McKinnon2017}.
The dust mass growth in each gas cell is regulated by the abundance of five chemical elements, (C, O, Mg, Si and Fe) and gas temperature, while destruction accounts for processes of sputtering and shocks \citep{Kannan2025,Garaldi2026}.

The high resolution of the \thz simulations together with the description of cooling and heating of the gas at low temperatures ($T\lesssim10^4$\,K), implemented considering molecular hydrogen, dust-gas collisions, and metal lines \citep{Kannan2025} enables a detailed treatment of gas inside and around galaxies, making these simulations an ideal framework for our purposes.

\subsection{DLA-like gas selection}
\label{sec:selection}

\begin{figure*}
    \centering
    \includegraphics[width=1\linewidth]{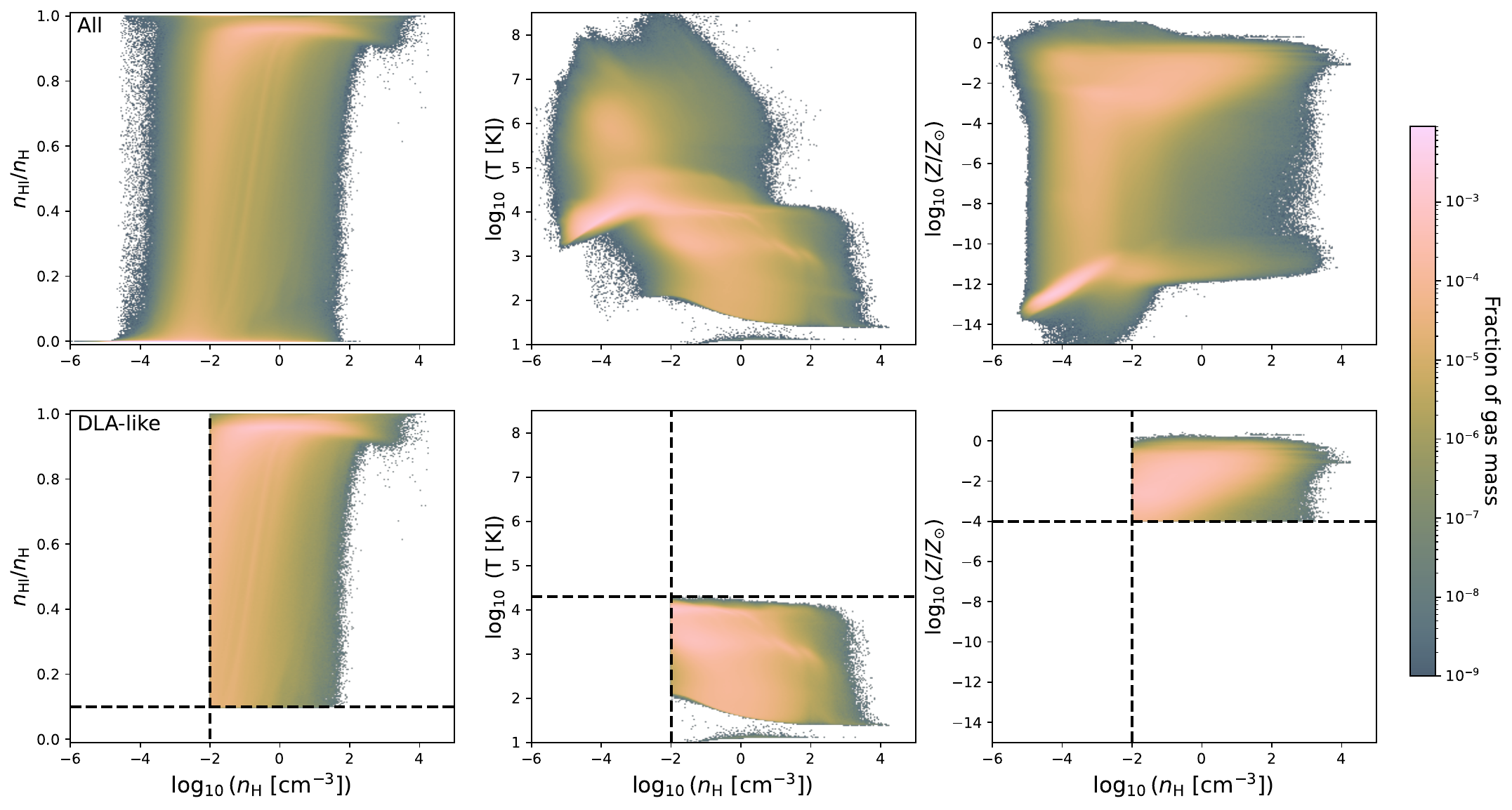}
    \caption{Comparison between properties of the overall gas (upper panels) and DLA-like gas cells at $z=6$ (bottom panels), with colours indicating the gas mass fraction of the considered subset.
    DLA-like gas cells are selected to have a large neutral fraction ($n_{\rm HI}/n_{\rm H}>0.1$), high density ($n_{\rm H}>10^{-2}$\,cm$^{-3}$), low temperature ($T<2\times10^4$\,K), and a metallicity threshold ($\log (Z/\Zsun)> -4$) to ensure the possibility of detecting metal absorbers. While constraints on neutral fraction, density and metallicity exclude a large fraction of the overall gas cells population, the temperature limit is almost always satisfied.}
    \label{fig:dla_cuts}
\end{figure*}

Observational constraints on metal abundances of gas during the epoch of reionization are primarily derived from the column density ratios of low-ionisation elements in intervening self-shielded absorption systems along quasar sight lines. 
To directly compare our simulations with these observations, we would ideally need to generate mock spectra by casting lines of sight through our simulations. However, as we are analysing zoom simulations, the volume over which we simulate the gas at high resolution is limited to gas in the vicinity of galaxies.
These shorter lines of sight would have a lower probability of intersecting observable absorbers and their resulting properties might be affected by artificial discontinuities at the boundaries between the high- and low-resolution regions.
To avoid these numerical artifacts while still enabling a comparison with observational measurements, we adopt a different approach.
Instead of having random lines of sight through the simulations to identify absorbers, we select individual gas cells that could potentially make up metal-enriched DLAs.
Although this method is less precise than a like-for-like comparison of individual observed absorption features, it is still a step forward compared with previous theoretical analyses of these Pop~III candidate absorbers, which have so far been centred on comparisons with semi-analytic models \citep[e.g.][]{Vanni2024}.
The latter, in fact, do not account for inhomogeneous metal distribution around galaxies driven by galaxy evolution processes captured in hydrodynamical simulations, such as bursty star formation histories, the contribution of multiple enrichment channels and metal mixing. We therefore proceed in this fashion and leave a full study of synthetic absorption lines to a future work.

We construct our selection function for DLA-like gas based on observational constraints of their properties.
Studies of very metal-poor DLAs at redshifts $z \sim 2$ -- $3$ indicate that the neutral hydrogen density of these systems is $\log (n_{\rm HI}/\rm{cm}^{-3}) \approx -1.6$ \citep[e.g.,][]{Cooke2015}. Their characteristic temperatures are typically $\sim 10^3$\,K, although some systems can reach $10^4$\,K \citep[e.g.,][]{Pettini2012, Carswell2012, Noterdaeme2021}.
The metallicity of DLAs observed at $z<5$ shows a decline while approaching higher redshift, with an average value of $\log(Z/\Zsun)\approx-2$ at $z=5$ \citep[][]{Rafelski2014}, and very metal-poor systems at $z\geq 3$ showing iron abundances close to $-3$ \citep[][]{Cooke2011, Dutta2014}. At $z>5$, because of the large density of absorption lines in the Ly$\alpha$ forest, evaluating the \ion{H}{i} column density of single absorption systems becomes challenging. 
Nevertheless, new methods based on the detection of low-ionisation metal transitions have been successfully employed to identify DLA analogues \citep[][]{Becker2011}.
The absorption systems at $z\simeq6$ presented in \citet[][S24]{Sodini2024}, the primary observational dataset to which we compare our simulations in this study, come from the analysis of the E-XQR-30 sample \citep[][]{DOdorico2023_xqr}, an ensemble of $42$ spectra of luminous quasars during reionization observed with X-SHOOTER \citep[][]{XSHOOTER}.
In particular, absorption systems are identified along quasar lines of sight through the detection of \ion{O}{i} absorption accompanied by \ion{C}{ii} and \ion{Si}{ii} transitions at the same redshift. 
However, the relation between \ion{O}{i} and \ion{H}{i} column densities is complicated by the overall metallicity of the gas, which can vary substantially depending on the galactic environment.
As we demonstrated in \citet{Pruto2026}, detectable \ion{O}{i} absorbers usually trace Lyman-Limit Systems (LLSs) or DLAs, with associated \ion{H}{i} column densities that could vary up to $2$ orders of magnitude, depending on the metallicity and the neutral fraction of the gas cloud.
In \citetalias{Sodini2024}, \ion{H}{i} column densities are evaluated only for a subset of systems representing proximate DLAs, from the analysis of the corresponding damping wing profiles.
Although the \ion{O}{i} absorbers in their catalogue exhibit lower column densities, up to 1 dex lower than those observed in metal-poor DLAs at lower redshift, the inferred \ion{H}{i} column density values are consistent with DLA-like systems. 
This finding corroborates the idea that the metallicity of gas probed by high-redshift DLA analogues tends to be lower ([Fe/H] $\lesssim -2.5$, with some systems showing values $\lesssim -3$) than the ones found at lower redshift \citepalias[see Fig. 2 in][]{Sodini2024}. 

In our simulations, we impose a metallicity threshold to identify the gas cells that could host similar observable metal absorbers. 
However, instead of applying the selection based on iron abundance, we use the overall metallicity of each gas cell. 
This choice avoids introducing a bias toward gas that has been preferentially enriched by SN~Ia events, the dominant source of iron enrichment, while more faithfully reproducing the selection of high-redshift observations. 
In fact, the observed absorbers are selected by the detectability of low-ionisation transitions of O, C and Si, rather than through Fe abundances.
We have also verified the presence of an approximately linear relation between the logarithm of gas metallicity and [Fe/H] in our simulations, with the adopted threshold of $\log_{10}(\rm{Z}/\Zsun) > -4$ approximately corresponding to [Fe/H]$\gtrsim-6$.

Based on these observational constraints, in our simulations we select gas cells that simultaneously satisfy these conditions:
\begin{enumerate}[leftmargin=*, itemindent=0pt,  itemsep=4pt]
    \item $n_{\rm HI}/n_{\rm H} > 0.1$
    \item $n_{\rm H}> 10^{-2}$\,cm$^{-3}$
    \item $T< 2 \times 10^4$\,K
    \item $\log_{10}(\rm{Z}/\Zsun) > -4$
\end{enumerate}

The applied selection criteria can be visualised in Fig.~\ref{fig:dla_cuts}. 
The 2D histograms in the upper panels show the distribution of all the gas cells in the analysed zoom-in regions at $z=6$, while the histograms in the bottom panels represent the corresponding distribution for the subset of gas cells satisfying our selection criteria and therefore identified as DLA-like.
As shown in the figure, the adopted cuts on neutral fraction, density and metallicity exclude a substantial amount of the simulated gas cells, while the temperature criterion is almost always satisfied by gas with the required neutral content.
These DLA-like gas cells occupy between $0.01$ and $0.1\%$ of the total volume in the zoom-in regions, with no significant evolution between $z= 5 - 8$ and higher values generally presented by denser galactic environments.
In these environments, earlier structure formation is expected, resulting in more high density neutral filamentary structures between haloes and an overall higher metal enrichment.
Their spatial distribution can be observed in Fig.~\ref{fig:first_map} and Fig.~\ref{fig:second_map}, and will be discussed more in depth in Section~\ref{sec:metab}.

\begin{figure}
    \centering
    \includegraphics[width=1\linewidth]{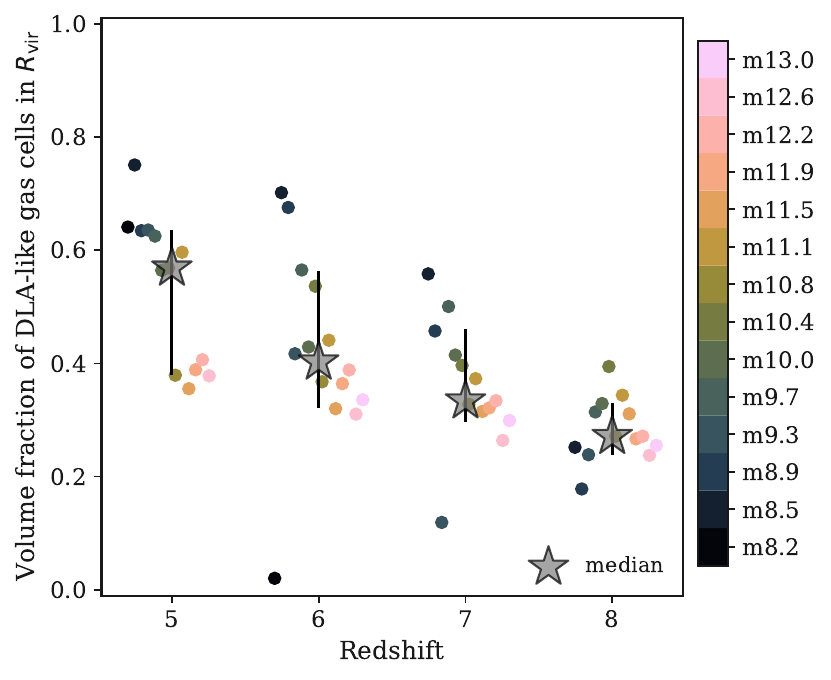}
    \caption{Volume fraction of DLA-like gas cells inside the virial radius of haloes, as expressed in Eq.~\ref{eq:inrvir}. Results for individual zoom-in regions are represented by points coloured according to the simulated target halo and slightly shifted on the $x$-axis for visualisation purposes. Stars mark the median value at each redshift, with errorbars defining the $16^{\rm th}$ and $84^{\rm th}$ percentiles. The volume fraction of DLA-like gas cells bound to haloes increases at low redshifts, probably due to the effects of reionization and cosmic expansion.}
    \label{fig:volinrvir}
\end{figure}

\begin{figure*}
    \includegraphics[width=1\linewidth]{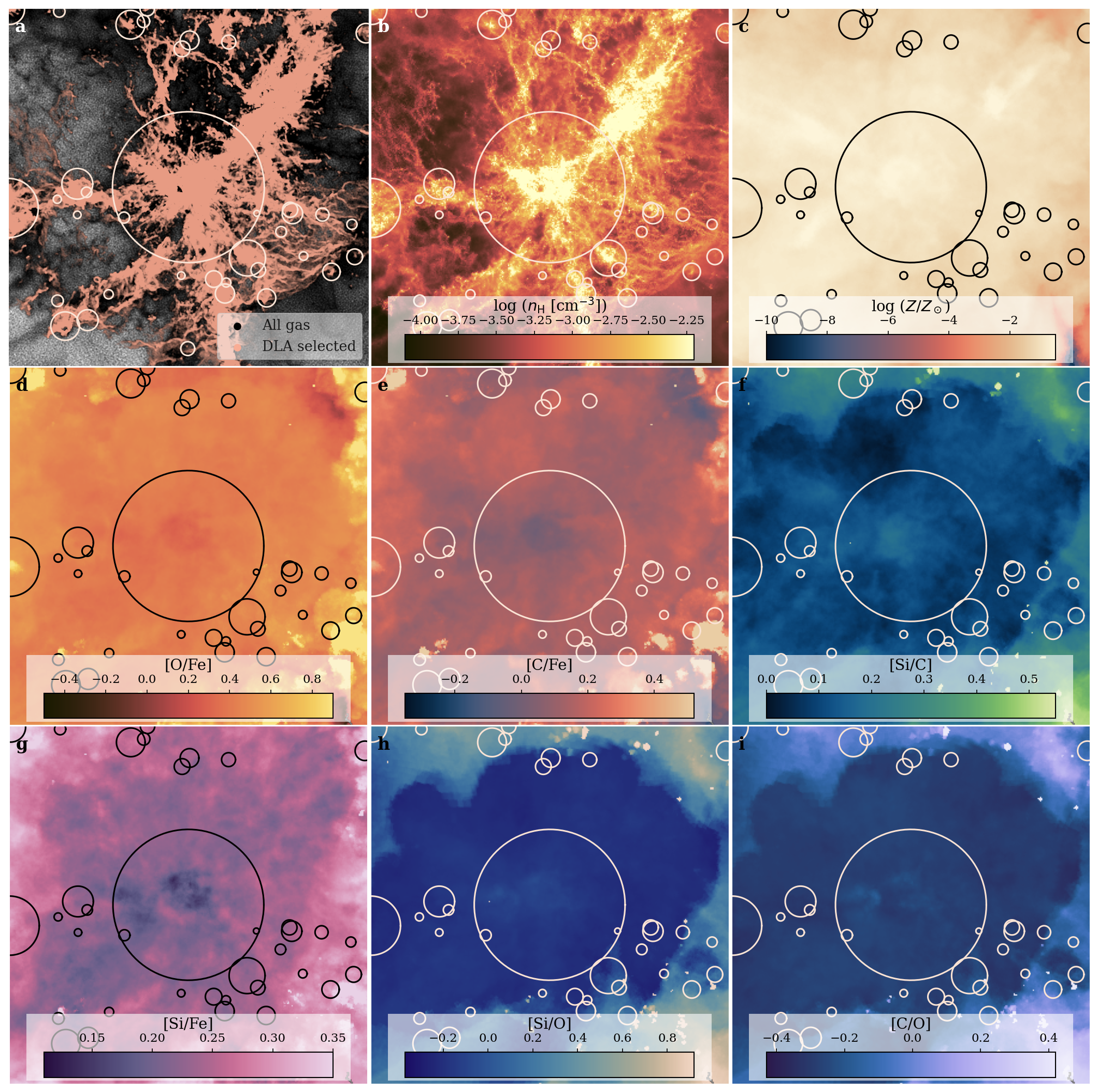}
    \caption{Maps of gas properties in a cubic region of size $300$\,ckpc/h around the target halo of m12.6 ($\textrm{M}_h = 6.7 \times 10^{11}~\Msun$) at $z=6$. All the quantities shown represent mass-weighted average values along the line of sight inside the selected cubic box. The circles mark the virial radii of haloes with mass above $10^8~\Msun$. From the upper left: (\textit{a}) Position of DLA-like gas cells with respect to the overall gas distribution. (\textit{b}) Hydrogen density. (\textit{c}) Metallicity. (\textit{d})--(\textit{i}) Metal abundance ratios: [O/Fe], [C/Fe], [Si/C], [Si/Fe], [Si/O] and [C/O], respectively. Relationships between spatial distributions of different abundances can be noted: [O/Fe] displays a pattern similar to [C/Fe], while [Si/C], similarly to [Si/O], anticorrelates with [C/Fe]. The different timescales associated with distinct enrichment channels are needed to explain these relations.}
    \label{fig:first_map}
\end{figure*}

\begin{figure*}
    \includegraphics[width=1\linewidth]{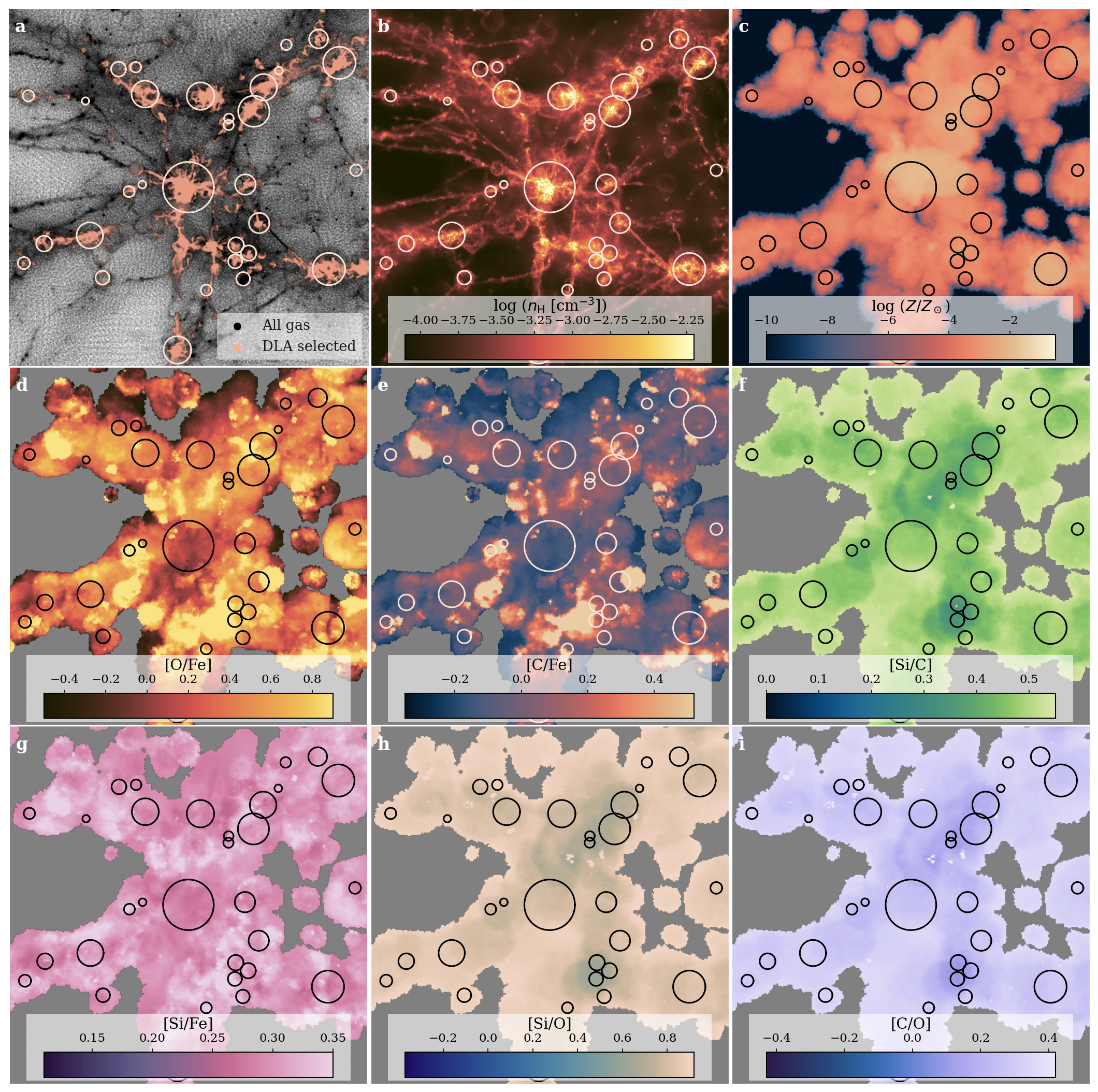}
    \caption{Maps of gas properties as in Fig.~\ref{fig:first_map} in a cubic region of size $300$\,ckpc/h around the target halo of m11.1 ($\textrm{M}_h = 1.2 \times 10^{10} \ \Msun$) at $z=6$. The gray regions indicate non-enriched areas with $\log(Z/\Zsun)<-9$. In contrast with the more homogeneous distribution shown in Fig.~\ref{fig:first_map}, metal enriched regions around this lower-mass galaxy, approximately trace the cosmic structures linking the surrounding haloes.}
    \label{fig:second_map}
\end{figure*}

In Fig.~\ref{fig:volinrvir}, we show the volume fraction of DLA-like gas cells inside the virial radius of simulated haloes, as indicated in Eq.~\ref{eq:inrvir}.
\begin{equation}
\label{eq:inrvir}
    f = \frac{\sum_i V_i \ | \ i \in R_{200, c}}{\sum_j V_j}
\end{equation}
where $R_{200, c}$ is defined as the virial radius of haloes, i.e., the radius enclosing a mean density of $200$ times the critical density of the Universe at the corresponding epoch, and $V_i$ ($V_j$) represents the volume of the $i$-th ($j$-th) DLA-like gas cell.
The condition $i \in  R_{200, c}$ is satisfied if the centre of mass of the $i$-th gas cell lies within the virial radius of haloes identified in the simulations ($M_h \gtrsim 10^7~\Msun$).
As shown in Fig.~\ref{fig:volinrvir}, a fraction between $0.2$ and $0.7$ of the total volume occupied by the selected DLA-like gas cells resides within the virial radius of haloes.
This fraction increases with cosmic time, probably also influenced by the pseudo-growth of the virial radii following cosmic expansion \citep[e.g.,][]{Diemer2013}.
Interestingly, the zoom-in regions hosting less dense galactic environments, marked by darker points, exhibit a stronger evolution and higher values at lower redshift.
This might reflect the rapid evolution of metal enrichment in these regions over the considered redshift range.
Moreover, the intergalactic medium in these low-density regions tends to reionize later than in high-density ones, probably playing a role in decreasing the amount of neutral DLA-like gas particles outside the virial radius between $z=8$ and $z=5$.

\subsection{Metal Abundance Ratios}
\label{sec:metab}

To compare our simulations with observations of high-redshift absorption systems, in the selected DLA-like gas cells we measure metal abundance ratios.
Specifically, these are defined as the ratio between the number densities of two chemical elements, $X$ and $Y$, relative to the solar abundances from \citet{Asplund2009}, as represented by Eq.~\ref{eq:si_o_ab}.

\begin{equation}
    [X/Y] = \log \left( \frac{n_X}{n_Y} \right)  - \log \left( \frac{n_X}{n_Y} \right)_{\astrosun} \, .
    \label{eq:si_o_ab}
\end{equation}

To visualise the spatial distribution of these abundances and how they relate to galaxies, we show in Fig.~\ref{fig:first_map} and Fig.~\ref{fig:second_map} maps of gas properties over two different cubic regions of size $300$~ckpc/h at $z=6$.
The first figure is centred on the target halo of the zoom-in region m12.6 ($\textrm{M}_h = 6.7 \times 10^{11} \ \Msun$ at $z=6$), while the second is centred on the target halo of m11.1 ($\textrm{M}_h = 1.2 \times 10^{10} \ \Msun$ at $z=6$). 
The overall metallicity around the high-mass halo shows values of $\approx 3$ orders of magnitude higher than in the lower-mass halo, as can be seen from panel (\textit{c}) of the two figures.
The difference in the environments around the two haloes can be directly appreciated by observing the patterns formed by the spatial distribution of hydrogen, shown in the second panel.
Around the higher-mass halo, gas forms denser, large, metal enriched filaments, which provide an ideal location for DLA-like gas, indicated by the coloured points in panel (\textit{a}) of the two figures.
In contrast, the filaments surrounding low-mass haloes tend to be thinner and less dense, which causes a larger fraction of DLA-like gas cells to reside within the virial radius of haloes, marked by circles.
Overall, this is consistent with the quantitative results presented in Fig.~\ref{fig:volinrvir} for DLA-like gas cells from different zoom-in regions.

Focusing on panels (\textit{d})--(\textit{i}), we notice that the distribution of metal abundances appears more homogeneous around the higher-mass halo and it seems to be primarily affected by the stellar activity of the target halo itself, found at the centre of this extended, uniformly enriched bubble.
On the other hand, in the second figure, metal abundance ratios display more patchy structures, probably influenced by the feedback activity of multiple galaxies in the region, which results in enriched areas approximately tracing the cosmic web connecting haloes.
In both cases, similar relationships between abundance patterns emerge, providing a qualitative means of tracing star formation history of the region.
For example, [O/Fe] and [C/Fe] show similar distributions within the same environments, and higher values can be seen outside the target halo. 
This can be due to the effects of the delayed SN~Ia with respect to SN~II. 
In fact, while oxygen and carbon are ejected by short-lived massive stars through SN~II events, common during star-forming phases, iron is mainly expelled into the gas by SN~Ia explosions, which require lower-mass stars with longer main sequence lifetimes. 
Thus, the delay in SN~Ia events might cause iron to be detected more centrally concentrated within the halo, reflecting the different enrichment timescales associated with the two supernovae channels.
A similar argument can be also used to explain the anticorrelation between [C/Fe] and [Si/C] abundances that can be appreciated especially at the centre of the target halo in the first figure. 
SN~Ia explosions, in fact, contribute significantly to the production of both iron and silicon, while leaving the carbon abundance unaffected.
Therefore, the distributions observed in the target halo of m12.6 could be explained by assuming a constant enrichment caused by core-collapse SNe, caused by a long-lasting star formation activity, with the addition of recent powerful SN~Ia explosions.

\section{Comparison with observed absorption systems}
\label{sec:results}

\subsection{Metal abundances in the fiducial \thz model}

\begin{figure*}
    \includegraphics[width=1.8\columnwidth]{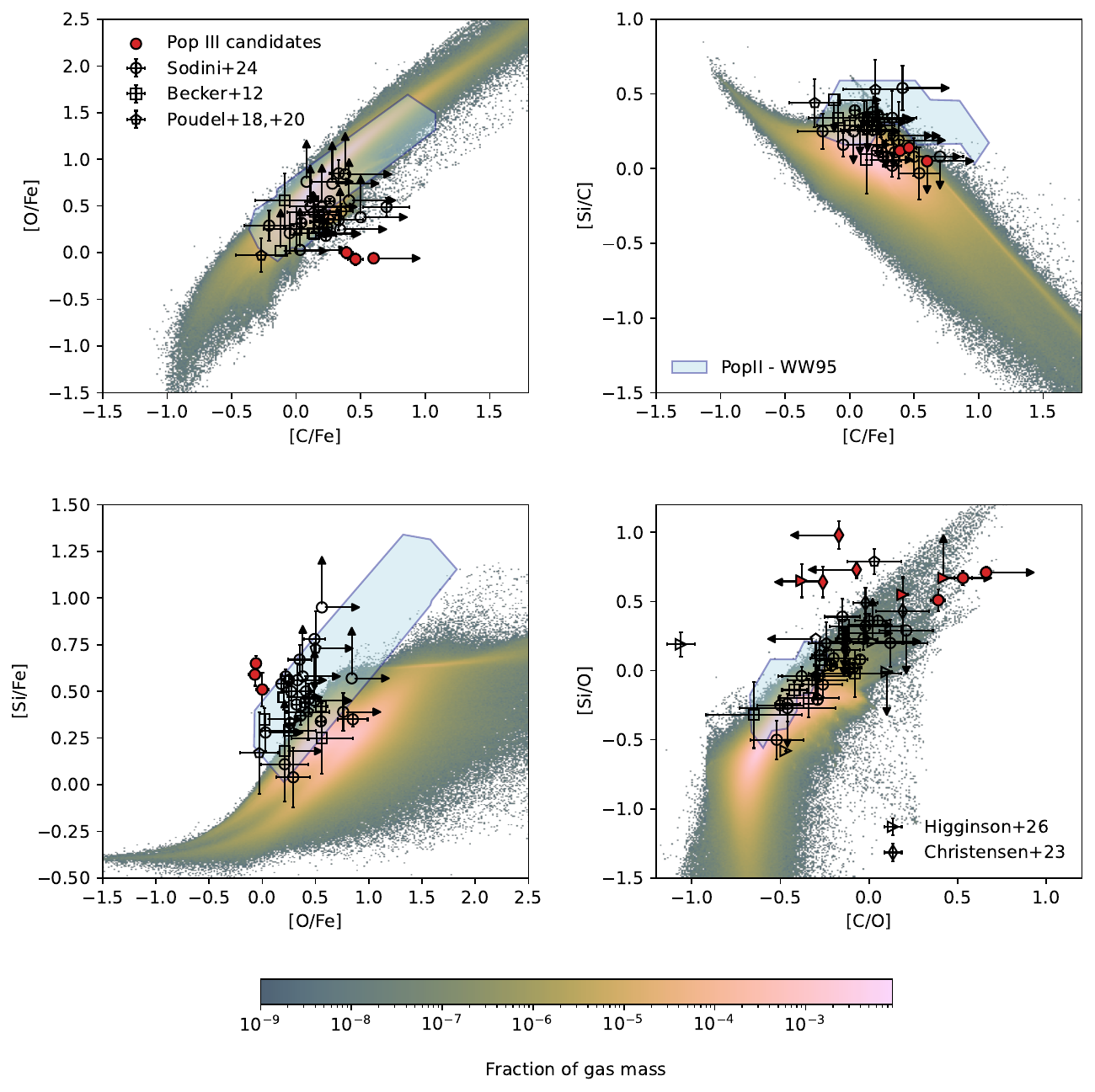}
    \caption{Metal abundances in the selected enriched DLA-like gas cells in the $14$ zoom-in regions of the fiducial \thz simulations at $z=6$. The 2D histograms show the distribution of selected gas cells in our simulations as a function of their metal abundance ratios, with lighter colours marking denser areas. The blue shaded region represents metal abundances from Pop~II yields from \citet{WoosleyWeaver1995}. The data points are obtained from observations of metal absorption systems identified at the end of reionization in quasar absorption spectroscopy. Specifically, we report measurements from \citet{Sodini2024}, \citet{Becker2012} and \citet{Poudel2018, Poudel2020}, marked by circles, squares and pentagons respectively. The arrows indicate abundances for which only upper or lower limits are available. In the last panel, we add measurements from {\textit{JWST}} observations reported in \citet{Higginson2025} with triangles and \citet{Christensen2023} with diamonds. The filled markers show observations previously identified as gas potentially enriched by Pop~III stars when compared to semi-analytical models. Overall, the abundances obtained in our simulations broadly agree with observational data, even if they suggest a systematic underproduction of silicon relative to oxygen in our models. Some of the observed Pop~III candidates in the [Si/O]--[C/O] plane lie in regions where our Pop~II-only model predicts the presence of similar metal absorbers.}
    \label{fig:hist2d_z6}
\end{figure*}

We next compare metal abundance ratios in DLA-like gas cells with those observed in absorption systems during reionization.
The results are presented in Fig.~\ref{fig:hist2d_z6}, where the 2D histograms show the fraction of gas mass at $z=6$, selected as explained in Section~\ref{sec:selection}, as a function of different metal abundance ratios, with lighter colours identifying highly populated regions. We investigated in our simulation how metal abundance ratios evolve from $z=3$ to $z=9$, finding no strong evolutionary trends, as reported in Appendix~\ref{app:redev}.
Clear correlations between abundances emerge: [O/Fe], [C/Fe] and [Si/Fe] are overall positively correlated, as also [Si/O] and [C/O].
In contrast, [Si/C] displays an inverse correlation with [C/Fe], as already observed more qualitatively in Fig.~\ref{fig:first_map} and Fig.~\ref{fig:second_map}.
These patterns are mainly due to the different formation channels of these elements. 
While C and O are predominantly produced by core-collapse SNe, a significant fraction of Fe and Si is ejected through SN~Ia explosions, as already discussed in Section~\ref{sec:metab}. 
The data points show measurements of metal abundance ratios in absorption systems primarily identified through the detection of low-ionisation metal transitions during the epoch of reionization.
Specifically, we report measurements of $z\approx 6$ systems presented in \citet{Sodini2024} with circles and in \citet{Becker2012} with squares.
We also add results obtained from metal-poor ([Fe/H]$\leq -2$) neutral hydrogen absorbers (log ($N_{\rm HI}$/[cm$^{-2}$]) $\geq 20.1$) in the redshift range $4.5\leq z \leq 5.5$ described in \citet{Poudel2018, Poudel2020}, marking them with pentagons.
In the fourth panel, we add  results from {\textit{JWST}} observations, showing the [Si/O] and [C/O] abundances of absorption systems
at $z\approx 6$ \citep[triangles,][]{Higginson2025} and at slightly higher redshift ($6.2 \leq z \leq 7.4$) \citep[diamonds,][]{Christensen2023}.

Overall, the abundance patterns predicted by our simulations are broadly consistent with observational data, with particularly good agreement in the [O/Fe]--[C/Fe] relation shown in the first panel.
However, in the third panel especially, our [Si/Fe] values seem to be systematically underestimated with respect to observational data.
This tension could be reduced by multiplying the silicon abundances in our simulations by a factor of $\sim2$, which would also result in an improved agreement with observations in the [Si/C]--[C/Fe] (second panel) and [Si/O]--[C/O] relation (fourth panel).
The abundances considered are only those in the gas phase and not in the dust phase, consistently with observations.
However, we also note that since silicon is one of the elements that is trapped into dust particles, assuming different dust models might slightly affect the results, but a change by a factor of $2$ is unlikely to be recovered through a different model of dust.

The same conclusion about silicon is true even when comparing our abundances to predictions from Pop~II yields, such as those presented in \citet{WoosleyWeaver1995} indicated by the light blue shaded regions. 
Semi-analytic models incorporating yields such as these typically provide predictions on chemical enrichment based on stellar evolution models of stars in a wide range of masses and metallicities. 
Some of them include a certain level of gas dilution \citep[e.g.,][]{Vanni2024}, taking into account that metals ejected by stars will mix with pristine gas in the surrounding of the star forming region.
However, they do not account for the complex star-formation history and feedback processes that favour the expulsion, mixing and infall of gas at different metallicities, for which we need a hydrodynamic self-consistent description.
Among these models, the ones from \citet{Vanni2024} include Pop~III stars and subsequent Pop~II stars. 
Their main results show that with an increasing Pop~III pollution the scatter in metal abundances increases.
Similar models are usually employed to identify Pop~III-enriched gas among observed absorbers.

In Fig.~\ref{fig:hist2d_z6}, we mark candidates selected through comparison with such semi-analytic models with red-filled markers, noticing that, in general, all these systems display high [Si/O] values.
Among these, absorbers with lower [C/O] ratios, such as those reported in \citet{Christensen2023}, could represent signatures of enrichment by Pop~III pair-instability supernovae (PISN).
These are high-mass stars ($140 \lesssim M/\Msun \lesssim 260$) in which the loss of pressure due to pair-production results in explosive oxygen and silicon burning \citep[][]{Barkat1967}, leaving in the gas very high [Si/O] at low [C/O] abundances.
However, no robust detection of such systems has been made so far, as it would require values of [C/O]$\approx -0.65$ \citep[][]{Vanni2024}, while only compatible [C/O] upper limits have been observed.
Analytical models also agree that gas polluted by Pop~III SN or hypernovae explosions should contribute to a higher dispersion in metal abundance ratios \citep[e.g.,][]{HegerWoosley2010, Vanni2024}.
Three absorption systems in \citetalias{Sodini2024} have been shown to be compatible with gas enriched by Pop~III stars with a percentage ranging from 60 to 90\%, and two detections from \citet{Higginson2025} show [Si/O]--[C/O] abundances consistent with Pop~III hypernovae.
When comparing these detections with our models, we notice that some of the potential Pop~III-enriched systems lie in regions of high [Si/O] and [C/O] ratios, where a small fraction of metal-enriched DLA-like gas cells in our simulations also reside. 
A more detailed discussion on this comparison is presented in Section~\ref{sec:possiblepopIII}.

\subsection{Comparison with $z>7$ observations}

\begin{figure*}
    \includegraphics[width=1.\linewidth]{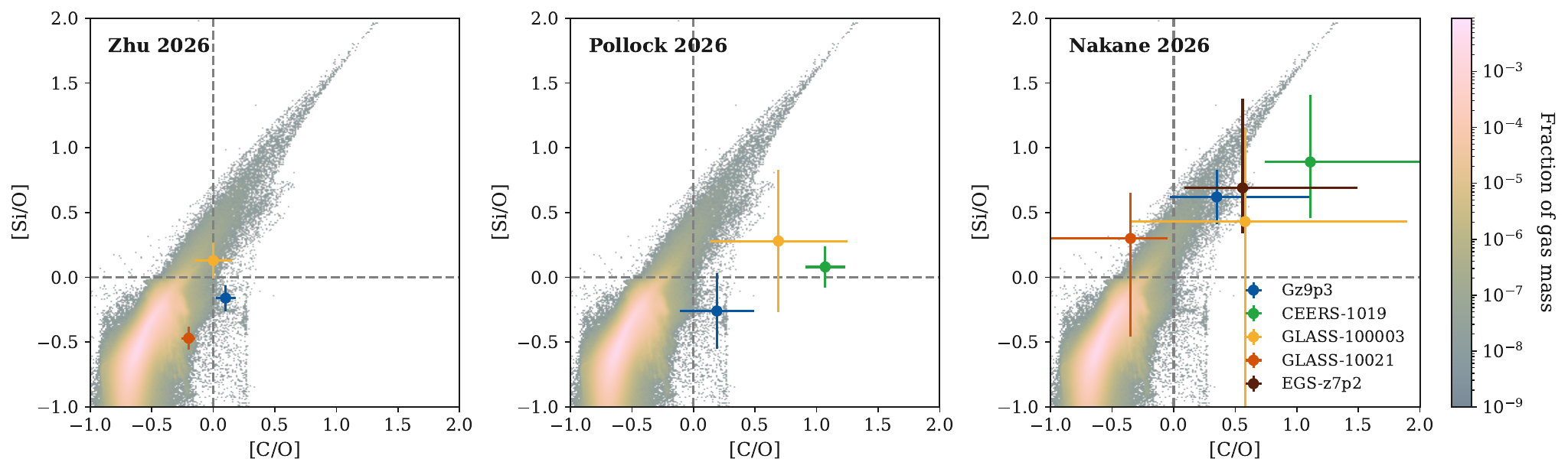}
    \caption{[Si/O]--[C/O] relation in the selected enriched DLA-like gas cells of the analysed $14$ zoom-in regions at $z=8$. 2D histograms show the distribution of the gas cells in the fiducial \thz simulations, with lighter colours marking denser regions. The dashed lines indicate solar abundances. In each panel we compare our distribution to measurements of high-redshift ($7.2 \leq z \leq 10.6$) absorption systems obtained by different studies: \citet{Zhu2026}, \citet{Pollock2026} and \citet{Nakane2026}, respectively. The observational results are represented as coloured points, with each colour associated with a different high-redshift source. Overall, metal enrichment was present in the early Universe, but given the large uncertainties and the different results obtained by different analyses, understanding how much of it is due to Pop~III stars is currently challenging. CEERS-1019 seems to be an ideal candidate for follow-up investigations with higher resolution spectroscopy. }
    \label{fig:highzabs}
\end{figure*}

\textit{JWST} observations have extended the study of metal absorbers to previously unexplored redshifts, with multiple absorption systems detected in the spectra of luminous galaxies at $z\geq7$.
This progress has been mainly enabled by the Cycle 4 program SPectroscopic Ultra-deep Reionization-era Survey \citep[SPURS; GO-9214, PIs: Mason, Stark;][]{Chen2026}, which collected spectroscopic data for more than $150$ galaxies at $z>5$, with an analysis of the identified absorption lines presented in \citet{Vasan2026}. 
Unlike the systems previously introduced, these are not intervening absorbers, but they constitute absorbers within the galaxy itself, probably arising from the CGM \citep[][]{Pollock2026, Vasan2026}.
These observations demonstrate that chemically enriched gas, potentially associated with galactic feedback was already present in the early Universe \citep[][]{Vasan2026, Zhu2026}.
We compare the [Si/O]--[C/O] abundance ratios inferred from these observations with predictions from our fiducial \thz simulations at $z=8$, as represented in Fig.~\ref{fig:highzabs}.
We note that a more accurate comparison could be made by selecting only gas cells in the CGM around galaxies with properties comparable to those of the observed sources, though we leave this refinement to future work.
Here, in each of the three panels in the figure, we show the same 2D histogram of the distribution of our metal enriched DLA-like gas cells at $z=8$, similarly to Fig.~\ref{fig:hist2d_z6}. 
We then compare this distribution with measurements presented in \citet{Zhu2026}, \citet{Pollock2026} and \citet{Nakane2026}, obtained from the collected SPURS data, shown as coloured points in the first, second and third panels, respectively.
Each data point corresponds to a metal absorption system detected within the galaxy, with colours indicating the different sources, as specified by the label.
A given system may therefore appear in more than one panel when it has been independently analysed by multiple studies. 

However, despite being based on the same observational data, Fig.~\ref{fig:highzabs} illustrates how different approaches in the analysis could lead to substantially different interpretations.
For example, in \citet{Zhu2026} the analysed systems are classified by the authors as galaxies enriched by mainly subsequent stellar populations, with only Gz9p3 showing the possibility of a mild ($<30\%$) contamination from Pop~III stars. 
When comparing these observations to the \thz simulations, all of the three absorbers reside in the regions reproduced by our Pop~II-only model, corroborating the idea that these galaxies mainly contain subsequent stellar populations.
However, one of them, the one associated with GLASS-100003, in the analysis presented by \citet{Pollock2026} (middle panel), is found to exhibit a higher [C/O] value. 
This places the system in a region where the abundance ratios are consistent with gas enriched by either Pop~II or Pop~III supernovae with a comparable likelihood, as the authors of the work conclude.
In the same study, the absorber associated with CEERS-1019, which shows the highest [C/O] value, is classified as Pop~III-enriched gas.
The same galaxy, when analysed by \citet{Nakane2026}, still exhibits similarly high [C/O] values that are not reproduced by the fiducial \thz model.
However, it also shows a higher value of [Si/O], which despite the large uncertainties, is not compatible within 1$\sigma$ with the value reported in \citet{Pollock2026}.

The larger uncertainties that affect measurements in \citet{Nakane2026}, shown in the third panel, primarily arise from differences in the analysis methodology. 
All three studies infer metal abundance ratios from the measured column densities of low-ionisation elements without applying ionization corrections, under the assumption that the absorbing gas is predominantly neutral.
However, while \citet{Zhu2026} and \citet{Pollock2026} derive these column densities through standard line-profile fitting, yielding different but overall statistically compatible results, in \citet{Nakane2026} absorption line fitting is performed by simultaneously varying column densities and covering fractions. 
These two factors both contribute to the strength of absorption line features, and they are degenerate.
This means that gas with a lower covering fraction and high column density could potentially produce the same absorption feature that would be observed in a high covering fraction but low column density cloud.
This degeneracy results in larger uncertainties in the inferred column densities and, consequently, in the derived abundance ratios.

Overall, current observations of high-redshift absorption systems remain subject to significant uncertainties, which might be improved by future high-resolution spectroscopic data.
For this reason, robust conclusions on possible Pop~III-enriched gas at $z>7$ remain challenging.
Among the systems considered here, the absorber associated with CEERS-1019 currently provides the most plausible candidate for Pop~III enrichment, with high [C/O] values reported by two independent studies.

\subsection{Pop~III identification criteria}
\label{sec:possiblepopIII}

\begin{figure*}
    \includegraphics[width=1.8\columnwidth]{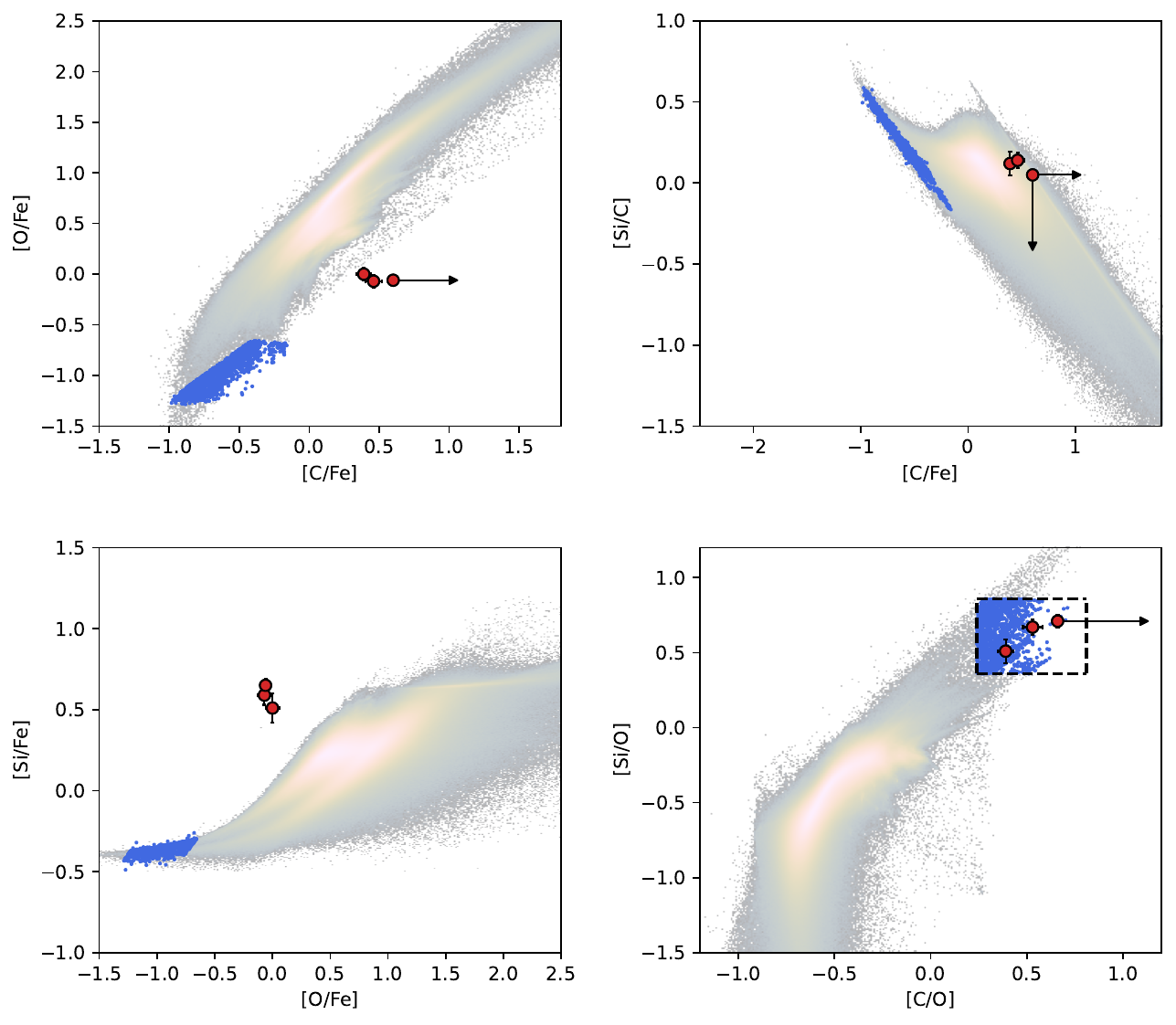}
    \caption{Metal abundance ratios of DLA-like gas cells with [Si/O] and [C/O] values similar to Pop~III-enriched candidates. The selected gas cells are marked as coloured points and they are required to reside within the area outlined by the dashed rectangle in the fourth panel, where the red points represent possible Pop~III-enriched gas identified in S24. Despite showing metal abundances similar to Pop~III candidates in the fourth panel, these gas cells tend to occupy different specific regions when comparing different element abundance ratios. This suggests that [Si/O] and [C/O] abundances might not be sufficient to robustly identify Pop~III candidates, as some values can be reproduced by Pop~II stars in a hydrodynamical context.}
    \label{fig:hist2d_popIII}
\end{figure*}

\begin{figure*}
    \includegraphics[width=1.\linewidth]{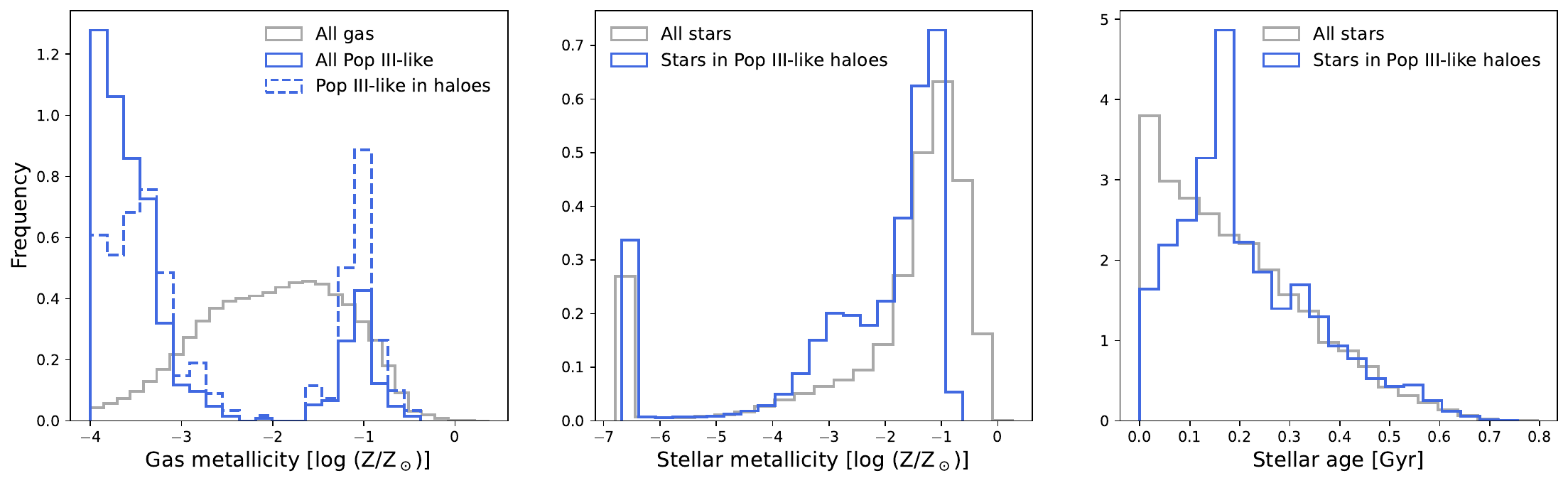}
    \caption{Properties of gas cells showing [Si/O] and [C/O] abundances similar to Pop~III-enriched candidates, as selected in Fig.~\ref{fig:hist2d_popIII}. The left plot compares the metallicity distribution of these gas cells (blue solid lines) to the overall distribution obtained from the larger sample of DLA-like gas (gray solid line). While the latter is more similar to a normal distribution, the former is closer to a double-peaked distribution. Out of these selected gas cells, the distribution of metallicities of those found inside haloes is represented by the dashed histogram. In the middle and right panels, we compare properties (stellar metallicity and age, respectively) of stellar particles found inside haloes where the selected gas cells reside. These stars seem to display lower values of metallicities and higher ages, suggesting that slightly more metal-poor, and intermediate-mass stars might be mainly responsible for the enrichment able to cause confusion between Pop~II and Pop~III models.}
    \label{fig:fraction_popIII}
\end{figure*}

Motivated by the findings presented in the previous section, which show how some of the observed Pop~III-enriched gas candidates occupy regions of abundance space also populated by gas reproduced by our Pop~II-only models, we proceed with a more detailed investigation. 
Specifically, we compare the abundance ratios found in our simulated gas with those of the three candidates presented in \citetalias{Sodini2024}, which are compatible with being enriched by Pop~III core-collapse SN events and for which values of different element ratios are available.
We initially focus on the [Si/O]--[C/O] relation, since it represents one of the most common ones, used across different redshifts to identify Pop~III-enriched gas candidates \citep[e.g.,][]{Christensen2023, Pollock2026}.
In Fig.~\ref{fig:hist2d_popIII}, we identify all the DLA-like gas cells whose values of [Si/O]--[C/O] ratios are consistent with those measured in \citetalias{Sodini2024}, and that we will refer to as Pop~III-like gas throughout this section.
Given that the typical uncertainties on the observed abundance ratios are $\approx 0.05$, we define a selection region extending from $-0.15$ ($\approx 3\sigma$) from the lowest abundance ratio value to $+0.15$ from the highest value. 
Within the considered region, marked by the dashed lines in the fourth panel, we highlight all the DLA-like gas cells as coloured points. 
At $z=6$, we find $\approx 1500$ gas cells satisfying this condition. Of these, $66\%$ are located in the zoom-in region hosting the most massive galaxy, whereas none of them are found in the $6$ lowest density zoom-in regions.
The volume occupied by Pop~III-like gas in each region varies between $10^{-6} - 10^{-5}$ of the total volume of the selected DLA-like gas cells.
In zoom-in regions with low-mass target haloes, the number of total DLA-like gas cells is normally below $10^{5}$.
This small number makes it difficult to determine whether the lack of Pop~III-like gas is primarily a statistical effect, arising from the limited number of DLA-like gas cells available, or whether Pop~III-like abundance ratios are intrinsically associated with more evolved, denser environments. 
In other words, the apparent concentration of Pop~III-like gas in the most massive zoom-in region may reflect the larger amount of DLA-like gas cells available for selection, but we cannot exclude an intrinsically higher occurrence of Pop~III-like abundance patterns in denser, more massive environments.

Overall, Fig.~\ref{fig:hist2d_popIII} shows that the simulated gas with [Si/O] and [C/O] abundances consistent with Pop~III candidates identified in metal absorbers at $z=6$, does not exhibit similar abundance patterns across the other abundance-ratio planes.
Because our fiducial models only consider enrichment from Pop~II stars, the fact that some of the simulated gas can reproduce the [Si/O] and [C/O] abundances of Pop~III candidates without matching the rest of the observed metal abundance ratios demonstrates that [Si/O] and [C/O] alone are not sufficient to uniquely identify gas enriched by Pop~III core-collapse SN events.
Instead, additional elemental abundance ratios are required to break this degeneracy and provide a more robust diagnostic of Pop~III enrichment, as investigated in \citetalias{Sodini2024}.
The need for multiple metal abundance ratios in order to identify Pop~III-enriched gas has also been proposed by models used for the analysis of metal poor DLAs at lower redshift, as presented in \citet{Welsh2019}.
Future analysis will help to quantify how likely it is to find similar Pop~III-like absorbers along lines of sight through the simulations.
In particular, by comparing the metal abundance ratio distribution of absorbers, i.e., spatially confined groups of DLA-like gas cells, to each observed Pop~III candidate, we could understand how probable this confusion between Pop~II and Pop~III-enriched gas is to arise in observations of [Si/O] and [C/O] abundance ratios.

We further investigate the origin of this degeneracy by examining physical properties of the simulated Pop~III-like gas.
We find that $~45\%$ of these gas cells reside within the virial radius of a halo, suggesting no strong preference for these cells to be either bound or unbound to galaxies.
Their metallicity distribution, shown in the first panel of Fig.~\ref{fig:fraction_popIII}, exhibits a bimodal structure with one peak at very low values ($\log (Z/\Zsun) \lesssim -3.5$) and the second peak at $\log (Z/\Zsun) \approx -1$.
Gas cells located within haloes preferentially populate the second peak, indicating that they have on average higher metallicities than their counterparts outside haloes. 
More generally, the metallicity distribution of Pop~III-like gas differs from that of DLA-like population, represented by the light-coloured histogram, which is closer to a lognormal function.
The combination of the bimodal distribution and the tendency of gas bound to haloes to show higher metallicity values may suggest that Pop~III-like gas originates from a specific enrichment channel by stellar populations with characteristic metallicities inside haloes, and part of it, transported further out through feedback processes, gets diluted by more pristine gas, populating the low-metallicity part of the distribution.
However, we cannot exclude an alternative scenario in which this bimodal distribution arises from contributions of two distinct enrichment channels associated with stellar populations at different metallicities.
We also note that there may be an effect due to the additional physical metal diffusion or feedback behavior that depends on the model choice in the simulations.

From Fig.~\ref{fig:hist2d_popIII}, we observe that although all the Pop~III-like gas cells are selected to only show similar [Si/O] and [C/O] abundances, they also reside in confined regions in all the other abundance ratio planes, favouring the scenario in which all the selected gas cells might be enriched by similar types of stars.
To understand which enrichment channels could produce these abundances in our simulations, we examine the metal yields from AGB stars and SN~II used in the simulations and obtained from the tables reported in \citet{Karakas2010} and \citet{Portinari1998}, respectively, as we present in Appendix~\ref{app:thzmetyields}.
These tables provide yields for a discrete set of stellar masses and metallicities, which in the simulations are interpolated to obtain a continuous yield function and incorporated into the galaxy evolution model.
We find that the regions covered by Pop~III-like gas across the different abundance ratio planes are consistent with yields of SN~II of $\approx 9\,\Msun$ stars at $Z = 0.008$, corresponding to $\approx 0.6\,\Zsun$.
They are also compatible with enrichment from AGB stars with mass $5 - 6\,\Msun$ and lower metallicity $0.004$ ($\approx 0.3\,\Zsun$), which appear to be a more plausible mechanism given that the metallicity of Pop~III-like gas inside haloes peaks at $\log (Z/\Zsun) \approx -1$.
However, understanding the processes contributing to the enrichment of a given gas region in the simulations is challenging. 
Metal yield tables indicate the amount of ejected mass per element given the initial mass and metallicity of a star. 
In the simulation these abundances need to be interpolated over the chosen initial mass function (IMF), and scaled depending on the total mass of the stellar particle, while metals are released at discrete intervals of time depending on the age of the simulated stars.
Gas surrounding a stellar particle is then progressively enriched by different sources acting on different timescales, with the relative contribution of each channel depending on the age of the population. 
Furthermore, individual haloes can host multiple stellar particles with different ages, metallicities and masses, and feedback processes together with infall of pristine gas facilitate metal mixing, complicating the picture. 
Despite these complexities, the fact that Pop~III-like gas preferentially occupies regions of abundance space where the yields of intermediate-mass stars are found might suggest that it is mainly influenced by contributions of evolved stellar populations at sub-solar metallicities.

Following these hypotheses, we investigate the properties of stellar populations residing within haloes hosting Pop~III-like gas. 
Their metallicity and age distributions are shown in the middle and right panels of Fig.~\ref{fig:fraction_popIII}, respectively, and they are compared to the distributions of the overall stellar population in our zoom-in regions at $z=6$. 
The haloes in which Pop~III-like gas is found have only a few stars with metallicities  $\textrm{Z}>0.1 \,\Zsun$, and show an enhancement in the number of stellar populations with $\textrm{Z} \approx 0.001 \,\Zsun$.
These haloes also contain a lower number of very young stars, which would rapidly pollute the gas with powerful feedback from SN explosions, potentially diluting or obscuring signatures of enrichment due to lower-mass stars present in the same halo.
These results suggest that the metal abundances in Pop~III-like gas might mainly arise because of enrichment from metal-poor intermediate mass stars ($M_\star \approx 5 -9~\Msun$) which could be more easily observed in galaxies undergoing a quiescent phase of star formation, as suggested by the lack of very young stars in these haloes when compared to the overall population. 

In conclusion, our analysis shows that to identify robust candidates enriched by Pop~III SN events, it is necessary to explore multiple different metal abundance ratios.
This is because mainly intermediate-mass AGB stars in older stellar populations with $\textrm{Z}<0.1\,\Zsun$ in haloes undergoing quiescent star formation phases tend to enrich the gas around them with [Si/O] and [C/O] abundances similar to what semi-analytical models predict for enrichment by Pop~III core-collapse SNe.

\section{Interpreting Pop~III Candidates}
\label{sec:discussion}

Metal abundance ratio distributions in the simulated gas are influenced by several factors, including the adopted stellar evolution models, feedback prescriptions, and subgrid treatment of dust.
In this section, we first investigate how properties of gas and galaxy environment shape the resulting distributions in the fiducial \thz models.
We then assess the sensitivity of these distributions to the metal yields and stellar population employed in the simulations, testing whether this could help to explain some of the observational results. 

\subsection{Within the fiducial simulations}

\begin{figure*}
    \centering
    \includegraphics[width=1\linewidth]{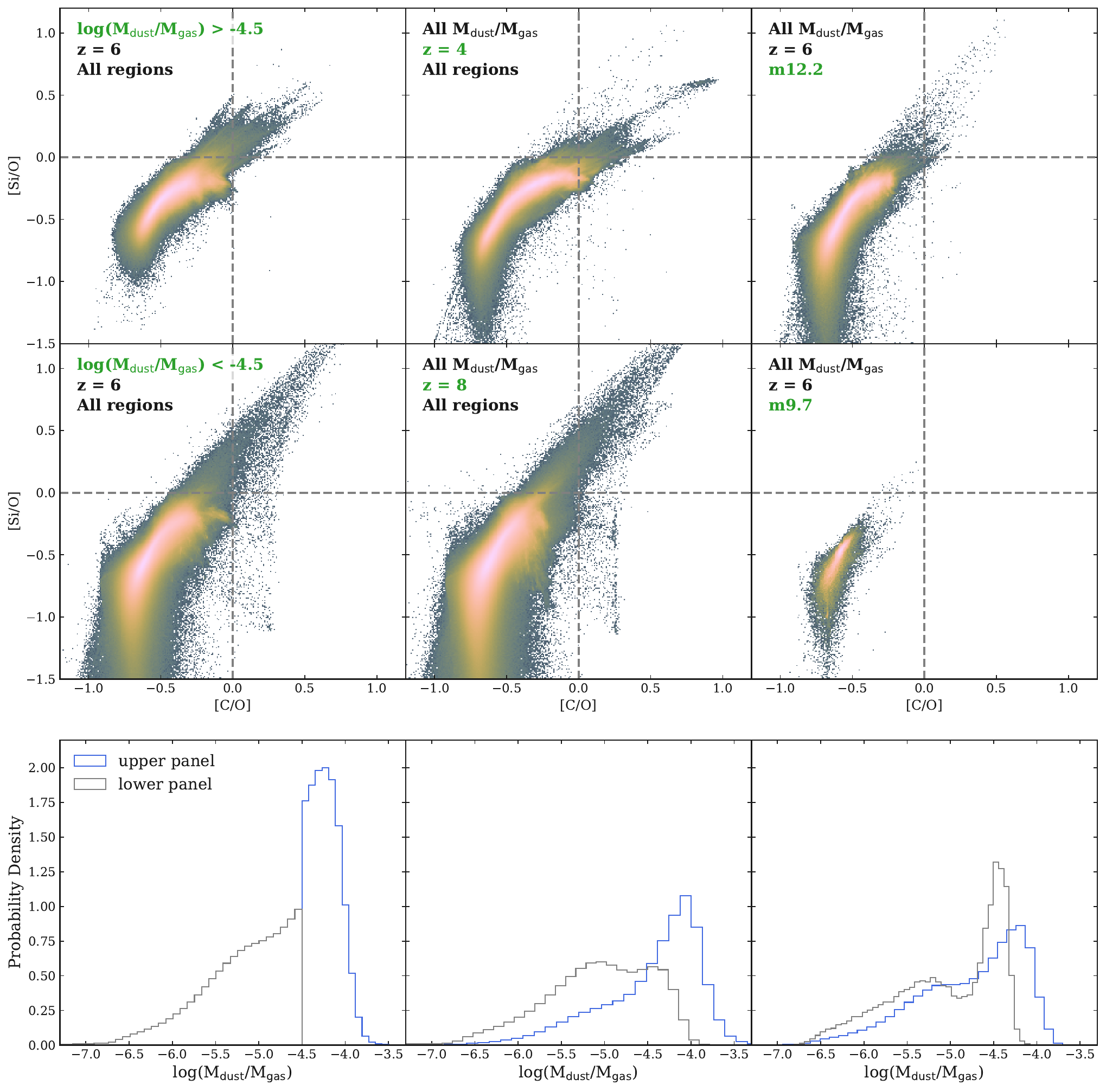}
    \caption{[Si/O]--[C/O] distribution when considering different subsets of DLA-like gas cells. The first column shows the effects of dust, with DLA-like gas particles at $z=6$ selected to have a high (upper panel) and low (lower panel) dust-to-gas ratio ($\log (\textrm{M}_{\rm dust}/\textrm{M}_{\rm gas})\gtrless-4.5$). Panels in the second column show the effect of redshift, with abundance ratios from DLA-like gas particles from all the zoom-in regions at $z=4$ in the top panel, and at $z=8$ in the bottom one. In the third column, we illustrate differences arising when considering two distinct zoom-in regions at $z=6$, with gas cells from m12.2 in the top panel and m9.7 in the bottom one. The histograms in the bottom row show the dust-to-gas mass distributions of the subsets represented in the two panels above them, respectively. Both gas with high dust content and low-redshift gas show a similar narrower distribution, suggesting that the evolution of metal abundance ratios with redshift might be mainly influenced by the increasing amount of metallicity and dust with cosmic time. }
    \label{fig:dtgcut}
\end{figure*}

Within the fiducial \thz simulations, we identify several key factors that shape the distribution of metal abundance ratios, including dust content, redshift and galactic environment.
In Fig.~\ref{fig:dtgcut}, we show how these three factors influence the [Si/O]--[C/O] distributions by selecting different subsets of DLA-like gas cells. 
We focus on the [Si/O]--[C/O] relation since this is the most commonly employed ratio in observations, especially at high redshift, however, similar conclusions are drawn when extending the analysis to all the other abundance ratios represented in Fig.~\ref{fig:hist2d_z6}.

In the first column, we apply a cut based on the dust-to-gas mass ratio in DLA-like gas cells at $z=6$. 
Dust content is expected to affect the gas-phase metallicity in our simulations, since a fraction of the metals, in particular C, O, Mg, Si and Fe, are incorporated into dust grains \citep[][]{Draine2003}. 
The upper panel shows that gas with high dust-to-gas mass ratio ($\log (\textrm{M}_{\rm dust}/\textrm{M}_{\rm gas}) > -4.5$) exhibits a narrower [Si/O]--[C/O] distribution when compared to gas with lower dust content, shown in the lower panel.
Similar distributions are shown in the middle panels, which illustrate results from all the zoom-in regions at $z=4$ (top panel) and at $z=8$ (bottom panel). 
This similarity suggests that the redshift evolution of abundance ratio distributions might be influenced by the increasing values of dust content with cosmic time.
This interpretation is supported by the histograms shown in the middle panel of the bottom row, where we can appreciate how the peak of the dust-to-gas ratio shifts from $\approx -5$ at $z=8$ (in gray) to $\approx -4$ at $z=4$ (in blue).
However, the dust content is tightly correlated to the metallicity of the gas, and they both increase with cosmic time.
To further investigate the role of metallicity and dust in the redshift evolution of metal abundance ratio distribution, we analyse how the latter changes for gas with similar metallicities but different dust content.
At $z=6$, we divide the DLA-like gas in metallicity bins and split each of them in gas with high and low dust-to-gas mass ratio.
We observe that the increasing metallicity leads to a thinning of the low [Si/O]--[C/O] tail. 
However, only in gas with high dust-to-gas mass ratio, which is more likely to be found in correspondence of high metallicities, we observe the characteristic features at high [Si/O] and [C/O] values that can be observed in the upper right corner of the first two top panels in Fig.~\ref{fig:dtgcut}.
These features arise because of a linear relation between the [Si/O] and [C/O] ratios with a slope lower than the one observed in dust-poor gas (bottom left panel) or at higher redshift (bottom middle panel), and they seem to appear only in dust-rich gas.

Differences arising in abundance ratios from different zoom-in regions can also be investigated, and we show them in the third column of Fig.~\ref{fig:dtgcut}. 
In the upper panel, the [Si/O]--[C/O] distribution of DLA-like gas cells in the m12.2 region at $z=6$ is displayed, while the lower panel shows the corresponding distribution in the m9.7 region.
At $z=6$, the mass of the target halo in the first region is $9.5 \times 10^{10} \,\Msun$, while in the second region the target halo has a mass which is $\approx 50$ times lower ($2.15 \times 10^9 \,\Msun$).
As a consequence, the total amount of DLA-like gas cells in the region surrounding the lower-mass halo is smaller, and the associated metal abundance distribution is narrower.
This might be simply due to a statistical effect, since the shape of the distribution in the m9.7 region mirrors the one in the more massive region.

Overall, our analysis shows that dispersion in abundance ratio distributions, which semi-analytical models have suggested as a possible signature of Pop~III enrichment, can be influenced by factors such as the metallicity and dust content of the analysed gas and the galactic environment considered. 
Specifically, both the increasing amount of metallicity and dust play a role in reducing the dispersion of the distribution at lower redshift, with dust mostly affecting the tail at high values and metallicity the one at lower values.

\subsection{Changing metal yields}
\label{sec:differentyields}

\begin{figure*}
    \includegraphics[width=0.75\linewidth]{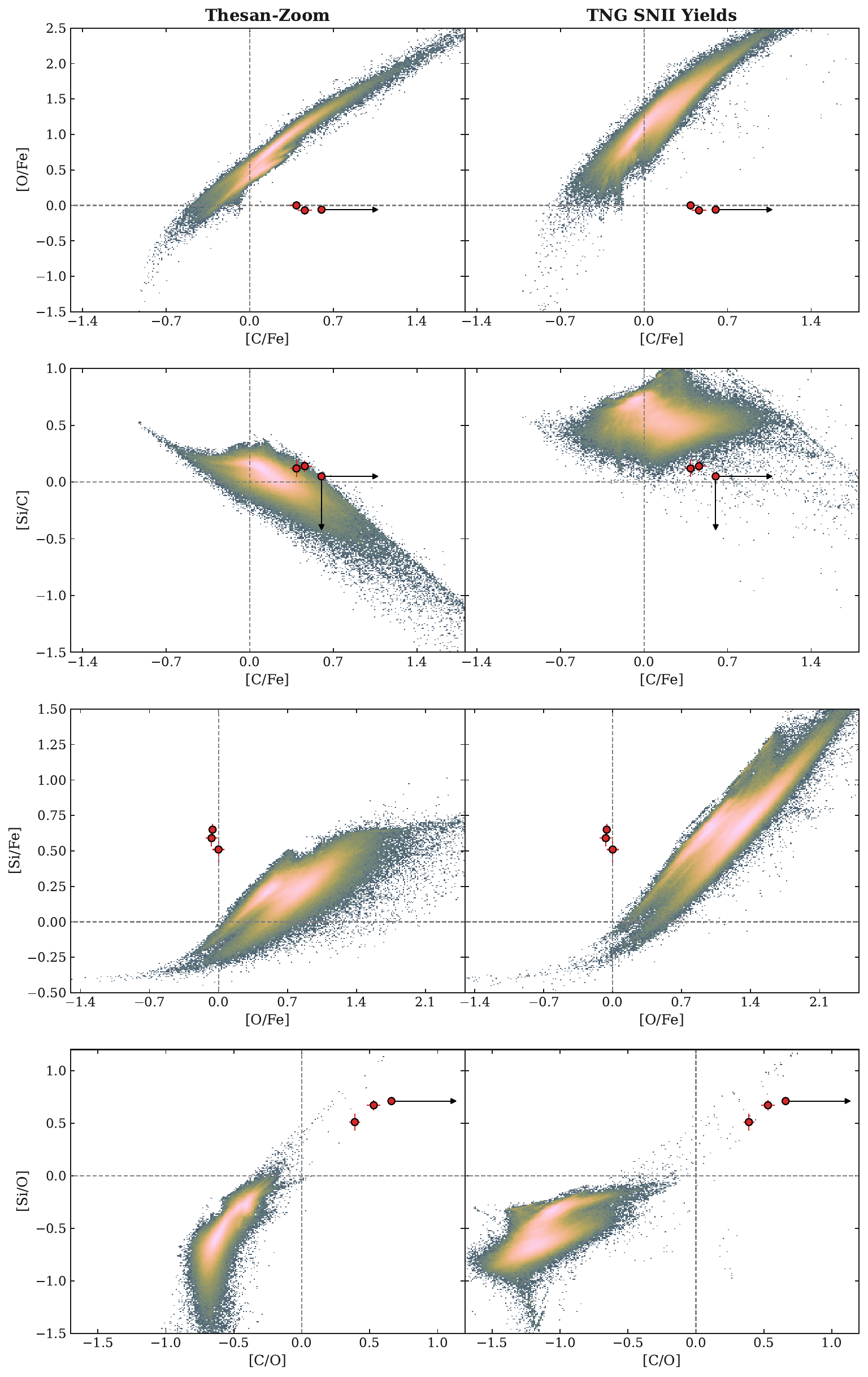}
    \caption{Metal abundance ratio distribution of DLA-like gas cells in the zoom-in region m10.8 at $z=6$ when assuming different metal yields. The first column shows abundances from the fiducial \thz model, while in the second column SN~II yields from the IllustrisTNG simulations are used \citep{Pillepich2018}. The coloured circles mark potential Pop~III-enriched gas candidates as identified by \citet[][]{Sodini2024}. Abundance ratio distributions vary substantially depending on the adopted metal yields.
    In particular, in the IllustrisTNG models, higher amounts of Si and O and lower amount of C and Fe are released by stars, significantly affecting the resulting distribution. Despite these variations induced by the choice of the metal yields, observed abundance ratios seem unlikely to be simultaneously reproduced within the \thz models.
    }
    \label{fig:Kobayashi}
\end{figure*}

Metal yields are prescriptions used in hydrodynamical simulations to quantify the amount of heavy elements that stars with different initial mass and metallicity eject into the gas surrounding them during SN explosions or through AGB winds.
In this section, we test how sensitive our results are to the assumed enrichment model by comparing two simulations of the same zoom-in region (m10.8).
The first is part of our fiducial \thz set, while in the second we replace the SN~II yields with those used in the IllustrisTNG simulations \citep{Pillepich2018}, adapted for the \thz model by removing the contribution from OB winds which is explicitly followed in our model.
IllustrisTNG adopts SN~II yields from \citet{Kobayashi2006} for stars with initial masses between $13 - 40\,\Msun$ and from \citet{Portinari1998}, the same yields used in \thz, for masses outside this range.

The resulting abundance ratio distributions at $z=6$ in the m10.8 zoom-in region simulated with IllustrisTNG SN~II yields, shown in the right column of Fig.~\ref{fig:Kobayashi}, exhibit significant differences when compared to the fiducial \thz model (left column).
In particular, the displayed [Si/C] values are higher, resembling the distribution predicted by SN~II yields of \citet{WoosleyWeaver1995}, as shown by the light blue shaded area in Fig.~\ref{fig:hist2d_z6}.
The [Si/Fe] ratio, represented in the third row, also shows values of up to a factor of $3$ higher than the ones we find in our fiducial simulations.
Moreover, these higher values are associated with higher [O/Fe], producing a more extended distribution, which again is closer to the predictions from the Pop~II SN yields.
Finally, the [C/O] distribution extends to values of more than a factor of $2$ lower when compared to \thz.
These differences in the distributions can be mainly attributed to an enhanced enrichment by silicon and oxygen together with a lower production of carbon and iron in IllustrisTNG with respect to our models (see Fig.~\ref{app:fig:metalyieldcomp} for more details).

Although the IllustrisTNG SN~II yields produce distinct metal abundance ratios overall, gas cells with [Si/O]--[C/O] abundances similar to Pop~III-enriched gas candidates, as analysed in Section~\ref{sec:possiblepopIII}, should theoretically persist.
This is because the metal yields for AGB stars of mass $5 - 6\,\Msun$ and even SN~II with mass $\approx 9\,\Msun$ are unchanged from the \thz model. 
However, as previously noted, the probability of detecting this specific enrichment pattern in DLA-like gas cells is low, and the identification of such systems requires simulating additional zoom-in regions.
Nevertheless, the consistency observed in the high [Si/O]--[C/O] tails between the fiducial \thz and the run with the IllustrisTNG yield supports our hypothesis, despite the current very low numbers of individual matching gas cells due to the selection of a single zoom-in region with an intermediate-mass target halo.

The analysis presented here shows that differences in the metal yields can lead to substantial changes in the abundance ratio distributions. 
This highlights the importance of an accurate description and implementation of stellar evolution and the associated metal ejection when studying feedback processes at high redshift.
Despite the obtained different metal abundance distributions when changing metal yields in our models, we still have not recovered all of the abundance ratios displayed by the possible Pop~III-enriched gas identified in observations.
In particular, higher values of [Si/Fe] and lower [O/Fe] are required to match observations.
However, we note that here we analysed a single zoom-in region, and more robust results might be drawn by collecting a more statistically significant sample by applying changes to the metal yields in all the \thz zoom-in regions.
In the next subsection, we investigate whether alternative assumptions regarding stellar populations might help in the interpretation of these observations.

\subsection{Varying the stellar initial mass function}

\begin{figure*}
    \includegraphics[width=0.98\linewidth]{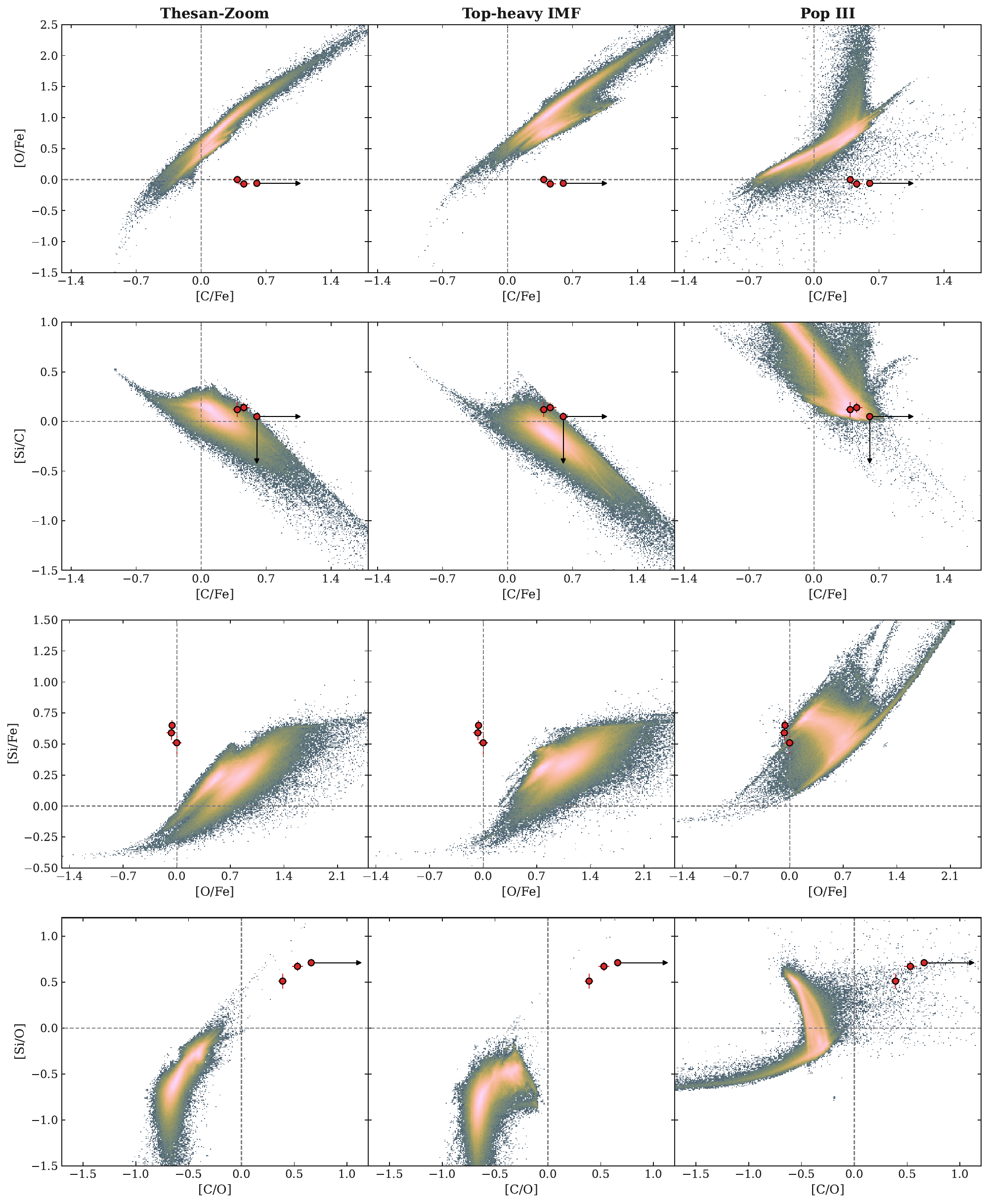}
    \caption{Metal abundance ratio distributions of DLA-like gas cells in the region m10.8 at $z=6$ when assuming different stellar populations. The first column shows abundances from the fiducial \thz model. The second and third columns display results from the same zoom-in region resimulated with different assumptions about the stellar populations. The second column depicts a more top-heavy IMF. The third column is performed with the latest version of the \thz model, which among other changes incorporates Pop~III star formation based on \citet{Saha2026}. Differences in the distribution are not particularly significant when allowing for more numerous massive stars. In contrast, the introduction of Pop~III enrichment with the new model heavily affects the results. The observed abundances of potential Pop~III-enriched gas lie in regions of the plots where our newest model predicts the presence of simulated gas cells.}
    \label{fig:changedstars}
\end{figure*}

The set of abundance ratios of the observed Pop~III-enriched gas candidates, represented by the coloured points in Fig.~\ref{fig:hist2d_z6}, cannot be reproduced simultaneously by the fiducial \thz model, nor by adopting SN~II metal yields from IllustrisTNG in our simulations, as explained in the previous sections.
To investigate the physical origin of these abundance patterns, we run two further simulations of the zoom-in region m10.8 varying the properties of the stellar populations, first by allowing for a more top-heavy IMF and then by incorporating a Pop~III stellar model developed for the most recent version of the \thz framework \citep[][]{Saha2026}.

The choice of a top-heavy IMF is motivated by the observation that, in Fig.~\ref{app:fig:snIIyields}, the metal yields of massive SNe~II at solar metallicity occupy a similar region in the abundance ratio planes as the Pop~III-enriched candidates observed in \citetalias{Sodini2024}.
We therefore reproduce the m10.8 zoom-in region with a different IMF, allowing for a higher number of massive stars.
Specifically, in the fiducial \thz simulations, every star particle represents a stellar population with a Chabrier IMF extending from $0.1$ to $100\,\Msun$, a slope of $-2.3$ at $\textrm{M} > 1 \, \Msun$ and a log-normal distribution with $\mu = \log(0.079)$ and $\sigma = 0.69$.
In our new run, we change the high-mass end slope to $-1.5$, while keeping the same log-normal distribution at low masses and ensuring that the total integral of the curve is held constant.
The resulting metal abundance ratio distributions at $z=6$ are represented in the second column of Fig.~\ref{fig:changedstars}, which shows how no significant differences from the fiducial model are present.
Although the peaks of the distributions are slightly shifted, the change in their dispersion is not large enough for the simulated gas to reproduce observational results.
Moreover, the bottom panel shows how the region at high [Si/O] and [C/O] values appears less populated than in the fiducial \thz model. 
The fact that this specific enrichment pattern is more rarely detected when allowing for a higher fraction of massive stars favours the interpretation that the gas reproducing possible Pop~III observations in the [Si/O]--[C/O] plane is predominantly enriched by intermediate-mass stars, as suggested in Section~\ref{sec:possiblepopIII}. 
However, our analysis considers only one zoom-in region, and the distributions might differ, potentially becoming broader, when the full set of zoom-in regions available in the \thz simulations is included.

\subsection{Including enrichment by Pop~III stars}

Another possibility is that the observed abundance patterns can only be reproduced by including enrichment from Pop~III stars. 
To test this hypothesis, we ran a simulation of the same zoom-in region with the latest model of \thz (Kannan et al., in prep.) which among other updates incorporates rotating Pop~II stars and Pop~III stellar model presented in \citet{Saha2026}.
In this model, the Pop~II stellar yields have been updated using SN~II tables from \citet{LimongiChieffi2018}, which account for stellar rotation with the distributions of the rotational velocities of stars computed from \citet{Prantzos2018}.
AGB ejecta are modelled by combining multiple tables  \citep{Karakas2010, Doherty2014, Fishlock2014, KarakasLugaro, Cinquegrana2022}, while SN~Ia yields are obtained from \citet{Seitanzahl2013}.
The model also involves an updated thermochemical network to include H$_2^+$ and H$^-$, essential for an accurate description of cooling in pristine gas.
If star forming gas has a metallicity lower than $10^{-4}\,\Zsun$, the derived stellar particles will be treated as Pop~III stellar populations.
Each Pop~III population is assumed to follow a log-flat IMF with slope $-1$, covering the mass range $2 - 250\,\Msun$.
Stellar evolution tracks are obtained using Modules for Experiments in Stellar Astrophysics \citep[\textsc{mesa};][]{Paxton2011, Paxton2013, Paxton2015, Paxton2018, Paxton2019, Jermyn2023} and their radiation spectra are computed employing \textsc{tlusty} \citep[][]{Hubeny1988} and are implemented self-consistently in the \textsc{arepo-rt} radiative transfer module.
Pop~III stars in the mass range $10 - 100\,\Msun$ and $140 - 250\,\Msun$ are predicted to explode as SNe, accounting for both core-collapse SNe events and PISNe, which are predicted to eject more energy into the surrounding gas, together with a specific chemical signature in the [Si/O]--[C/O] abundance ratio plane.
The metal yields adopted in this model come from \citet{HegerWoosley2010} for SNe with initial mass between $10 - 100\,\Msun$ and from \citet{HegerWoosley2002} for stars with higher initial mass.

The abundance ratio distributions obtained with this new model are shown in the third column of Fig.~\ref{fig:changedstars}.
The figure illustrates how with the new model of rotating Pop~II stars and the addition of Pop~III stars, most of the distributions are substantially modified.
The dispersion in the [O/Fe] -- [C/Fe] plane increases substantially, and higher values of [Si/C] and lower [C/Fe] are displayed, as shown in the second row, with the lower right corner of the plot becoming less populated than in the fiducial \thz simulation.
The implementation of this new model also seems to improve the underestimation of [Si/Fe] abundance ratio that we pointed out in Fig.~\ref{fig:hist2d_z6}, with the overall distribution occupying regions more similar to those indicated by observations and the SN~II yields from \citet{WoosleyWeaver1995}.
Tails of gas cells deviating from the main correlations found in the \thz fiducial model can be observed in the first panel, at high [O/Fe] values, and in the last one, occupying the upper left corner of the plot, at high [Si/O] and low [C/O] ratios.
These gas cells are probably mainly enriched by PISN events, as also other semi-analytical models based on the same Pop~III yield tables used in our simulations suggest \citep{Vanni2024}.
Overall, with this model, we find that in all four panels the observed Pop~III-enriched gas candidates seem to fall closer to regions where some of the simulated gas cells reside.
However, to robustly test whether the observed abundances are simultaneously reproduced by the same gas cells and their occurrence, simulating more zoom-in regions is necessary, and we plan to address this analysis with future work.

\section{Conclusions}
\label{sec:conclusions}

In this paper, we have used the \thz simulations to interpret observations of metal abundance patterns in absorbing gas during reionization.
To evaluate gas metal abundances, observations mainly rely on data obtained through quasar absorption spectroscopy, primarily focusing on the analysis of low-ionization metal absorbers in systems exhibiting detectable \ion{O}{i} absorption lines.
In \citet{Sodini2024}, three of these absorbers have been identified as compatible with being enriched by Pop~III stars, based on comparison with predictions from semi-analytical models of stellar evolution.
Crucially, semi-analytical approaches do not account for the spatial distribution of metals, which is governed by hydrodynamical processes, such as feedback and gas inflow, that regulate galaxy evolution. 
These complex dynamics are captured more accurately only through hydrodynamical simulations.
We have therefore employed the high-resolution \thz simulations to first select gas with physical properties compatible with observations of gas in absorption and compare their metal abundance ratios with results from observations. 
In this way, we provided a more physically consistent framework to robustly infer detection of Pop~III-enriched gas.

Our main findings are as follows:
\vspace{-5pt}

\begin{itemize}[leftmargin=*, itemindent=0pt,  itemsep=4pt]
    \item The abundance ratios in the fiducial \thz simulations are generally in good agreement with observed metal-absorbers that do not show signs of Pop~III-enrichment. The only exception is the silicon abundance, which the simulations seem to underestimate by a factor of $\approx 2$, as illustrated in Fig.~\ref{fig:hist2d_z6}.
    \item In the [Si/O]--[C/O] plane, about a fraction of $10^{-5}$ of the volume occupied by total absorber-like gas exhibits abundances compatible with observed Pop~III-enriched candidates. However, the same simulated gas fails to simultaneously reproduce the other metal abundance ratios associated with the same candidates. This indicates that identifying Pop~III-enriched gas in absorption requires analysing multiple abundance ratios beyond [Si/O] and [C/O] to solve this degeneracy, as in \citet{Sodini2024}.
    \item Large uncertainties in the observed abundance ratios of high-redshift ($z>7$) absorbers complicate the identification of potential Pop~III host galaxies. CEERS-1019 appears to be the only candidate analysed by multiple independent studies that might host gas enriched by the first generation of stars, as found by \citet{Pollock2026} and highlighted in our work in Fig.~\ref{fig:highzabs}.
    \item Towards higher redshift, the simulations present a larger dispersion in abundance ratio distributions. This is usually associated with Pop~III pollution signatures in semi-analytical models \citep{Vanni2024}. In our simulations however, this is driven by the increasingly lower dust and metallicity content. 
    \item Abundance ratio distributions are highly sensitive to the adopted metal yields, highlighting the importance of an accurate description and implementation of stellar evolution and its associated metal ejection. However, changing the Pop~II yield model, or our assumed IMF, did not reproduce the abundance ratio patterns seen in the Pop~III candidate absorbers. 
    \item Resimulating one of the \thz haloes with an updated model which among other changes includes Pop~III stars, produced the metal abundance ratios more consistent with observations of potential Pop~III-enriched gas. However, conclusive tests of this model will require resimulating more than one zoom region, to sample a wider variety of star formation histories, which will be the object of future work.
\end{itemize}

While our work has helped to refine the constraints required for the robust identification of Pop~III-enriched gas during reionization, it also highlights how current observations continue to challenge existing models. 
We demonstrate that resolving these discrepancies requires a multi-scale approach: accurate metal yields from individual stars must be integrated into theoretical models alongside precise descriptions of stellar populations and feedback mechanisms. We also note that the galaxy evolution model shapes these abundances, as feedback predictions vary greatly between simulations. 
For this reason, comparative studies of high-redshift simulations, such as the latest FIRE models \citep[][]{Samuel2026}, are needed to build a clearer broader picture.
Future analysis will also aim at refining the comparison with observations through the use of construction of mock observational data from the simulated boxes, to facilitate direct comparison with the observations.

Future surveys aiming at understanding the cradles of the first generation of stars will provide essential additional constraints on our models. Characterizing the galactic environments of metal absorbers remains a primary focus of ongoing \textit{JWST} programs \citep[e.g.,][]{Kashino2023, Zou2024, Li2026}. More follow-up of the environments of interesting absorption line systems will provide useful points of reference for the simulations. However, measurements, especially those at $z>7$, are currently affected by high uncertainties and high-resolution spectroscopy is necessary to provide more precise abundance ratio measurements.
In the future, this will be available thanks to the spectrographs High Angular Resolution Monolithic Optical and Near-infrared Integral field spectrograph \citep[HARMONI;][]{harmoni2021} and ArmazoNes high Dispersion Echelle Spectrograph \citep[ANDES;][]{DOdorico2023} on the Extremely Large Telescope (ELT), which will be decisive in advancing our understanding of feedback processes and stellar populations across cosmic time.

\section*{Acknowledgements}
We thank Annalisa Pillepich for sharing the metal yields developed for the IllustrisTNG simulations. We also thank Alessandra Venditti, Alessandro Trinca, Louise Welsh, Chiaki Kobayashi  and Valentina D'Odorico for interesting discussions.

The authors gratefully acknowledge the Gauss Centre for Supercomputing e.V. (\url{www.gauss-centre.eu}) for funding this project by providing computing time on the GCS Supercomputer SuperMUC-NG at Leibniz Supercomputing Centre (\url{www.lrz.de}), under project pn29we. This work used the DiRAC@Durham facility managed by the Institute for Computational Cosmology on behalf of the STFC DiRAC HPC Facility (\url{www.dirac.ac.uk}) as part of project dp436. The equipment was funded by BEIS capital funding via STFC capital grants ST/P002293/1, ST/R002371/1 and ST/S002502/1, Durham University and STFC operations grant ST/R000832/1. DiRAC is part of the National e-Infrastructure.

LK acknowledges the support of a Royal Society University Research Fellowship (grant number URF$\backslash$R1$\backslash$251793). RK acknowledges support of the Natural Sciences and Engineering Research Council of Canada (NSERC) through a Discovery Grant and a Discovery Launch Supplement, funding reference numbers RGPIN-2024-06222 and DGECR-2024-00144. Support for programs JWST-AR-08709 (AS) and JWST-AR-04814 (XS, MV) was provided by NASA through a grant from the Space Telescope Science Institute, which is operated by the Association of Universities for Research in Astronomy, Inc., under NASA contract NAS 5-03127.
MV acknowledges support through NASA ATP Grant 23-ATP23-149 and NSF AAG Grant AST-2307699.
This work was supported by the LOEWE programme of the State of Hesse (LOEWE-Spitzen-Professur).

\section*{Data Availability}

All simulation data from the fiducial \thz runs, including snapshots, group, subhalo catalogues and merger trees will be made publicly available in the near
future via \href{www.thesan-project.com}{www.thesan-project.com}. Before the public data release, data underlying this paper will be shared on reasonable request to the corresponding author.



\bibliographystyle{mnras}
\bibliography{bibliography} 

@ARTICLE{Wang2026,
       author = {{Wang}, Feige and {Champagne}, Jaclyn B. and {Huang}, Jiamu and {Yang}, Jinyi and {Hennawi}, Joseph F. and {Fan}, Xiaohui and {Zhang}, Haowen and {Costa}, Tiago and {Decarli}, Roberto and {Habouzit}, Melanie and et al.},
        title = "{ASPIRE: The Environments and Dark Matter Halos of Luminous Quasars in the Epoch of Reionization}",
      journal = {\apj},
         year = 2026,
        month = jul,
       volume = {1006},
       number = {1},
          eid = {39},
        pages = {39},
          doi = {10.3847/1538-4357/ae7bfa},
archivePrefix = {arXiv},
       eprint = {2602.04979},
 primaryClass = {astro-ph.GA},
       adsurl = {https://scixplorer.org/abs/2026ApJ..1006...39W}
}

@ARTICLE{Sodini2024,
       author = {{Sodini}, Alessio and {D'Odorico}, Valentina and {Salvadori}, Stefania and {Vanni}, Irene and {Bischetti}, Manuela and {Cupani}, Guido and {Davies}, Rebecca and {Becker}, George D. and {Ba{\~n}ados}, Eduardo and {Bosman}, Sarah and {Davies}, Frederick and {Paolo Farina}, Emanuele and {Ferrara}, Andrea and {Keating}, Laura and {Kulkarni}, Girish and {Lai}, Samuel and {Ryan-Weber}, Emma and {Sebastian}, Alma Maria and {Walter}, Fabian},
      journal = {\aap},
         year = 2024,
        month = jul,
       volume = {687},
          eid = {A314},
        pages = {A314},
          doi = {10.1051/0004-6361/202349062},
archivePrefix = {arXiv},
       eprint = {2404.10722},
 primaryClass = {astro-ph.GA},
       adsurl = {https://ui.adsabs.harvard.edu/abs/2024A&A...687A.314S}
}

@ARTICLE{Becker2012,
       author = {{Becker}, George D. and {Sargent}, Wallace L.~W. and {Rauch}, Michael and {Carswell}, Robert F.},
        title = "{Iron and {\ensuremath{\alpha}}-element Production in the First One Billion Years after the Big Bang}",
      journal = {\apj},
         year = 2012,
        month = jan,
       volume = {744},
       number = {2},
          eid = {91},
        pages = {91},
          doi = {10.1088/0004-637X/744/2/91},
archivePrefix = {arXiv},
       eprint = {1111.4843},
 primaryClass = {astro-ph.CO},
       adsurl = {https://ui.adsabs.harvard.edu/abs/2012ApJ...744...91B}
}

@ARTICLE{Poudel2018,
       author = {{Poudel}, Suraj and {Kulkarni}, Varsha P. and {Morrison}, Sean and {P{\'e}roux}, C{\'e}line and {Som}, Debopam and {Rahmani}, Hadi and {Quiret}, Samuel},
        title = "{Early metal enrichment of gas-rich galaxies at z {\ensuremath{\sim}} 5}",
      journal = {\mnras},
         year = 2018,
        month = jan,
       volume = {473},
       number = {3},
        pages = {3559-3572},
          doi = {10.1093/mnras/stx2607},
archivePrefix = {arXiv},
       eprint = {1710.03315},
 primaryClass = {astro-ph.GA},
       adsurl = {https://ui.adsabs.harvard.edu/abs/2018MNRAS.473.3559P}
}

@ARTICLE{Poudel2020,
       author = {{Poudel}, Suraj and {Kulkarni}, Varsha P. and {Cashman}, Frances H. and {Frye}, Brenda and {P{\'e}roux}, C{\'e}line and {Rahmani}, Hadi and {Quiret}, Samuel},
        title = "{Metal-enriched galaxies in the first {\ensuremath{\sim}}1 billion years: evidence of a smooth metallicity evolution at z {\ensuremath{\sim}} 5}",
      journal = {\mnras},
         year = 2020,
        month = jan,
       volume = {491},
       number = {1},
        pages = {1008-1025},
          doi = {10.1093/mnras/stz3000},
archivePrefix = {arXiv},
       eprint = {1910.10023},
 primaryClass = {astro-ph.GA},
       adsurl = {https://ui.adsabs.harvard.edu/abs/2020MNRAS.491.1008P}
}

@ARTICLE{Asplund2009,
       author = {{Asplund}, Martin and {Grevesse}, Nicolas and {Sauval}, A. Jacques and {Scott}, Pat},
        title = "{The Chemical Composition of the Sun}",
      journal = {\araa},
         year = 2009,
        month = sep,
       volume = {47},
       number = {1},
        pages = {481-522},
          doi = {10.1146/annurev.astro.46.060407.145222},
archivePrefix = {arXiv},
       eprint = {0909.0948},
 primaryClass = {astro-ph.SR},
       adsurl = {https://ui.adsabs.harvard.edu/abs/2009ARA&A..47..481A}
}

@ARTICLE{Kannan2025,
       author = {{Kannan}, Rahul and {Puchwein}, Ewald and {Smith}, Aaron and {Borrow}, Josh and {Garaldi}, Enrico and {Keating}, Laura and {Vogelsberger}, Mark and {Zier}, Oliver and {McClymont}, William and {Shen}, Xuejian and {Popovic}, Filip and {Tacchella}, Sandro and {Hernquist}, Lars and {Springel}, Volker},
        title = "{Introducing the THESAN-ZOOM project: radiation-hydrodynamic simulations of high-redshift galaxies with a multi-phase interstellar medium}",
      journal = {The Open Journal of Astrophysics},
         year = 2025,
        month = oct,
       volume = {8},
          eid = {153},
        pages = {153},
          doi = {10.33232/001c.145804},
archivePrefix = {arXiv},
       eprint = {2502.20437},
 primaryClass = {astro-ph.GA},
       adsurl = {https://ui.adsabs.harvard.edu/abs/2025OJAp....8E.153K}
}

@ARTICLE{Kannan2020,
       author = {{Kannan}, Rahul and {Marinacci}, Federico and {Vogelsberger}, Mark and {Sales}, Laura V. and {Torrey}, Paul and {Springel}, Volker and {Hernquist}, Lars},
        title = "{Simulating the interstellar medium of galaxies with radiative transfer, non-equilibrium thermochemistry, and dust}",
      journal = {MNRAS},
         year = 2020,
        month = dec,
       volume = {499},
       number = {4},
        pages = {5732-5748},
          doi = {10.1093/mnras/staa3249},
archivePrefix = {arXiv},
       eprint = {1910.14041},
 primaryClass = {astro-ph.GA},
       adsurl = {https://ui.adsabs.harvard.edu/abs/2020MNRAS.499.5732K}
}

@ARTICLE{Smith2022,
       author = {{Smith}, A. and {Kannan}, R. and {Garaldi}, E. and {Vogelsberger}, M. and {Pakmor}, R. and {Springel}, V. and {Hernquist}, L.},
        title = "{The THESAN project: Lyman-{\ensuremath{\alpha}} emission and transmission during the Epoch of Reionization}",
      journal = {\mnras},
         year = 2022,
        month = may,
       volume = {512},
       number = {3},
        pages = {3243-3265},
          doi = {10.1093/mnras/stac713},
archivePrefix = {arXiv},
       eprint = {2110.02966},
 primaryClass = {astro-ph.CO},
       adsurl = {https://ui.adsabs.harvard.edu/abs/2022MNRAS.512.3243S}
}

@ARTICLE{Weinberger2020,
       author = {{Weinberger}, Rainer and {Springel}, Volker and {Pakmor}, R{\"u}diger},
        title = "{The AREPO Public Code Release}",
      journal = {ApJS},
         year = 2020,
        month = jun,
       volume = {248},
       number = {2},
          eid = {32},
        pages = {32},
          doi = {10.3847/1538-4365/ab908c},
archivePrefix = {arXiv},
       eprint = {1909.04667},
 primaryClass = {astro-ph.IM},
       adsurl = {https://ui.adsabs.harvard.edu/abs/2020ApJS..248...32W}
}

@ARTICLE{Marinacci2019,
       author = {{Marinacci}, Federico and {Sales}, Laura V. and {Vogelsberger}, Mark and {Torrey}, Paul and {Springel}, Volker},
        title = "{Simulating the interstellar medium and stellar feedback on a moving mesh: implementation and isolated galaxies}",
      journal = {MNRAS},
         year = 2019,
        month = nov,
       volume = {489},
       number = {3},
        pages = {4233-4260},
          doi = {10.1093/mnras/stz2391},
archivePrefix = {arXiv},
       eprint = {1905.08806},
 primaryClass = {astro-ph.GA},
       adsurl = {https://ui.adsabs.harvard.edu/abs/2019MNRAS.489.4233M}
}

@ARTICLE{Becker2011,
       author = {{Becker}, George D. and {Sargent}, Wallace L.~W. and {Rauch}, Michael and {Calverley}, Alexander P.},
        title = "{High-redshift Metals. II. Probing Reionization Galaxies with Low-ionization Absorption Lines at Redshift Six}",
      journal = {ApJ},
         year = 2011,
        month = jul,
       volume = {735},
       number = {2},
          eid = {93},
        pages = {93},
          doi = {10.1088/0004-637X/735/2/93},
archivePrefix = {arXiv},
       eprint = {1101.4399},
 primaryClass = {astro-ph.CO},
       adsurl = {https://ui.adsabs.harvard.edu/abs/2011ApJ...735...93B}
}

@ARTICLE{Sebastian2024,
       author = {{Sebastian}, Alma Maria and {Ryan-Weber}, Emma and {Davies}, Rebecca L. and {Becker}, George D. and {Keating}, Laura C. and {D'Odorico}, Valentina and {Meyer}, Romain A. and {Bosman}, Sarah E.~I. and {Cupani}, Guido and {Kulkarni}, Girish and {Haehnelt}, Martin G. and {Lai}, Samuel and {Eilers}, Anna-Christina and {Bischetti}, Manuela and {Gallerani}, Simona},
        title = "{E-XQR-30: The evolution of Mg II, C II, and O I across 2 < z < 6}",
      journal = {MNRAS},
         year = 2024,
        month = may,
       volume = {530},
       number = {2},
        pages = {1829-1848},
          doi = {10.1093/mnras/stae789},
archivePrefix = {arXiv},
       eprint = {2403.10072},
 primaryClass = {astro-ph.GA},
       adsurl = {https://ui.adsabs.harvard.edu/abs/2024MNRAS.530.1829S}
}

@ARTICLE{DOdorico2023,
       author = {{D'Odorico}, Valentina and {Bolton}, James S. and {Christensen}, Lise and {De Cia}, Annalisa and {Zackrisson}, Erik and {Kordt}, Aron and {Izzo}, Luca and {Li}, Jiangtao and {Maiolino}, Roberto and {Marconi}, Alessandro and {Richter}, Philipp and {Saccardi}, Andrea and {Salvadori}, Stefania and {Vanni}, Irene and {Feruglio}, Chiara and {Fumagalli}, Michele and {Fynbo}, Johan P.~U. and {Noterdaeme}, Pasquier and {Papaderos}, Polychronis and {Peroux}, Celine and {Verma}, Aprajita and {Di Marcantonio}, Paolo and {Origlia}, Livia and {Zanutta}, Alessio},
        title = "{Galaxy Formation and Symbiotic Evolution with the Inter-Galactic Medium in the Age of ELT-ANDES}",
      journal = {arXiv e-prints},
         year = 2023,
        month = nov,
          eid = {arXiv:2311.16803},
        pages = {arXiv:2311.16803},
          doi = {10.48550/arXiv.2311.16803},
archivePrefix = {arXiv},
       eprint = {2311.16803},
 primaryClass = {astro-ph.GA},
       adsurl = {https://ui.adsabs.harvard.edu/abs/2023arXiv231116803D}
}

@ARTICLE{Becker2019,
       author = {{Becker}, George D. and {Pettini}, Max and {Rafelski}, Marc and {D'Odorico}, Valentina and {Boera}, Elisa and {Christensen}, Lise and {Cupani}, Guido and {Ellison}, Sara L. and {Farina}, Emanuele Paolo and {Fumagalli}, Michele and {L{\'o}pez}, Sebastian and {Neeleman}, Marcel and {Ryan-Weber}, Emma V. and {Worseck}, G{\'a}bor},
        title = "{The Evolution of O I over 3.2 < z < 6.5: Reionization of the Circumgalactic Medium}",
      journal = {ApJ},
         year = 2019,
        month = oct,
       volume = {883},
       number = {2},
          eid = {163},
        pages = {163},
          doi = {10.3847/1538-4357/ab3eb5},
archivePrefix = {arXiv},
       eprint = {1907.02983},
 primaryClass = {astro-ph.GA},
       adsurl = {https://ui.adsabs.harvard.edu/abs/2019ApJ...883..163B}
}

@ARTICLE{Christensen2023,
       author = {{Christensen}, L. and {Jakobsen}, P. and {Willott}, C. and {Arribas}, S. and {Bunker}, A. and {Charlot}, S. and {Maiolino}, R. and {Marshall}, M. and {Perna}, M. and {{\"U}bler}, H.},
        title = "{Metal enrichment and evolution in four z > 6.5 quasar sightlines observed with JWST/NIRSpec}",
      journal = {A&A},
         year = 2023,
        month = dec,
       volume = {680},
          eid = {A82},
        pages = {A82},
          doi = {10.1051/0004-6361/202347943},
archivePrefix = {arXiv},
       eprint = {2309.06470},
 primaryClass = {astro-ph.GA},
       adsurl = {https://ui.adsabs.harvard.edu/abs/2023A&A...680A..82C}
}

@ARTICLE{Springel2010,
       author = {{Springel}, Volker},
        title = "{E pur si muove: Galilean-invariant cosmological hydrodynamical simulations on a moving mesh}",
      journal = {MNRAS},
         year = 2010,
        month = jan,
       volume = {401},
       number = {2},
        pages = {791-851},
          doi = {10.1111/j.1365-2966.2009.15715.x},
archivePrefix = {arXiv},
       eprint = {0901.4107},
 primaryClass = {astro-ph.CO},
       adsurl = {https://ui.adsabs.harvard.edu/abs/2010MNRAS.401..791S}
}

@ARTICLE{Kannan2022,
       author = {{Kannan}, R. and {Garaldi}, E. and {Smith}, A. and {Pakmor}, R. and {Springel}, V. and {Vogelsberger}, M. and {Hernquist}, L.},
        title = "{Introducing the THESAN project: radiation-magnetohydrodynamic simulations of the epoch of reionization}",
      journal = {MNRAS},
         year = 2022,
        month = apr,
       volume = {511},
       number = {3},
        pages = {4005-4030},
          doi = {10.1093/mnras/stab3710},
archivePrefix = {arXiv},
       eprint = {2110.00584},
 primaryClass = {astro-ph.GA},
       adsurl = {https://ui.adsabs.harvard.edu/abs/2022MNRAS.511.4005K}
}

@ARTICLE{Kannan2019,
       author = {{Kannan}, Rahul and {Vogelsberger}, Mark and {Marinacci}, Federico and {McKinnon}, Ryan and {Pakmor}, R{\"u}diger and {Springel}, Volker},
        title = "{AREPO-RT: radiation hydrodynamics on a moving mesh}",
      journal = {MNRAS},
         year = 2019,
        month = may,
       volume = {485},
       number = {1},
        pages = {117-149},
          doi = {10.1093/mnras/stz287},
archivePrefix = {arXiv},
       eprint = {1804.01987},
 primaryClass = {astro-ph.IM},
       adsurl = {https://ui.adsabs.harvard.edu/abs/2019MNRAS.485..117K}
}

@ARTICLE{Vogelsberger2013,
       author = {{Vogelsberger}, Mark and {Genel}, Shy and {Sijacki}, Debora and {Torrey}, Paul and {Springel}, Volker and {Hernquist}, Lars},
        title = "{A model for cosmological simulations of galaxy formation physics}",
      journal = {MNRAS},
         year = 2013,
        month = dec,
       volume = {436},
       number = {4},
        pages = {3031-3067},
          doi = {10.1093/mnras/stt1789},
archivePrefix = {arXiv},
       eprint = {1305.2913},
 primaryClass = {astro-ph.CO},
       adsurl = {https://ui.adsabs.harvard.edu/abs/2013MNRAS.436.3031V}
}

@ARTICLE{Garaldi2022,
       author = {{Garaldi}, E. and {Kannan}, R. and {Smith}, A. and {Springel}, V. and {Pakmor}, R. and {Vogelsberger}, M. and {Hernquist}, L.},
        title = "{The THESAN project: properties of the intergalactic medium and its connection to reionization-era galaxies}",
      journal = {MNRAS},
         year = 2022,
        month = jun,
       volume = {512},
       number = {4},
        pages = {4909-4933},
          doi = {10.1093/mnras/stac257},
archivePrefix = {arXiv},
       eprint = {2110.01628},
 primaryClass = {astro-ph.CO},
       adsurl = {https://ui.adsabs.harvard.edu/abs/2022MNRAS.512.4909G}
}

@ARTICLE{McKinnon2016,
       author = {{McKinnon}, Ryan and {Torrey}, Paul and {Vogelsberger}, Mark},
        title = "{Dust formation in Milky Way-like galaxies}",
      journal = {\mnras},
         year = 2016,
        month = apr,
       volume = {457},
       number = {4},
        pages = {3775-3800},
          doi = {10.1093/mnras/stw253},
archivePrefix = {arXiv},
       eprint = {1505.04792},
 primaryClass = {astro-ph.GA},
       adsurl = {https://ui.adsabs.harvard.edu/abs/2016MNRAS.457.3775M}
}

@ARTICLE{McKinnon2017,
       author = {{McKinnon}, Ryan and {Torrey}, Paul and {Vogelsberger}, Mark and {Hayward}, Christopher C. and {Marinacci}, Federico},
        title = "{Simulating the dust content of galaxies: successes and failures}",
      journal = {\mnras},
         year = 2017,
        month = jun,
       volume = {468},
       number = {2},
        pages = {1505-1521},
          doi = {10.1093/mnras/stx467},
archivePrefix = {arXiv},
       eprint = {1606.02714},
 primaryClass = {astro-ph.GA},
       adsurl = {https://ui.adsabs.harvard.edu/abs/2017MNRAS.468.1505M}
}

@ARTICLE{Zier2025b,
       author = {{Zier}, Oliver and {Kannan}, Rahul and {Smith}, Aaron and {Puchwein}, Ewald and {Vogelsberger}, Mark and {Borrow}, Josh and {Garaldi}, Enrico and {Keating}, Laura and {McClymont}, William and {Shen}, Xuejian and et al.},
        title = "{The THESAN-ZOOM project: Population III star formation continues until the end of reionization}",
      journal = {\mnras},
         year = 2025,
        month = nov,
       volume = {544},
       number = {1},
        pages = {410-429},
          doi = {10.1093/mnras/staf1053},
archivePrefix = {arXiv},
       eprint = {2503.03806},
 primaryClass = {astro-ph.GA},
       adsurl = {https://ui.adsabs.harvard.edu/abs/2025MNRAS.544..410Z}
}

@ARTICLE{Zier2024,
       author = {{Zier}, Oliver and {Kannan}, Rahul and {Smith}, Aaron and {Vogelsberger}, Mark and {Verbeek}, Erkin},
        title = "{Adapting AREPO-RT for exascale computing: GPU acceleration and efficient communication}",
      journal = {\mnras},
         year = 2024,
        month = sep,
       volume = {533},
       number = {1},
        pages = {268-286},
          doi = {10.1093/mnras/stae1837},
archivePrefix = {arXiv},
       eprint = {2404.17630},
 primaryClass = {astro-ph.IM},
       adsurl = {https://ui.adsabs.harvard.edu/abs/2024MNRAS.533..268Z}
}

@ARTICLE{Fan2023,
       author = {{Fan}, Xiaohui and {Ba{\~n}ados}, Eduardo and {Simcoe}, Robert A.},
        title = "{Quasars and the Intergalactic Medium at Cosmic Dawn}",
      journal = {ARA&A},
         year = 2023,
        month = aug,
       volume = {61},
        pages = {373-426},
          doi = {10.1146/annurev-astro-052920-102455},
archivePrefix = {arXiv},
       eprint = {2212.06907},
 primaryClass = {astro-ph.GA},
       adsurl = {https://ui.adsabs.harvard.edu/abs/2023ARA&A..61..373F}
}

@ARTICLE{Eldridge2017,
       author = {{Eldridge}, J.~J. and {Stanway}, E.~R. and {Xiao}, L. and {McClelland}, L.~A.~S. and {Taylor}, G. and {Ng}, M. and {Greis}, S.~M.~L. and {Bray}, J.~C.},
        title = "{Binary Population and Spectral Synthesis Version 2.1: Construction, Observational Verification, and New Results}",
      journal = {PASA},
         year = 2017,
        month = nov,
       volume = {34},
          eid = {e058},
        pages = {e058},
          doi = {10.1017/pasa.2017.51},
archivePrefix = {arXiv},
       eprint = {1710.02154},
 primaryClass = {astro-ph.SR},
       adsurl = {https://ui.adsabs.harvard.edu/abs/2017PASA...34...58E}
}

@ARTICLE{Levermore1984,
       author = {{Levermore}, C.~D.},
        title = "{Relating Eddington factors to flux limiters.}",
      journal = {JQSRT},
         year = 1984,
        month = feb,
       volume = {31},
       number = {2},
        pages = {149-160},
          doi = {10.1016/0022-4073(84)90112-2},
       adsurl = {https://ui.adsabs.harvard.edu/abs/1984JQSRT..31..149L}
}

@ARTICLE{PlanckCosmo2016,
       author = {{Planck Collaboration} and {Ade}, P.~A.~R. and {Aghanim}, N. and {Arnaud}, M. and {Ashdown}, M. and {Aumont}, J. and {Baccigalupi}, C. and {Banday}, A.~J. and {Barreiro}, R.~B. and {Bartlett}, J.~G. and {Bartolo}, N. and {Battaner}, E. and {Battye}, R. and {Benabed}, K. and {Beno{\^\i}t}, A. and {Benoit-L{\'e}vy}, A. and {Bernard}, J. -P. and {Bersanelli}, M. and {Bielewicz}, P. and {Bock}, J.~J. and {Bonaldi}, A. and {Bonavera}, L. and {Bond}, J.~R. and {Borrill}, J. and {Bouchet}, F.~R. and {Boulanger}, F. and {Bucher}, M. and {Burigana}, C. and {Butler}, R.~C. and {Calabrese}, E. and {Cardoso}, J. -F. and {Catalano}, A. and {Challinor}, A. and {Chamballu}, A. and {Chary}, R. -R. and {Chiang}, H.~C. and {Chluba}, J. and {Christensen}, P.~R. and {Church}, S. and {Clements}, D.~L. and {Colombi}, S. and {Colombo}, L.~P.~L. and {Combet}, C. and {Coulais}, A. and {Crill}, B.~P. and {Curto}, A. and {Cuttaia}, F. and {Danese}, L. and {Davies}, R.~D. and {Davis}, R.~J. and {de Bernardis}, P. and {de Rosa}, A. and {de Zotti}, G. and {Delabrouille}, J. and {D{\'e}sert}, F. -X. and {Di Valentino}, E. and {Dickinson}, C. and {Diego}, J.~M. and {Dolag}, K. and {Dole}, H. and {Donzelli}, S. and {Dor{\'e}}, O. and {Douspis}, M. and {Ducout}, A. and {Dunkley}, J. and {Dupac}, X. and {Efstathiou}, G. and {Elsner}, F. and {En{\ss}lin}, T.~A. and {Eriksen}, H.~K. and {Farhang}, M. and {Fergusson}, J. and {Finelli}, F. and {Forni}, O. and {Frailis}, M. and {Fraisse}, A.~A. and {Franceschi}, E. and {Frejsel}, A. and {Galeotta}, S. and {Galli}, S. and {Ganga}, K. and {Gauthier}, C. and {Gerbino}, M. and {Ghosh}, T. and {Giard}, M. and {Giraud-H{\'e}raud}, Y. and {Giusarma}, E. and {Gjerl{\o}w}, E. and {Gonz{\'a}lez-Nuevo}, J. and {G{\'o}rski}, K.~M. and {Gratton}, S. and {Gregorio}, A. and {Gruppuso}, A. and {Gudmundsson}, J.~E. and {Hamann}, J. and {Hansen}, F.~K. and {Hanson}, D. and {Harrison}, D.~L. and {Helou}, G. and {Henrot-Versill{\'e}}, S. and {Hern{\'a}ndez-Monteagudo}, C. and {Herranz}, D. and {Hildebrandt}, S.~R. and {Hivon}, E. and {Hobson}, M. and {Holmes}, W.~A. and {Hornstrup}, A. and {Hovest}, W. and {Huang}, Z. and {Huffenberger}, K.~M. and {Hurier}, G. and {Jaffe}, A.~H. and {Jaffe}, T.~R. and {Jones}, W.~C. and {Juvela}, M. and {Keih{\"a}nen}, E. and {Keskitalo}, R. and {Kisner}, T.~S. and {Kneissl}, R. and {Knoche}, J. and {Knox}, L. and {Kunz}, M. and {Kurki-Suonio}, H. and {Lagache}, G. and {L{\"a}hteenm{\"a}ki}, A. and {Lamarre}, J. -M. and {Lasenby}, A. and {Lattanzi}, M. and {Lawrence}, C.~R. and {Leahy}, J.~P. and {Leonardi}, R. and {Lesgourgues}, J. and {Levrier}, F. and {Lewis}, A. and {Liguori}, M. and {Lilje}, P.~B. and {Linden-V{\o}rnle}, M. and {L{\'o}pez-Caniego}, M. and {Lubin}, P.~M. and {Mac{\'\i}as-P{\'e}rez}, J.~F. and {Maggio}, G. and {Maino}, D. and {Mandolesi}, N. and {Mangilli}, A. and {Marchini}, A. and {Maris}, M. and {Martin}, P.~G. and {Martinelli}, M. and {Mart{\'\i}nez-Gonz{\'a}lez}, E. and {Masi}, S. and {Matarrese}, S. and {McGehee}, P. and {Meinhold}, P.~R. and {Melchiorri}, A. and {Melin}, J. -B. and {Mendes}, L. and {Mennella}, A. and {Migliaccio}, M. and {Millea}, M. and {Mitra}, S. and {Miville-Desch{\^e}nes}, M. -A. and {Moneti}, A. and {Montier}, L. and {Morgante}, G. and {Mortlock}, D. and {Moss}, A. and {Munshi}, D. and {Murphy}, J.~A. and {Naselsky}, P. and {Nati}, F. and {Natoli}, P. and {Netterfield}, C.~B. and {N{\o}rgaard-Nielsen}, H.~U. and {Noviello}, F. and {Novikov}, D. and {Novikov}, I. and {Oxborrow}, C.~A. and {Paci}, F. and {Pagano}, L. and {Pajot}, F. and {Paladini}, R. and {Paoletti}, D. and {Partridge}, B. and {Pasian}, F. and {Patanchon}, G. and {Pearson}, T.~J. and {Perdereau}, O. and {Perotto}, L. and {Perrotta}, F. and {Pettorino}, V. and {Piacentini}, F. and {Piat}, M. and {Pierpaoli}, E. and {Pietrobon}, D. and {Plaszczynski}, S. and {Pointecouteau}, E. and {Polenta}, G. and {Popa}, L. and {Pratt}, G.~W. and {Pr{\'e}zeau}, G.},
        title = "{Planck 2015 results. XIII. Cosmological parameters}",
      journal = {\aap},
         year = 2016,
        month = sep,
       volume = {594},
          eid = {A13},
        pages = {A13},
          doi = {10.1051/0004-6361/201525830},
archivePrefix = {arXiv},
       eprint = {1502.01589},
 primaryClass = {astro-ph.CO},
       adsurl = {https://ui.adsabs.harvard.edu/abs/2016A&A...594A..13P}
}

@ARTICLE{GunnPeterson1965,
       author = {{Gunn}, James E. and {Peterson}, Bruce A.},
        title = "{On the Density of Neutral Hydrogen in Intergalactic Space.}",
      journal = {ApJ},
         year = 1965,
        month = nov,
       volume = {142},
        pages = {1633-1636},
          doi = {10.1086/148444},
       adsurl = {https://ui.adsabs.harvard.edu/abs/1965ApJ...142.1633G}
}

@ARTICLE{Chabrier2003,
       author = {{Chabrier}, Gilles},
        title = "{Galactic Stellar and Substellar Initial Mass Function}",
      journal = {\pasp},
         year = 2003,
        month = jul,
       volume = {115},
       number = {809},
        pages = {763-795},
          doi = {10.1086/376392},
archivePrefix = {arXiv},
       eprint = {astro-ph/0304382},
 primaryClass = {astro-ph},
       adsurl = {https://ui.adsabs.harvard.edu/abs/2003PASP..115..763C}
}

@ARTICLE{Garaldi2024,
       author = {{Garaldi}, Enrico and {Kannan}, Rahul and {Smith}, Aaron and {Borrow}, Josh and {Vogelsberger}, Mark and {Pakmor}, R{\"u}diger and {Springel}, Volker and {Hernquist}, Lars and {Gal{\'a}rraga-Espinosa}, Daniela and {Yeh}, Jessica Y. -C. and {Shen}, Xuejian and {Xu}, Clara and {Neyer}, Meredith and {Spina}, Benedetta and {Almualla}, Mouza and {Zhao}, Yu},
        title = "{The THESAN project: public data release of radiation-hydrodynamic simulations matching reionization-era JWST observations}",
      journal = {\mnras},
         year = 2024,
        month = jun,
       volume = {530},
       number = {4},
        pages = {3765-3786},
          doi = {10.1093/mnras/stae839},
archivePrefix = {arXiv},
       eprint = {2309.06475},
 primaryClass = {astro-ph.CO},
       adsurl = {https://ui.adsabs.harvard.edu/abs/2024MNRAS.530.3765G}
}

@ARTICLE{Kashino2023,
       author = {{Kashino}, Daichi and {Lilly}, Simon J. and {Matthee}, Jorryt and {Eilers}, Anna-Christina and {Mackenzie}, Ruari and {Bordoloi}, Rongmon and {Simcoe}, Robert A.},
        title = "{EIGER. I. A Large Sample of [O III]-emitting Galaxies at 5.3 < z < 6.9 and Direct Evidence for Local Reionization by Galaxies}",
      journal = {ApJ},
         year = 2023,
        month = jun,
       volume = {950},
       number = {1},
          eid = {66},
        pages = {66},
          doi = {10.3847/1538-4357/acc588},
archivePrefix = {arXiv},
       eprint = {2211.08254},
 primaryClass = {astro-ph.GA},
       adsurl = {https://ui.adsabs.harvard.edu/abs/2023ApJ...950...66K}
}

@ARTICLE{SpringelHernquist2003,
       author = {{Springel}, Volker and {Hernquist}, Lars},
        title = "{Cosmological smoothed particle hydrodynamics simulations: a hybrid multiphase model for star formation}",
      journal = {\mnras},
         year = 2003,
        month = feb,
       volume = {339},
       number = {2},
        pages = {289-311},
          doi = {10.1046/j.1365-8711.2003.06206.x},
archivePrefix = {arXiv},
       eprint = {astro-ph/0206393},
 primaryClass = {astro-ph},
       adsurl = {https://ui.adsabs.harvard.edu/abs/2003MNRAS.339..289S}
}

@ARTICLE{Zou2024,
       author = {{Zou}, Siwei and {Cai}, Zheng and {Wang}, Feige and {Fan}, Xiaohui and {Champagne}, Jaclyn B. and {Hennawi}, Joseph F. and {Schindler}, Jan-Torge and {Farina}, Emanuele Paolo and {Yang}, Jinyi and {Inayoshi}, Kohei and {Ba{\~n}ados}, Eduardo and {Bosman}, Sarah E.~I. and {Li}, Zihao and {Lin}, Xiaojing and {Wu}, Yunjing and {Sun}, Fengwu and {Guo}, Ziyi and {Kulkuarni}, Girish and {Habouzit}, M{\'e}lanie and {Charlot}, Stephane and {Chevallard}, Jacopo and {Connor}, Thomas and {Eilers}, Anna-Christina and {Jiang}, Linhua and {Jin}, Xiangyu and {Kakiichi}, Koki and {Li}, Mingyu and {Meyer}, Romain A. and {Walter}, Fabian and {Zhang}, Huanian},
        title = "{A SPectroscopic survey of biased halos In the Reionization Era (ASPIRE): Impact of Galaxies on the Circumgalactic Medium Metal Enrichment at z > 6 Using the JWST and VLT}",
      journal = {ApJL},
         year = 2024,
        month = mar,
       volume = {963},
       number = {1},
          eid = {L28},
        pages = {L28},
          doi = {10.3847/2041-8213/ad23e7},
archivePrefix = {arXiv},
       eprint = {2402.00113},
 primaryClass = {astro-ph.GA},
       adsurl = {https://ui.adsabs.harvard.edu/abs/2024ApJ...963L..28Z}
}

@ARTICLE{harmoni2021,
       author = {{Thatte}, N. and {Tecza}, M. and {Schnetler}, H. and {Neichel}, B. and {Melotte}, D. and {Fusco}, T. and {Ferraro-Wood}, V. and {Clarke}, F. and {Bryson}, I. and {O'Brien}, K. and {Mateo}, M. and {Garcia Lorenzo}, B. and {Evans}, C. and {Bouch{\'e}}, N. and {Arribas}, S. and {HARMONI Consortium}},
        title = "{HARMONI: the ELT's First-Light Near-infrared and Visible Integral Field Spectrograph}",
      journal = {The Messenger},
         year = 2021,
        month = mar,
       volume = {182},
        pages = {7-12},
          doi = {10.18727/0722-6691/5215},
archivePrefix = {arXiv},
       eprint = {2103.11215},
 primaryClass = {astro-ph.IM},
       adsurl = {https://ui.adsabs.harvard.edu/abs/2021Msngr.182....7T}
}

@ARTICLE{Higginson2025,
       author = {{Higginson}, Jack and {Bordoloi}, Rongmon and {Simcoe}, Robert A. and {Matthee}, Jorryt and {Kashino}, Daichi and {Mackenzie}, Ruari and {Kramarenko}, Ivan and {Lilly}, Simon J. and {Eilers}, Anna-Christina and {Naidu}, Rohan P. and {Yue}, Minghao},
        title = "{EIGER. VIII. First Stars Signatures in the Connection between O I Absorption and Galaxies in the Epoch of Reionization}",
      journal = {\apj},
         year = 2026,
        month = mar,
       volume = {999},
       number = {1},
          eid = {49},
        pages = {49},
          doi = {10.3847/1538-4357/ae3e81},
archivePrefix = {arXiv},
       eprint = {2510.05220},
 primaryClass = {astro-ph.GA},
       adsurl = {https://ui.adsabs.harvard.edu/abs/2026ApJ...999...49H}
}

@ARTICLE{Portinari1998,
       author = {{Portinari}, L. and {Chiosi}, C. and {Bressan}, A.},
        title = "{Galactic chemical enrichment with new metallicity dependent stellar yields}",
      journal = {\aap},
         year = 1998,
        month = jun,
       volume = {334},
        pages = {505-539},
          doi = {10.48550/arXiv.astro-ph/9711337},
archivePrefix = {arXiv},
       eprint = {astro-ph/9711337},
 primaryClass = {astro-ph},
       adsurl = {https://ui.adsabs.harvard.edu/abs/1998A&A...334..505P}
}

@ARTICLE{Pollock2026,
       author = {{Pollock}, Clara L. and {Heintz}, Kasper E. and {Rusta}, Elka and {Watson}, Darach and {Salvadori}, Stefania and {Witten}, Callum and {Witstok}, Joris and {Koutsouridou}, Ioanna and {Gelli}, Viola and {Oesch}, Pascal A. and {Brammer}, Gabriel B. and {Saccardi}, Andrea and {Ito}, Kei and {Gottumukkala}, Rashmi and {Bergholt}, Dagmar and {Brooksby}, Kasper R. and {Terp}, Chamilla and {Valentino}, Francesco},
        title = "{Chemical signatures from the first stars embedded in metal-poor gas in galaxies at cosmic dawn}",
      journal = {arXiv e-prints},
         year = 2026,
        month = jun,
          eid = {arXiv:2606.13078},
        pages = {arXiv:2606.13078},
          doi = {10.48550/arXiv.2606.13078},
archivePrefix = {arXiv},
       eprint = {2606.13078},
 primaryClass = {astro-ph.GA},
       adsurl = {https://ui.adsabs.harvard.edu/abs/2026arXiv260613078P}
}

@ARTICLE{Klessen2023,
       author = {{Klessen}, Ralf S. and {Glover}, Simon C.~O.},
        title = "{The First Stars: Formation, Properties, and Impact}",
      journal = {\araa},
         year = 2023,
        month = aug,
       volume = {61},
        pages = {65-130},
          doi = {10.1146/annurev-astro-071221-053453},
archivePrefix = {arXiv},
       eprint = {2303.12500},
 primaryClass = {astro-ph.CO},
       adsurl = {https://ui.adsabs.harvard.edu/abs/2023ARA&A..61...65K}
}

@ARTICLE{Venditti2026,
       author = {{Venditti}, Alessandra and {Schaerer}, Daniel and {Zackrisson}, Erik and {Asada}, Yoshihisa and {Katz}, Harley and {Salvadori}, Stefania and {Vanzella}, Eros and {Mu{\~n}oz}, Julian B. and {Storck}, Anatole and {Bunker}, Andrew J. and {Trinca}, Alessandro and {Scholte}, Dirk and {Pacucci}, Fabio and {P{\'e}rez-Gonz{\'a}lez}, Pablo G. and {Fujimoto}, Seiji and {Charbonnel}, Corinne and {Maiolino}, Roberto and {Ferrara}, Andrea and {Giavalisco}, Mauro and {Schneider}, Raffaella and {Baggen}, Josephine and {Atek}, Hakim and {Bromm}, Volker and {Caputi}, Karina and {Ciesla}, Laure and {Dayal}, Pratika and {Kobayashi}, Chiaki and {Castellano}, Marco and {Santini}, Paola},
        title = "{How can we finally see the first light? Status and perspective in the search for Population III stars}",
      journal = {arXiv e-prints},
         year = 2026,
        month = jun,
          eid = {arXiv:2607.00167},
        pages = {arXiv:2607.00167},
          doi = {10.48550/arXiv.2607.00167},
archivePrefix = {arXiv},
       eprint = {2607.00167},
 primaryClass = {astro-ph.GA},
       adsurl = {https://ui.adsabs.harvard.edu/abs/2026arXiv260700167V}
}

@ARTICLE{Harikane2026,
       author = {{Harikane}, Yuichi and {Perez-Gonzalez}, Pablo G. and {Alvarez-Marquez}, Javier and {Ouchi}, Masami and {Nakazato}, Yurina and {Ono}, Yoshiaki and {Nakajima}, Kimihiko and {Umeda}, Hiroya and {Isobe}, Yuki and {Xu}, Yi and {Zhang}, Yechi},
        title = "{A UV-Luminous Galaxy at z=11 with Surprisingly Weak Star Formation Activity}",
      journal = {arXiv e-prints},
         year = 2026,
        month = jan,
          eid = {arXiv:2601.21833},
        pages = {arXiv:2601.21833},
          doi = {10.48550/arXiv.2601.21833},
archivePrefix = {arXiv},
       eprint = {2601.21833},
 primaryClass = {astro-ph.GA},
       adsurl = {https://ui.adsabs.harvard.edu/abs/2026arXiv260121833H}
}

@ARTICLE{Witstok2026,
       author = {{Witstok}, Joris and {Smit}, Renske and {Baker}, William M. and {Rinaldi}, Pierluigi and {Hainline}, Kevin N. and {Algera}, Hiddo S.~B. and {Arribas}, Santiago and {Bakx}, Tom J.~L.~C. and {Bunker}, Andrew J. and {Carniani}, Stefano and {Charlot}, St{\'e}phane and {Chevallard}, Jacopo and {Curti}, Mirko and {Curtis-Lake}, Emma and {Eisenstein}, Daniel J. and {Heintz}, Kasper E. and {Helton}, Jakob M. and {Jones}, Gareth C. and {Maiolino}, Roberto and {Maseda}, Michael V. and {P{\'e}rez-Gonz{\'a}lez}, Pablo G. and {Pollock}, Clara L. and {Robertson}, Brant E. and {Saxena}, Aayush and {Scholtz}, Jan and {Shivaei}, Irene and {Sun}, Fengwu and {Tacchella}, Sandro and {{\"U}bler}, Hannah and {Watson}, Darach and {Willott}, Chris J. and {Wu}, Zihao},
        title = "{On the origins of oxygen: ALMA and JWST characterise the multi-phase, metal-enriched, star-bursting medium within a 'normal' z > 11 galaxy}",
      journal = {The Open Journal of Astrophysics},
         year = 2026,
        month = jan,
       volume = {9},
        pages = {55261},
          doi = {10.33232/001c.155261},
archivePrefix = {arXiv},
       eprint = {2507.22888},
 primaryClass = {astro-ph.GA},
       adsurl = {https://ui.adsabs.harvard.edu/abs/2026OJAp....955261W}
}

@ARTICLE{Nakane2026,
       author = {{Nakane}, Minami and {Ouchi}, Masami},
        title = "{JWST Absorption-Line Analysis of UV-Bright Galaxies at $z=7.2-10.6$: Early Chemical Enrichment Traced by C, O, Mg, Al, Si, and Fe}",
      journal = {arXiv e-prints},
         year = 2026,
        month = jun,
          eid = {arXiv:2606.08782},
        pages = {arXiv:2606.08782},
          doi = {10.48550/arXiv.2606.08782},
archivePrefix = {arXiv},
       eprint = {2606.08782},
 primaryClass = {astro-ph.GA},
       adsurl = {https://ui.adsabs.harvard.edu/abs/2026arXiv260608782N}
}

@ARTICLE{Cooke2015,
       author = {{Cooke}, Ryan J. and {Pettini}, Max and {Jorgenson}, Regina A.},
        title = "{The Most Metal-poor Damped Ly{\ensuremath{\alpha}} Systems: An Insight into Dwarf Galaxies at High-redshift}",
      journal = {\apj},
         year = 2015,
        month = feb,
       volume = {800},
       number = {1},
          eid = {12},
        pages = {12},
          doi = {10.1088/0004-637X/800/1/12},
archivePrefix = {arXiv},
       eprint = {1406.7003},
 primaryClass = {astro-ph.GA},
       adsurl = {https://ui.adsabs.harvard.edu/abs/2015ApJ...800...12C}
}

@ARTICLE{Noterdaeme2021,
       author = {{Noterdaeme}, P. and {Balashev}, S. and {Ledoux}, C. and {Duchoquet}, G. and {L{\'o}pez}, S. and {Telikova}, K. and {Boiss{\'e}}, P. and {Krogager}, J.-K. and {De Cia}, A. and {Bergeron}, J.},
        title = "{Sharpening quasar absorption lines with ESPRESSO. Temperature of warm gas at z {\ensuremath{\sim}} 2, constraints on the Mg isotopic ratio, and structure of cold gas at z {\ensuremath{\sim}} 0.5}",
      journal = {\aap},
         year = 2021,
        month = jul,
       volume = {651},
          eid = {A78},
        pages = {A78},
          doi = {10.1051/0004-6361/202140501},
archivePrefix = {arXiv},
       eprint = {2105.00697},
 primaryClass = {astro-ph.GA},
       adsurl = {https://ui.adsabs.harvard.edu/abs/2021A&A...651A..78N}
}

@ARTICLE{Carswell2012,
       author = {{Carswell}, R.~F. and {Becker}, G.~D. and {Jorgenson}, R.~A. and {Murphy}, M.~T. and {Wolfe}, A.~M.},
        title = "{The kinetic temperature in a damped Lyman {\ensuremath{\alpha}} absorption system in Q2206-199 - an example of the warm neutral medium}",
      journal = {\mnras},
         year = 2012,
        month = may,
       volume = {422},
       number = {2},
        pages = {1700-1708},
          doi = {10.1111/j.1365-2966.2012.20746.x},
archivePrefix = {arXiv},
       eprint = {1202.3012},
 primaryClass = {astro-ph.GA},
       adsurl = {https://ui.adsabs.harvard.edu/abs/2012MNRAS.422.1700C}
}

@ARTICLE{Pettini2012,
       author = {{Pettini}, Max and {Cooke}, Ryan},
        title = "{A new, precise measurement of the primordial abundance of deuterium}",
      journal = {\mnras},
         year = 2012,
        month = oct,
       volume = {425},
       number = {4},
        pages = {2477-2486},
          doi = {10.1111/j.1365-2966.2012.21665.x},
archivePrefix = {arXiv},
       eprint = {1205.3785},
 primaryClass = {astro-ph.CO},
       adsurl = {https://ui.adsabs.harvard.edu/abs/2012MNRAS.425.2477P}
}

@ARTICLE{Cooke2011,
       author = {{Cooke}, Ryan and {Pettini}, Max and {Steidel}, Charles C. and {Rudie}, Gwen C. and {Nissen}, Poul E.},
        title = "{The most metal-poor damped Ly{\ensuremath{\alpha}} systems: insights into chemical evolution in the very metal-poor regime}",
      journal = {\mnras},
         year = 2011,
        month = oct,
       volume = {417},
       number = {2},
        pages = {1534-1558},
          doi = {10.1111/j.1365-2966.2011.19365.x},
archivePrefix = {arXiv},
       eprint = {1106.2805},
 primaryClass = {astro-ph.CO},
       adsurl = {https://ui.adsabs.harvard.edu/abs/2011MNRAS.417.1534C}
}

@ARTICLE{Pruto2026,
       author = {{Pruto}, Giulia and {Keating}, Laura and {Kannan}, Rahul and {Puchwein}, Ewald and {Smith}, Aaron and {Borrow}, Josh and {Garaldi}, Enrico and {Vogelsberger}, Mark and {Zier}, Oliver and {McClymont}, William and {Shen}, Xuejian and {Tacchella}, Sandro},
        title = "{The THESAN-ZOOM project: the hidden neighbours of O I absorbers during reionization}",
      journal = {\mnras},
         year = 2026,
        month = feb,
       volume = {546},
       number = {2},
          eid = {stag064},
        pages = {stag064},
          doi = {10.1093/mnras/stag064},
archivePrefix = {arXiv},
       eprint = {2510.13977},
 primaryClass = {astro-ph.GA},
       adsurl = {https://ui.adsabs.harvard.edu/abs/2026MNRAS.546ag064P}
}

@ARTICLE{Dutta2014,
       author = {{Dutta}, R. and {Srianand}, R. and {Rahmani}, H. and {Petitjean}, P. and {Noterdaeme}, P. and {Ledoux}, C.},
        title = "{A study of low-metallicity DLAs at high redshift and C II* as a probe of their physical conditions}",
      journal = {\mnras},
         year = 2014,
        month = may,
       volume = {440},
       number = {1},
        pages = {307-326},
          doi = {10.1093/mnras/stu260},
archivePrefix = {arXiv},
       eprint = {1402.2975},
 primaryClass = {astro-ph.GA},
       adsurl = {https://ui.adsabs.harvard.edu/abs/2014MNRAS.440..307D}
}

@ARTICLE{Vanni2024,
       author = {{Vanni}, Irene and {Salvadori}, Stefania and {D'Odorico}, Valentina and {Becker}, George D. and {Cupani}, Guido},
        title = "{Chemical Diagnostics to Unveil Environments Enriched by First Stars}",
      journal = {\apjl},
         year = 2024,
        month = jun,
       volume = {967},
       number = {2},
          eid = {L22},
        pages = {L22},
          doi = {10.3847/2041-8213/ad46fa},
archivePrefix = {arXiv},
       eprint = {2402.18640},
 primaryClass = {astro-ph.GA},
       adsurl = {https://ui.adsabs.harvard.edu/abs/2024ApJ...967L..22V}
}

@ARTICLE{HegerWoosley2010,
       author = {{Heger}, Alexander and {Woosley}, S.~E.},
        title = "{Nucleosynthesis and Evolution of Massive Metal-free Stars}",
      journal = {\apj},
         year = 2010,
        month = nov,
       volume = {724},
       number = {1},
        pages = {341-373},
          doi = {10.1088/0004-637X/724/1/341},
archivePrefix = {arXiv},
       eprint = {0803.3161},
 primaryClass = {astro-ph},
       adsurl = {https://ui.adsabs.harvard.edu/abs/2010ApJ...724..341H}
}

@ARTICLE{HegerWoosley2002,
       author = {{Heger}, A. and {Woosley}, S.~E.},
        title = "{The Nucleosynthetic Signature of Population III}",
      journal = {\apj},
         year = 2002,
        month = mar,
       volume = {567},
       number = {1},
        pages = {532-543},
          doi = {10.1086/338487},
archivePrefix = {arXiv},
       eprint = {astro-ph/0107037},
 primaryClass = {astro-ph},
       adsurl = {https://ui.adsabs.harvard.edu/abs/2002ApJ...567..532H}
}

@ARTICLE{WoosleyWeaver1995,
       author = {{Woosley}, S.~E. and {Weaver}, Thomas A.},
        title = "{The Evolution and Explosion of Massive Stars. II. Explosive Hydrodynamics and Nucleosynthesis}",
      journal = {\apjs},
         year = 1995,
        month = nov,
       volume = {101},
        pages = {181},
          doi = {10.1086/192237},
       adsurl = {https://ui.adsabs.harvard.edu/abs/1995ApJS..101..181W}
}

@ARTICLE{Kobayashi2006,
       author = {{Kobayashi}, Chiaki and {Umeda}, Hideyuki and {Nomoto}, Ken'ichi and {Tominaga}, Nozomu and {Ohkubo}, Takuya},
        title = "{Galactic Chemical Evolution: Carbon through Zinc}",
      journal = {\apj},
         year = 2006,
        month = dec,
       volume = {653},
       number = {2},
        pages = {1145-1171},
          doi = {10.1086/508914},
archivePrefix = {arXiv},
       eprint = {astro-ph/0608688},
 primaryClass = {astro-ph},
       adsurl = {https://ui.adsabs.harvard.edu/abs/2006ApJ...653.1145K}
}

@ARTICLE{Wolfe1986,
       author = {{Wolfe}, A.~M. and {Turnshek}, D.~A. and {Smith}, H.~E. and {Cohen}, R.~D.},
        title = "{Damped Lyman-Alpha Absorption by Disk Galaxies with Large Redshifts. I. The Lick Survey}",
      journal = {\apjs},
         year = 1986,
        month = jun,
       volume = {61},
        pages = {249},
          doi = {10.1086/191114},
       adsurl = {https://ui.adsabs.harvard.edu/abs/1986ApJS...61..249W}
}

@ARTICLE{Rafelski2014,
       author = {{Rafelski}, Marc and {Neeleman}, Marcel and {Fumagalli}, Michele and {Wolfe}, Arthur M. and {Prochaska}, J. Xavier},
        title = "{The Rapid Decline in Metallicity of Damped Ly{\ensuremath{\alpha}} Systems at z \raisebox{-0.5ex}\textasciitilde 5}",
      journal = {\apjl},
         year = 2014,
        month = feb,
       volume = {782},
       number = {2},
          eid = {L29},
        pages = {L29},
          doi = {10.1088/2041-8205/782/2/L29},
archivePrefix = {arXiv},
       eprint = {1310.6042},
 primaryClass = {astro-ph.CO},
       adsurl = {https://ui.adsabs.harvard.edu/abs/2014ApJ...782L..29R}
}

@ARTICLE{DOdorico2023_xqr,
       author = {{D'Odorico}, Valentina and {Ba{\~n}ados}, E. and {Becker}, G.~D. and {Bischetti}, M. and {Bosman}, S.~E.~I. and {Cupani}, G. and {Davies}, R. and {Farina}, E.~P. and {Ferrara}, A. and {Feruglio}, C. and {Mazzucchelli}, C. and {Ryan-Weber}, E. and {Schindler}, J.-T. and {Sodini}, A. and {Venemans}, B.~P. and {Walter}, F. and {Chen}, H. and {Lai}, S. and {Zhu}, Y. and {Bian}, F. and {Campo}, S. and {Carniani}, S. and {Cristiani}, S. and {Davies}, F. and {Decarli}, R. and {Drake}, A. and {Eilers}, A.-C. and {Fan}, X. and {Gaikwad}, P. and {Gallerani}, S. and {Greig}, B. and {Haehnelt}, M.~G. and {Hennawi}, J. and {Keating}, L. and {Kulkarni}, G. and {Mesinger}, A. and {Meyer}, R.~A. and {Neeleman}, M. and {Onoue}, M. and {Pallottini}, A. and {Qin}, Y. and {Rojas-Ruiz}, S. and {Satyavolu}, S. and {Sebastian}, A. and {Tripodi}, R. and {Wang}, F. and {Wolfson}, M. and {Yang}, J. and {Zanchettin}, M.~V.},
        title = "{XQR-30: The ultimate XSHOOTER quasar sample at the reionization epoch}",
      journal = {\mnras},
         year = 2023,
        month = jul,
       volume = {523},
       number = {1},
        pages = {1399-1420},
          doi = {10.1093/mnras/stad1468},
archivePrefix = {arXiv},
       eprint = {2305.05053},
 primaryClass = {astro-ph.GA},
       adsurl = {https://ui.adsabs.harvard.edu/abs/2023MNRAS.523.1399D}
}

@ARTICLE{XSHOOTER,
       author = {{Vernet}, J. and {Dekker}, H. and {D'Odorico}, S. and {Kaper}, L. and {Kjaergaard}, P. and {Hammer}, F. and {Randich}, S. and {Zerbi}, F. and {Groot}, P.~J. and {Hjorth}, J. and {Guinouard}, I. and {Navarro}, R. and {Adolfse}, T. and {Albers}, P.~W. and {Amans}, J.-P. and {Andersen}, J.~J. and {Andersen}, M.~I. and {Binetruy}, P. and {Bristow}, P. and {Castillo}, R. and {Chemla}, F. and {Christensen}, L. and {Conconi}, P. and {Conzelmann}, R. and {Dam}, J. and {de Caprio}, V. and {de Ugarte Postigo}, A. and {Delabre}, B. and {di Marcantonio}, P. and {Downing}, M. and {Elswijk}, E. and {Finger}, G. and {Fischer}, G. and {Flores}, H. and {Fran{\c{c}}ois}, P. and {Goldoni}, P. and {Guglielmi}, L. and {Haigron}, R. and {Hanenburg}, H. and {Hendriks}, I. and {Horrobin}, M. and {Horville}, D. and {Jessen}, N.~C. and {Kerber}, F. and {Kern}, L. and {Kiekebusch}, M. and {Kleszcz}, P. and {Klougart}, J. and {Kragt}, J. and {Larsen}, H.~H. and {Lizon}, J.-L. and {Lucuix}, C. and {Mainieri}, V. and {Manuputy}, R. and {Martayan}, C. and {Mason}, E. and {Mazzoleni}, R. and {Michaelsen}, N. and {Modigliani}, A. and {Moehler}, S. and {M{\o}ller}, P. and {Norup S{\o}rensen}, A. and {N{\o}rregaard}, P. and {P{\'e}roux}, C. and {Patat}, F. and {Pena}, E. and {Pragt}, J. and {Reinero}, C. and {Rigal}, F. and {Riva}, M. and {Roelfsema}, R. and {Royer}, F. and {Sacco}, G. and {Santin}, P. and {Schoenmaker}, T. and {Spano}, P. and {Sweers}, E. and {Ter Horst}, R. and {Tintori}, M. and {Tromp}, N. and {van Dael}, P. and {van der Vliet}, H. and {Venema}, L. and {Vidali}, M. and {Vinther}, J. and {Vola}, P. and {Winters}, R. and {Wistisen}, D. and {Wulterkens}, G. and {Zacchei}, A.},
        title = "{X-shooter, the new wide band intermediate resolution spectrograph at the ESO Very Large Telescope}",
      journal = {\aap},
         year = 2011,
        month = dec,
       volume = {536},
          eid = {A105},
        pages = {A105},
          doi = {10.1051/0004-6361/201117752},
archivePrefix = {arXiv},
       eprint = {1110.1944},
 primaryClass = {astro-ph.IM},
       adsurl = {https://ui.adsabs.harvard.edu/abs/2011A&A...536A.105V}
}

@ARTICLE{Barkat1967,
       author = {{Barkat}, Z. and {Rakavy}, G. and {Sack}, N.},
        title = "{Dynamics of Supernova Explosion Resulting from Pair Formation}",
      journal = {\prl},
         year = 1967,
        month = mar,
       volume = {18},
       number = {10},
        pages = {379-381},
          doi = {10.1103/PhysRevLett.18.379},
       adsurl = {https://ui.adsabs.harvard.edu/abs/1967PhRvL..18..379B}
}

@ARTICLE{Bolamperti2026,
       author = {{Bolamperti}, A. and {Messa}, M. and {Zanella}, A. and {Vanzella}, E. and {Bergamini}, P. and {Loiacono}, F. and {Koekemoer}, A.~M. and {Vernet}, J. and {Windhorst}, R.~A. and {Adamo}, A. and {Annibali}, F. and {Calura}, F. and {Castellano}, M. and {Diego}, J.~M. and {Grillo}, C. and {Gronke}, M. and {Iani}, E. and {Meneghetti}, M. and {Mercurio}, A. and {Nakajima}, K. and {Ravindranath}, S. and {Ricotti}, M. and {Rosati}, P. and {Yan}, H.},
        title = "{A Cosmic Archipelago of lensed metal-poor galaxies at z {\ensuremath{\sim}} 6}",
      journal = {\aap},
         year = 2026,
        month = aug,
       volume = {712},
          eid = {A153},
        pages = {A153},
          doi = {10.1051/0004-6361/202558496},
archivePrefix = {arXiv},
       eprint = {2606.02720},
 primaryClass = {astro-ph.GA},
       adsurl = {https://ui.adsabs.harvard.edu/abs/2026A&A...712A.153B}
}

@ARTICLE{Morishita2025,
       author = {{Morishita}, Takahiro and {Liu}, Zhaoran and {Stiavelli}, Massimo and {Treu}, Tommaso and {Bergamini}, Pietro and {Zhang}, Yechi},
        title = "{Pristine Massive Star Formation Caught at the Break of Cosmic Dawn}",
      journal = {arXiv e-prints},
         year = 2025,
        month = jul,
          eid = {arXiv:2507.10521},
        pages = {arXiv:2507.10521},
          doi = {10.48550/arXiv.2507.10521},
archivePrefix = {arXiv},
       eprint = {2507.10521},
 primaryClass = {astro-ph.CO},
       adsurl = {https://ui.adsabs.harvard.edu/abs/2025arXiv250710521M}
}

@ARTICLE{Abel2000,
       author = {{Abel}, Tom and {Bryan}, Greg L. and {Norman}, Michael L.},
        title = "{The Formation and Fragmentation of Primordial Molecular Clouds}",
      journal = {\apj},
         year = 2000,
        month = sep,
       volume = {540},
       number = {1},
        pages = {39-44},
          doi = {10.1086/309295},
archivePrefix = {arXiv},
       eprint = {astro-ph/0002135},
 primaryClass = {astro-ph},
       adsurl = {https://ui.adsabs.harvard.edu/abs/2000ApJ...540...39A}
}

@ARTICLE{Bromm2001,
       author = {{Bromm}, V. and {Ferrara}, A. and {Coppi}, P.~S. and {Larson}, R.~B.},
        title = "{The fragmentation of pre-enriched primordial objects}",
      journal = {\mnras},
         year = 2001,
        month = dec,
       volume = {328},
       number = {3},
        pages = {969-976},
          doi = {10.1046/j.1365-8711.2001.04915.x},
archivePrefix = {arXiv},
       eprint = {astro-ph/0104271},
 primaryClass = {astro-ph},
       adsurl = {https://ui.adsabs.harvard.edu/abs/2001MNRAS.328..969B}
}

@ARTICLE{Bromm2002,
       author = {{Bromm}, Volker and {Coppi}, Paolo S. and {Larson}, Richard B.},
        title = "{The Formation of the First Stars. I. The Primordial Star-forming Cloud}",
      journal = {\apj},
         year = 2002,
        month = jan,
       volume = {564},
       number = {1},
        pages = {23-51},
          doi = {10.1086/323947},
archivePrefix = {arXiv},
       eprint = {astro-ph/0102503},
 primaryClass = {astro-ph},
       adsurl = {https://ui.adsabs.harvard.edu/abs/2002ApJ...564...23B}
}

@ARTICLE{Vanni2023,
       author = {{Vanni}, Irene and {Salvadori}, Stefania and {Sk{\'u}lad{\'o}ttir}, {\'A}sa and {Rossi}, Martina and {Koutsouridou}, Ioanna},
        title = "{Characterizing the true descendants of the first stars}",
      journal = {\mnras},
         year = 2023,
        month = dec,
       volume = {526},
       number = {2},
        pages = {2620-2644},
          doi = {10.1093/mnras/stad2910},
archivePrefix = {arXiv},
       eprint = {2309.07958},
 primaryClass = {astro-ph.GA},
       adsurl = {https://ui.adsabs.harvard.edu/abs/2023MNRAS.526.2620V}
}

@ARTICLE{Maiolino2026,
       author = {{Maiolino}, Roberto and {{\"U}bler}, Hannah and {Perna}, Michele and {Witstok}, Joris and {Jones}, Gareth C. and {Perez-Gonzalez}, Pablo G. and {Nakajima}, Kimihiko and {Rusta}, Elka and {Salvadori}, Stefania and {Tacchella}, Sandro and {Madau}, Piero and {Trussler}, James A.~A. and {D'Eugenio}, Francesco and {Ji}, Xihan and {Scholtz}, Jan and {Carniani}, Stefano and {Isobe}, Yuki and {Katz}, Harley and {Arribas}, Santiago and {Baker}, William M. and {B{\"o}ker}, Torsten and {Bromm}, Volker and {Bunker}, Andrew J. and {Charlot}, Stephane and {Chevallard}, Jacopo and {Curti}, Mirko and {Curtis-Lake}, Emma and {Eisenstein}, Daniel and {Egami}, Eiichi and {Ferrara}, Andrea and {Graziani}, Luca and {Hainline}, Kevin and {Helton}, Jakob M. and {Ivey}, Lucy and {Jonson}, Benjamin and {Koller}, Maria and {Kumari}, Nimisha and {Marconi}, Alessandro and {Mazzolari}, Giovanni and {Laporte}, Nicolas and {Parlanti}, Eleonora and {Pascalau}, Robert and {Pentericci}, Laura and {Rinaldi}, Pierluigi and {Robertson}, Brant and {Rodr{\'\i}guez Del Pino}, Bruno and {Schneider}, Raffaella and {Venditti}, Alessandra and {Venturi}, Giacomo and {Willmer}, Christopher N.~A. and {Witten}, Callum and {Zamora}, Sandra},
        title = "{The search for Population III: Confirmation of a HeII emitter with no metal lines at z=10.6}",
      journal = {arXiv e-prints},
         year = 2026,
        month = mar,
          eid = {arXiv:2603.20362},
        pages = {arXiv:2603.20362},
          doi = {10.48550/arXiv.2603.20362},
archivePrefix = {arXiv},
       eprint = {2603.20362},
 primaryClass = {astro-ph.GA},
       adsurl = {https://ui.adsabs.harvard.edu/abs/2026arXiv260320362M}
}

@ARTICLE{Tumlinson2017,
       author = {{Tumlinson}, Jason and {Peeples}, Molly S. and {Werk}, Jessica K.},
        title = "{The Circumgalactic Medium}",
      journal = {\araa},
         year = 2017,
        month = aug,
       volume = {55},
       number = {1},
        pages = {389-432},
          doi = {10.1146/annurev-astro-091916-055240},
archivePrefix = {arXiv},
       eprint = {1709.09180},
 primaryClass = {astro-ph.GA},
       adsurl = {https://ui.adsabs.harvard.edu/abs/2017ARA&A..55..389T}
}

@ARTICLE{Rusta2025,
       author = {{Rusta}, Elka and {Salvadori}, Stefania and {Gelli}, Viola and {Schaerer}, Daniel and {Marconi}, Alessandro and {Koutsouridou}, Ioanna and {Carniani}, Stefano},
        title = "{Metal-polluted Population III Galaxies and How to Find Them}",
      journal = {\apjl},
         year = 2025,
        month = aug,
       volume = {989},
       number = {2},
          eid = {L32},
        pages = {L32},
          doi = {10.3847/2041-8213/adf4e3},
archivePrefix = {arXiv},
       eprint = {2506.17400},
 primaryClass = {astro-ph.GA},
       adsurl = {https://ui.adsabs.harvard.edu/abs/2025ApJ...989L..32R}
}

@ARTICLE{Caffau2011,
       author = {{Caffau}, Elisabetta and {Bonifacio}, Piercarlo and {Fran{\c{c}}ois}, Patrick and {Sbordone}, Luca and {Monaco}, Lorenzo and {Spite}, Monique and {Spite}, Fran{\c{c}}ois and {Ludwig}, Hans-G. and {Cayrel}, Roger and {Zaggia}, Simone and {Hammer}, Fran{\c{c}}ois and {Randich}, Sofia and {Molaro}, Paolo and {Hill}, Vanessa},
        title = "{An extremely primitive star in the Galactic halo}",
      journal = {\nat},
         year = 2011,
        month = sep,
       volume = {477},
       number = {7362},
        pages = {67-69},
          doi = {10.1038/nature10377},
archivePrefix = {arXiv},
       eprint = {1203.2612},
 primaryClass = {astro-ph.GA},
       adsurl = {https://ui.adsabs.harvard.edu/abs/2011Natur.477...67C}
}

@ARTICLE{Keller2014,
       author = {{Keller}, S.~C. and {Bessell}, M.~S. and {Frebel}, A. and {Casey}, A.~R. and {Asplund}, M. and {Jacobson}, H.~R. and {Lind}, K. and {Norris}, J.~E. and {Yong}, D. and {Heger}, A. and {Magic}, Z. and {da Costa}, G.~S. and {Schmidt}, B.~P. and {Tisserand}, P.},
        title = "{A single low-energy, iron-poor supernova as the source of metals in the star SMSS J031300.36-670839.3}",
      journal = {\nat},
         year = 2014,
        month = feb,
       volume = {506},
       number = {7489},
        pages = {463-466},
          doi = {10.1038/nature12990},
archivePrefix = {arXiv},
       eprint = {1402.1517},
 primaryClass = {astro-ph.SR},
       adsurl = {https://ui.adsabs.harvard.edu/abs/2014Natur.506..463K}
}

@ARTICLE{Frebel2015,
       author = {{Frebel}, Anna and {Norris}, John E.},
        title = "{Near-Field Cosmology with Extremely Metal-Poor Stars}",
      journal = {\araa},
         year = 2015,
        month = aug,
       volume = {53},
        pages = {631-688},
          doi = {10.1146/annurev-astro-082214-122423},
archivePrefix = {arXiv},
       eprint = {1501.06921},
 primaryClass = {astro-ph.SR},
       adsurl = {https://ui.adsabs.harvard.edu/abs/2015ARA&A..53..631F}
}

@ARTICLE{Karakas2010,
       author = {{Karakas}, A.~I.},
        title = "{Updated stellar yields from asymptotic giant branch models}",
      journal = {\mnras},
         year = 2010,
        month = apr,
       volume = {403},
       number = {3},
        pages = {1413-1425},
          doi = {10.1111/j.1365-2966.2009.16198.x},
archivePrefix = {arXiv},
       eprint = {0912.2142},
 primaryClass = {astro-ph.SR},
       adsurl = {https://ui.adsabs.harvard.edu/abs/2010MNRAS.403.1413K}
}

@ARTICLE{KarakasLugaro,
       author = {{Karakas}, Amanda I. and {Lugaro}, Maria},
        title = "{Stellar Yields from Metal-rich Asymptotic Giant Branch Models}",
      journal = {\apj},
         year = 2016,
        month = jul,
       volume = {825},
       number = {1},
          eid = {26},
        pages = {26},
          doi = {10.3847/0004-637X/825/1/26},
archivePrefix = {arXiv},
       eprint = {1604.02178},
 primaryClass = {astro-ph.SR},
       adsurl = {https://scixplorer.org/abs/2016ApJ...825...26K}
}

@ARTICLE{Cinquegrana2022,
       author = {{Cinquegrana}, Giulia C. and {Karakas}, Amanda I.},
        title = "{The most metal-rich stars in the universe: chemical contributions of low- and intermediate-mass asymptotic giant branch stars with metallicities within 0.04 {\ensuremath{\leq}} Z {\ensuremath{\leq}} 0.10}",
      journal = {\mnras},
         year = 2022,
        month = feb,
       volume = {510},
       number = {2},
        pages = {1557-1576},
          doi = {10.1093/mnras/stab3379},
archivePrefix = {arXiv},
       eprint = {2111.09527},
 primaryClass = {astro-ph.SR},
       adsurl = {https://scixplorer.org/abs/2022MNRAS.510.1557C}
}

@ARTICLE{Seitanzahl2013,
       author = {{Seitenzahl}, Ivo R. and {Ciaraldi-Schoolmann}, Franco and {R{\"o}pke}, Friedrich K. and {Fink}, Michael and {Hillebrandt}, Wolfgang and {Kromer}, Markus and {Pakmor}, R{\"u}diger and {Ruiter}, Ashley J. and {Sim}, Stuart A. and {Taubenberger}, Stefan},
        title = "{Three-dimensional delayed-detonation models with nucleosynthesis for Type Ia supernovae}",
      journal = {\mnras},
         year = 2013,
        month = feb,
       volume = {429},
       number = {2},
        pages = {1156-1172},
          doi = {10.1093/mnras/sts402},
archivePrefix = {arXiv},
       eprint = {1211.3015},
 primaryClass = {astro-ph.SR},
       adsurl = {https://scixplorer.org/abs/2013MNRAS.429.1156S}
}

@ARTICLE{Saha2026,
       author = {{Saha}, Bipradeep and {Kannan}, Rahul and {Mirouh}, Giovanni M.},
        title = "{A framework for modelling Population III stars in cosmological simulations}",
      journal = {arXiv e-prints},
         year = 2026,
        month = may,
          eid = {arXiv:2605.26206},
        pages = {arXiv:2605.26206},
          doi = {10.48550/arXiv.2605.26206},
archivePrefix = {arXiv},
       eprint = {2605.26206},
 primaryClass = {astro-ph.GA},
       adsurl = {https://ui.adsabs.harvard.edu/abs/2026arXiv260526206S}
}

@ARTICLE{Rossi2021,
       author = {{Rossi}, Martina and {Salvadori}, Stefania and {Sk{\'u}lad{\'o}ttir}, {\'A}sa},
        title = "{Ultra-faint dwarf galaxies: unveiling the minimum mass of the first stars}",
      journal = {\mnras},
         year = 2021,
        month = jun,
       volume = {503},
       number = {4},
        pages = {6026-6044},
          doi = {10.1093/mnras/stab821},
archivePrefix = {arXiv},
       eprint = {2103.09834},
 primaryClass = {astro-ph.GA},
       adsurl = {https://ui.adsabs.harvard.edu/abs/2021MNRAS.503.6026R}
}

@ARTICLE{Welsh2024,
       author = {{Welsh}, Louise and {Cooke}, Ryan and {Fumagalli}, Michele and {Pettini}, Max and {Rudie}, Gwen C.},
        title = "{A survey of extremely metal-poor gas at cosmic noon: Evidence of elevated [O/Fe]}",
      journal = {\aap},
         year = 2024,
        month = nov,
       volume = {691},
          eid = {A285},
        pages = {A285},
          doi = {10.1051/0004-6361/202451147},
archivePrefix = {arXiv},
       eprint = {2409.07525},
 primaryClass = {astro-ph.GA},
       adsurl = {https://ui.adsabs.harvard.edu/abs/2024A&A...691A.285W}
}

@ARTICLE{Saccardi2023,
       author = {{Saccardi}, Andrea and {Salvadori}, Stefania and {D'Odorico}, Valentina and {Cupani}, Guido and {Fumagalli}, Michele and {Berg}, Trystyn A.~M. and {Becker}, George D. and {Ellison}, Sara and {Lopez}, Sebastian},
        title = "{Evidence of First Stars-enriched Gas in High-redshift Absorbers}",
      journal = {\apj},
         year = 2023,
        month = may,
       volume = {948},
       number = {1},
          eid = {35},
        pages = {35},
          doi = {10.3847/1538-4357/acc39f},
archivePrefix = {arXiv},
       eprint = {2305.02346},
 primaryClass = {astro-ph.GA},
       adsurl = {https://ui.adsabs.harvard.edu/abs/2023ApJ...948...35S}
}

@ARTICLE{Wolfe2005,
       author = {{Wolfe}, Arthur M. and {Gawiser}, Eric and {Prochaska}, Jason X.},
        title = "{Damped Ly {\ensuremath{\alpha}} Systems}",
      journal = {\araa},
         year = 2005,
        month = sep,
       volume = {43},
       number = {1},
        pages = {861-918},
          doi = {10.1146/annurev.astro.42.053102.133950},
archivePrefix = {arXiv},
       eprint = {astro-ph/0509481},
 primaryClass = {astro-ph},
       adsurl = {https://ui.adsabs.harvard.edu/abs/2005ARA&A..43..861W}
}

@ARTICLE{Travaglio2004,
       author = {{Travaglio}, C. and {Hillebrandt}, W. and {Reinecke}, M. and {Thielemann}, F.-K.},
        title = "{Nucleosynthesis in multi-dimensional SN Ia explosions}",
      journal = {\aap},
         year = 2004,
        month = oct,
       volume = {425},
        pages = {1029-1040},
          doi = {10.1051/0004-6361:20041108},
archivePrefix = {arXiv},
       eprint = {astro-ph/0406281},
 primaryClass = {astro-ph},
       adsurl = {https://ui.adsabs.harvard.edu/abs/2004A&A...425.1029T}
}

@INPROCEEDINGS{Thielemann2003,
       author = {{Thielemann}, F.-K. and {Argast}, D. and {Brachwitz}, F. and {Hix}, W.~R. and {H{\"o}flich}, P. and {Liebend{\"o}rfer}, M. and {Martinez-Pinedo}, G. and {Mezzacappa}, A. and {Nomoto}, K. and {Panov}, I.},
        title = "{Supernova Nucleosynthesis and Galactic Evolution}",
    booktitle = {From Twilight to Highlight: The Physics of Supernovae},
         year = 2003,
       editor = {{Hillebrandt}, Wolfgang and {Leibundgut}, Bruno},
        month = jan,
        pages = {331},
          doi = {10.1007/10828549_46},
       adsurl = {https://ui.adsabs.harvard.edu/abs/2003fthp.conf..331T}
}

@ARTICLE{Pillepich2018,
       author = {{Pillepich}, Annalisa and {Springel}, Volker and {Nelson}, Dylan and {Genel}, Shy and {Naiman}, Jill and {Pakmor}, R{\"u}diger and {Hernquist}, Lars and {Torrey}, Paul and {Vogelsberger}, Mark and {Weinberger}, Rainer and {Marinacci}, Federico},
        title = "{Simulating galaxy formation with the IllustrisTNG model}",
      journal = {\mnras},
         year = 2018,
        month = jan,
       volume = {473},
       number = {3},
        pages = {4077-4106},
          doi = {10.1093/mnras/stx2656},
archivePrefix = {arXiv},
       eprint = {1703.02970},
 primaryClass = {astro-ph.GA},
       adsurl = {https://ui.adsabs.harvard.edu/abs/2018MNRAS.473.4077P}
}

@ARTICLE{Paxton2011,
  author = {{Paxton}, B. and {Bildsten}, L. and {Dotter}, A. and {Herwig}, F. and {Lesaffre}, P. and {Timmes}, F.},
  title = {{Modules for Experiments in Stellar Astrophysics (MESA)}},
  journal = {\apjs},
  year = {2011},
  volume = {192},
  pages = {3},
  doi = {10.1088/0067-0049/192/1/3},
}

@ARTICLE{Paxton2013,
  author = {{Paxton}, B. and {Cantiello}, M. and {Arras}, P. and {Bildsten}, L. and {Brown}, E.~F. and {Dotter}, A. and {Mankovich}, C. and {Montgomery}, M.~H. and {Stello}, D. and {Timmes}, F.~X. and {Townsend}, R.},
  title = {{Modules for Experiments in Stellar Astrophysics (MESA): Planets, Oscillations, Rotation, and Massive Stars}},
  journal = {\apjs},
  year = {2013},
  volume = {208},
  pages = {4},
  doi = {10.1088/0067-0049/208/1/4},
}

@ARTICLE{Paxton2015,
  author = {{Paxton}, B. and {Marchant}, P. and {Schwab}, J. and {Bauer}, E.~B. and {Bildsten}, L. and {Cantiello}, M. and {Dessart}, L. and {Farmer}, R. and {Hu}, H. and {Langer}, N. and {Townsend}, R.~H.~D. and {Townsley}, D.~M. and {Timmes}, F.~X.},
  title = {{Modules for Experiments in Stellar Astrophysics (MESA): Binaries, Pulsations, and Explosions}},
  journal = {\apjs},
  year = {2015},
  volume = {220},
  pages = {15},
  doi = {10.1088/0067-0049/220/1/15},
}

@ARTICLE{Paxton2018,
  author = {{Paxton}, B. and {Schwab}, J. and {Bauer}, E.~B. and {Bildsten}, L. and {Blinnikov}, S. and {Duffell}, P. and {Farmer}, R. and {Goldberg}, J.~A. and {Marchant}, P. and {Sorokina}, E. and {Thoul}, A. and {Townsend}, R.~H.~D. and {Timmes}, F.~X.},
  title = {{Modules for Experiments in Stellar Astrophysics (MESA): Convective Boundaries, Element Diffusion, and Massive Star Explosions}},
  journal = {\apjs},
  year = {2018},
  volume = {234},
  pages = {34},
  doi = {10.3847/1538-4365/aaa5a8},
}

@ARTICLE{Paxton2019,
  author = {{Paxton}, Bill and {Smolec}, R. and {Schwab}, Josiah and {Gautschy}, A. and {Bildsten}, Lars and {Cantiello}, Matteo and {Dotter}, Aaron and {Farmer}, R. and {Goldberg}, Jared A. and {Jermyn}, Adam S. and {Kanbur}, S.~M. and {Marchant}, Pablo and {Thoul}, Anne and {Townsend}, Richard H.~D. and {Wolf}, William M. and {Zhang}, Michael and {Timmes}, F.~X.},
  title = "{Modules for Experiments in Stellar Astrophysics (MESA): Pulsating Variable Stars, Rotation, Convective Boundaries, and Energy Conservation}",
  journal = {\apjs},
  year = "2019",
  volume = {243},
  pages = {10},
  doi = {10.3847/1538-4365/ab2241},
}

@ARTICLE{Jermyn2023,
  author = {{Jermyn}, Adam S. and {Bauer}, Evan B. and {Schwab}, Josiah and {Farmer}, R. and {Ball}, Warrick H. and {Bellinger}, Earl P. and {Dotter}, Aaron and {Joyce}, Meridith and {Marchant}, Pablo and {Mombarg}, Joey S.~G. and {Wolf}, William M. and {Sunny Wong}, Tin Long and {Cinquegrana}, Giulia C. and {Farrell}, Eoin and {Smolec}, R. and {Thoul}, Anne and {Cantiello}, Matteo and {Herwig}, Falk and {Toloza}, Odette and {Bildsten}, Lars and {Townsend}, Richard H.~D. and {Timmes}, F.~X.},
  title = "{Modules for Experiments in Stellar Astrophysics (MESA): Time-dependent Convection, Energy Conservation, Automatic Differentiation, and Infrastructure}",
  journal = {\apjs},
  year = 2023,
  volume = {265},
  pages = {15},
  doi = {10.3847/1538-4365/acae8d},
}

@ARTICLE{Hubeny1988,
       author = {{Hubeny}, I.},
        title = "{A computer program for calculating non-LTE model stellar atmospheres}",
      journal = {Computer Physics Communications},
         year = 1988,
        month = dec,
       volume = {52},
       number = {1},
        pages = {103-132},
          doi = {10.1016/0010-4655(88)90177-4},
       adsurl = {https://ui.adsabs.harvard.edu/abs/1988CoPhC..52..103H}
}

@ARTICLE{Draine2003,
       author = {{Draine}, B.~T.},
        title = "{Interstellar Dust Grains}",
      journal = {\araa},
         year = 2003,
        month = jan,
       volume = {41},
        pages = {241-289},
          doi = {10.1146/annurev.astro.41.011802.094840},
archivePrefix = {arXiv},
       eprint = {astro-ph/0304489},
 primaryClass = {astro-ph},
       adsurl = {https://ui.adsabs.harvard.edu/abs/2003ARA&A..41..241D}
}

@ARTICLE{Doherty2014,
       author = {{Doherty}, Carolyn L. and {Gil-Pons}, Pilar and {Lau}, Herbert H.~B. and {Lattanzio}, John C. and {Siess}, Lionel},
        title = "{Super and massive AGB stars - II. Nucleosynthesis and yields - Z = 0.02, 0.008 and 0.004}",
      journal = {\mnras},
         year = 2014,
        month = jan,
       volume = {437},
       number = {1},
        pages = {195-214},
          doi = {10.1093/mnras/stt1877},
archivePrefix = {arXiv},
       eprint = {1310.2614},
 primaryClass = {astro-ph.SR},
       adsurl = {https://ui.adsabs.harvard.edu/abs/2014MNRAS.437..195D}
}

@ARTICLE{Fishlock2014,
       author = {{Fishlock}, Cherie K. and {Karakas}, Amanda I. and {Lugaro}, Maria and {Yong}, David},
        title = "{Evolution and Nucleosynthesis of Asymptotic Giant Branch Stellar Models of Low Metallicity}",
      journal = {\apj},
         year = 2014,
        month = dec,
       volume = {797},
       number = {1},
          eid = {44},
        pages = {44},
          doi = {10.1088/0004-637X/797/1/44},
archivePrefix = {arXiv},
       eprint = {1410.7457},
 primaryClass = {astro-ph.SR},
       adsurl = {https://ui.adsabs.harvard.edu/abs/2014ApJ...797...44F}
}

@ARTICLE{Li2026,
       author = {{Li}, Zihao and {Kakiichi}, Koki and {Christensen}, Lise and {Cai}, Zheng and {D'Odorico}, Valentina and {Matthee}, Jorryt and {Kashino}, Daichi and {Bordoloi}, Rongmon and {Mackenzie}, Ruari and {Berg}, Trystyn A.~M. and et al.},
        title = "{First-star imprints in a metal-poor galaxy overdensity near the end of reionization}",
      journal = {Nature Astronomy},
         year = 2026,
        month = sep,
          doi = {10.1038/s41550-026-02993-5},
archivePrefix = {arXiv},
       eprint = {2606.31638},
 primaryClass = {astro-ph.GA},
       adsurl = {https://ui.adsabs.harvard.edu/abs/2026NatAs.tmp..219L}
}

@ARTICLE{Samuel2026,
       author = {{Samuel}, Jenna and {Boylan-Kolchin}, Michael and {Feldmann}, Robert and {Hopkins}, Philip and {Sun}, Guochao and {Gandhi}, Pratik and {Venditti}, Alessandra and {Shen}, Xuejian and {Wetzel}, Andrew and {Moreno}, Jorge and {Munoz}, Julian and {Cochrane}, Rachel and {Faucher-Giguere}, Claude-Andre and {Bromm}, Volker and {Finkelstein}, Steven and {Straight}, Maria and {Painter}, Connor and {Stern}, Jonathan and {Bullock}, James},
        title = "{Resolving galaxy formation in the early Universe with BonFIRE and CampFIRE}",
      journal = {arXiv e-prints},
         year = 2026,
        month = may,
          eid = {arXiv:2605.24104},
        pages = {arXiv:2605.24104},
          doi = {10.48550/arXiv.2605.24104},
archivePrefix = {arXiv},
       eprint = {2605.24104},
 primaryClass = {astro-ph.GA},
       adsurl = {https://ui.adsabs.harvard.edu/abs/2026arXiv260524104S}
}

@ARTICLE{Garaldi2026,
       author = {{Garaldi}, Enrico and {Popovic}, Filip and {Kannan}, Rahul and {Smith}, Aaron and {Puchwein}, Ewald and {Yoshida}, Naoki and {Nagamine}, Kentaro and {Peroux}, Celine and {Keating}, Laura and {Vogelsberger}, Mark and {McClymont}, William and {Shen}, Xuejian and {Tacchella}, Sandro and {Hernquist}, Lars},
        title = "{The Thesan-Zoom project: bursty star formation is incompatible with prolonged dust survival}",
      journal = {arXiv e-prints},
         year = 2026,
        month = jul,
          eid = {arXiv:2607.08824},
        pages = {arXiv:2607.08824},
          doi = {10.48550/arXiv.2607.08824},
archivePrefix = {arXiv},
       eprint = {2607.08824},
 primaryClass = {astro-ph.GA},
       adsurl = {https://ui.adsabs.harvard.edu/abs/2026arXiv260708824G}
}

@ARTICLE{Simcoe2012,
       author = {{Simcoe}, Robert A. and {Sullivan}, Peter W. and {Cooksey}, Kathy L. and {Kao}, Melodie M. and {Matejek}, Michael S. and {Burgasser}, Adam J.},
        title = "{Extremely metal-poor gas at a redshift of 7}",
      journal = {\nat},
         year = 2012,
        month = dec,
       volume = {492},
       number = {7427},
        pages = {79-82},
          doi = {10.1038/nature11612},
archivePrefix = {arXiv},
       eprint = {1212.0548},
 primaryClass = {astro-ph.CO},
       adsurl = {https://ui.adsabs.harvard.edu/abs/2012Natur.492...79S}
}

@ARTICLE{Fujimoto2025,
       author = {{Fujimoto}, Seiji and {Naidu}, Rohan P. and {Chisholm}, John and {Atek}, Hakim and {Endsley}, Ryan and {Kokorev}, Vasily and {Furtak}, Lukas J. and {Pan}, Richard and {Liu}, Boyuan and {Bromm}, Volker and {Venditti}, Alessandra and {Visbal}, Eli and {Sarmento}, Richard and {Weibel}, Andrea and {Oesch}, Pascal A. and {Brammer}, Gabriel and {Schaerer}, Daniel and {Adamo}, Angela and {Berg}, Danielle A. and {Bezanson}, Rachel and {Bouwens}, Rychard and {Chemerynska}, Iryna and {Claeyssens}, Ad{\'e}la{\"\i}de and {Dessauges-Zavadsky}, Miroslava and {Frebel}, Anna and {Korber}, Damien and {Labbe}, Ivo and {Marques-Chaves}, Rui and {Matthee}, Jorryt and {McQuinn}, Kristen B.~W. and {Mu{\~n}oz}, Julian B. and {Natarajan}, Priyamvada and {Saldana-Lopez}, Alberto and {Suess}, Katherine A. and {Volonteri}, Marta and {Zitrin}, Adi},
        title = "{GLIMPSE: An Ultrafaint ≃{}10$^{5}$ M$_{{\ensuremath{\odot}}}$ Pop III Galaxy Candidate and First Constraints on the Pop III UV Luminosity Function at z ≃ 6─7}",
      journal = {\apj},
         year = 2025,
        month = aug,
       volume = {989},
       number = {1},
          eid = {46},
        pages = {46},
          doi = {10.3847/1538-4357/ade9a1},
archivePrefix = {arXiv},
       eprint = {2501.11678},
 primaryClass = {astro-ph.GA},
       adsurl = {https://ui.adsabs.harvard.edu/abs/2025ApJ...989...46F}
}

@ARTICLE{Fujimoto2025b,
       author = {{Fujimoto}, Seiji and {Asada}, Yoshihisa and {Naidu}, Rohan P. and {Chisholm}, John and {Atek}, Hakim and {Brammer}, Gabriel and {Berg}, Danielle A. and {Schaerer}, Daniel and {Kokorev}, Vasily and {Furtak}, Lukas J. and {Richard}, Johan and {Venditti}, Alessandra and {Bromm}, Volker and {Adamo}, Angela and {Claeyssens}, Adelaide and {Dessauges-Zavadsky}, Miroslava and {Fei}, Qinyue and {Hsiao}, Tiger Yu-Yang and {Korber}, Damien and {Munoz}, Julian B. and {Pan}, Richard and {Saldana-Lopez}, Alberto},
        title = "{GLIMPSE-D: An Exotic Balmer-Jump Object at z=6.20? Revisiting Photometric Selection and the Cosmic Abundance of Pop III Galaxies}",
      journal = {arXiv e-prints},
         year = 2025,
        month = dec,
          eid = {arXiv:2512.11790},
        pages = {arXiv:2512.11790},
          doi = {10.48550/arXiv.2512.11790},
archivePrefix = {arXiv},
       eprint = {2512.11790},
 primaryClass = {astro-ph.GA},
       adsurl = {https://ui.adsabs.harvard.edu/abs/2025arXiv251211790F}
}

@ARTICLE{Welsh2019,
       author = {{Welsh}, Louise and {Cooke}, Ryan and {Fumagalli}, Michele},
        title = "{Modelling the chemical enrichment of Population III supernovae: the origin of the metals in near-pristine gas clouds}",
      journal = {\mnras},
         year = 2019,
        month = aug,
       volume = {487},
       number = {3},
        pages = {3363-3376},
          doi = {10.1093/mnras/stz1526},
archivePrefix = {arXiv},
       eprint = {1906.00009},
 primaryClass = {astro-ph.SR},
       adsurl = {https://ui.adsabs.harvard.edu/abs/2019MNRAS.487.3363W}
}

@ARTICLE{Bosman2026,
       author = {{Bosman}, Sarah E.~I. and {Davies}, Frederick B. and {Welsh}, Louise and {Farina}, Emanuele P. and {Gaikwad}, Prakash and {Spina}, Benedetta and {Maity}, Barun and {Cooke}, Ryan and {Protu{\v{s}}ov{\'a}}, Klaudia and {D'Odorico}, Valentina and et al.},
        title = "{Nearly pristine gas in a galaxy 1 billion years after the Big Bang}",
      journal = {arXiv e-prints},
         year = 2026,
        month = oct,
          eid = {arXiv:2610.16509},
        pages = {arXiv:2610.16509},
          doi = {10.48550/arXiv.2610.16509},
archivePrefix = {arXiv},
       eprint = {2610.16509},
 primaryClass = {astro-ph.GA},
       adsurl = {https://ui.adsabs.harvard.edu/abs/2026arXiv261016509B}
}

@ARTICLE{Viegas1995,
       author = {{Viegas}, Sueli M.},
        title = "{Abundances at high redshift: ionization correction factors}",
      journal = {\mnras},
         year = 1995,
        month = sep,
       volume = {276},
       number = {1},
        pages = {268-272},
          doi = {10.1093/mnras/276.1.268},
       adsurl = {https://ui.adsabs.harvard.edu/abs/1995MNRAS.276..268V}
}

@ARTICLE{Diemer2013,
       author = {{Diemer}, Benedikt and {Kravtsov}, Andrey V. and {More}, Surhud},
        title = "{On the Evolution of Cluster Scaling Relations}",
      journal = {\apj},
         year = 2013,
        month = dec,
       volume = {779},
       number = {2},
          eid = {159},
        pages = {159},
          doi = {10.1088/0004-637X/779/2/159},
archivePrefix = {arXiv},
       eprint = {1306.3165},
 primaryClass = {astro-ph.CO},
       adsurl = {https://ui.adsabs.harvard.edu/abs/2013ApJ...779..159D}
}

@ARTICLE{Zhu2026,
       author = {{Zhu}, Yongda and {Ji}, Zhiyuan and {Becker}, George D. and {Ding}, Jiani and {Egami}, Eiichi and {Fan}, Xiaohui and {Jin}, Xiangyu and {Liu}, Weizhe and {Lyu}, Jianwei and {Ma}, Zheng and {Narisetty}, Suprabhas and {Rieke}, George H. and {Wu}, Yunjing and {Yue}, Minghao and {Zhang}, Junyu and {Rieke}, Marcia J.},
        title = "{Early metal-enriched baryon cycling before the midpoint of cosmic reionization}",
      journal = {Nature Astronomy},
         year = 2026,
        month = sep,
          doi = {10.1038/s41550-026-02988-2},
archivePrefix = {arXiv},
       eprint = {2604.21218},
 primaryClass = {astro-ph.GA},
       adsurl = {https://ui.adsabs.harvard.edu/abs/2026NatAs.tmp..217Z}
}

@ARTICLE{LimongiChieffi2018,
       author = {{Limongi}, Marco and {Chieffi}, Alessandro},
        title = "{Presupernova Evolution and Explosive Nucleosynthesis of Rotating Massive Stars in the Metallicity Range -3 {\ensuremath{\leq}} [Fe/H] {\ensuremath{\leq}} 0}",
      journal = {\apjs},
         year = 2018,
        month = jul,
       volume = {237},
       number = {1},
          eid = {13},
        pages = {13},
          doi = {10.3847/1538-4365/aacb24},
archivePrefix = {arXiv},
       eprint = {1805.09640},
 primaryClass = {astro-ph.SR},
       adsurl = {https://ui.adsabs.harvard.edu/abs/2018ApJS..237...13L}
}

@ARTICLE{Prantzos2018,
       author = {{Prantzos}, N. and {Abia}, C. and {Limongi}, M. and {Chieffi}, A. and {Cristallo}, S.},
        title = "{Chemical evolution with rotating massive star yields - I. The solar neighbourhood and the s-process elements}",
      journal = {\mnras},
         year = 2018,
        month = may,
       volume = {476},
       number = {3},
        pages = {3432-3459},
          doi = {10.1093/mnras/sty316},
archivePrefix = {arXiv},
       eprint = {1802.02824},
 primaryClass = {astro-ph.GA},
       adsurl = {https://ui.adsabs.harvard.edu/abs/2018MNRAS.476.3432P}
}

@ARTICLE{Vasan2026,
       author = {{Keerthi Vasan G.}, C. and {Senchyna}, Peter and {Mason}, Charlotte A. and {Chen}, Zuyi and {Stark}, Daniel P. and {Jones}, Tucker and {Whitler}, Lily and {Glazer}, Kelsey S. and {Aravena}, Manuel and {Gonzalez-Lopez}, Jorge and et al.},
        title = "{JWST absorption line spectroscopy with SPURS: ISM covering fractions and kinematics in individual galaxies at $z=5-9$}",
      journal = {arXiv e-prints},
         year = 2026,
        month = jun,
          eid = {arXiv:2606.06609},
        pages = {arXiv:2606.06609},
          doi = {10.48550/arXiv.2606.06609},
archivePrefix = {arXiv},
       eprint = {2606.06609},
 primaryClass = {astro-ph.GA},
       adsurl = {https://scixplorer.org/abs/2026arXiv260606609K}
}

@ARTICLE{Chen2026,
       author = {{Chen}, Zuyi and {Stark}, Daniel P. and {Mason}, Charlotte A. and {Plat}, Adele and {Gelli}, Viola and {Senchyna}, Peter and {Keerthi Vasan G.}, C. and {Endsley}, Ryan and {Tang}, Mengtao and {Topping}, Michael W. and et al.},
        title = "{SPURS: Bursty Star Formation in an Extremely Luminous Weak Emission Line Galaxy at $z=9.3$}",
      journal = {arXiv e-prints},
         year = 2026,
        month = apr,
          eid = {arXiv:2604.21516},
        pages = {arXiv:2604.21516},
          doi = {10.48550/arXiv.2604.21516},
archivePrefix = {arXiv},
       eprint = {2604.21516},
 primaryClass = {astro-ph.GA},
       adsurl = {https://scixplorer.org/abs/2026arXiv260421516C}
}



\appendix

\section{Metal yields in the thesan-zoom model}
\label{app:thzmetyields}

\begin{figure*}
    \includegraphics[width=1.8\columnwidth]{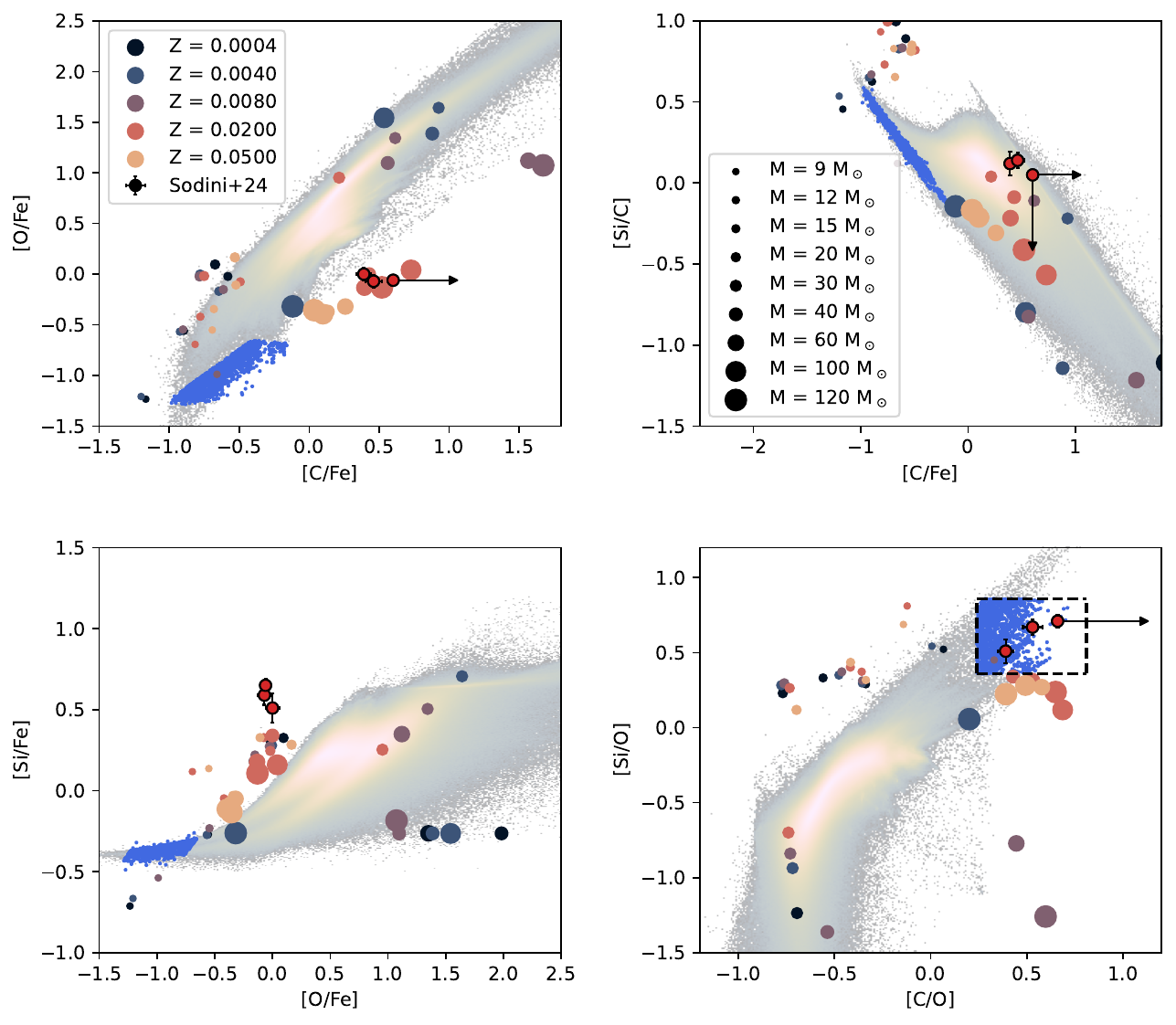}
    \caption{SN~II metal yields from \citet{Portinari1998} superimposed on the histogram shown in Fig.~\ref{fig:hist2d_popIII}. The filled circles mark the theoretical yields with colour indicating the metallicity of the star and size indicating the initial mass. Gas with [Si/O]--[C/O] abundances consistent with observed Pop~III-enriched gas candidates, marked by the red circles with errorbars, seem to show in every panel abundances similar to those released by intermediate-mass ($\approx 9\,\Msun$) SNe at metallicity $\approx 0.008$. On the other hand, the observed abundances seem closer to abundances produced by massive SNe~II at $Z = 0.02$. }
    \label{app:fig:snIIyields}
\end{figure*}

\begin{figure*}
    \includegraphics[width=1.8\columnwidth]{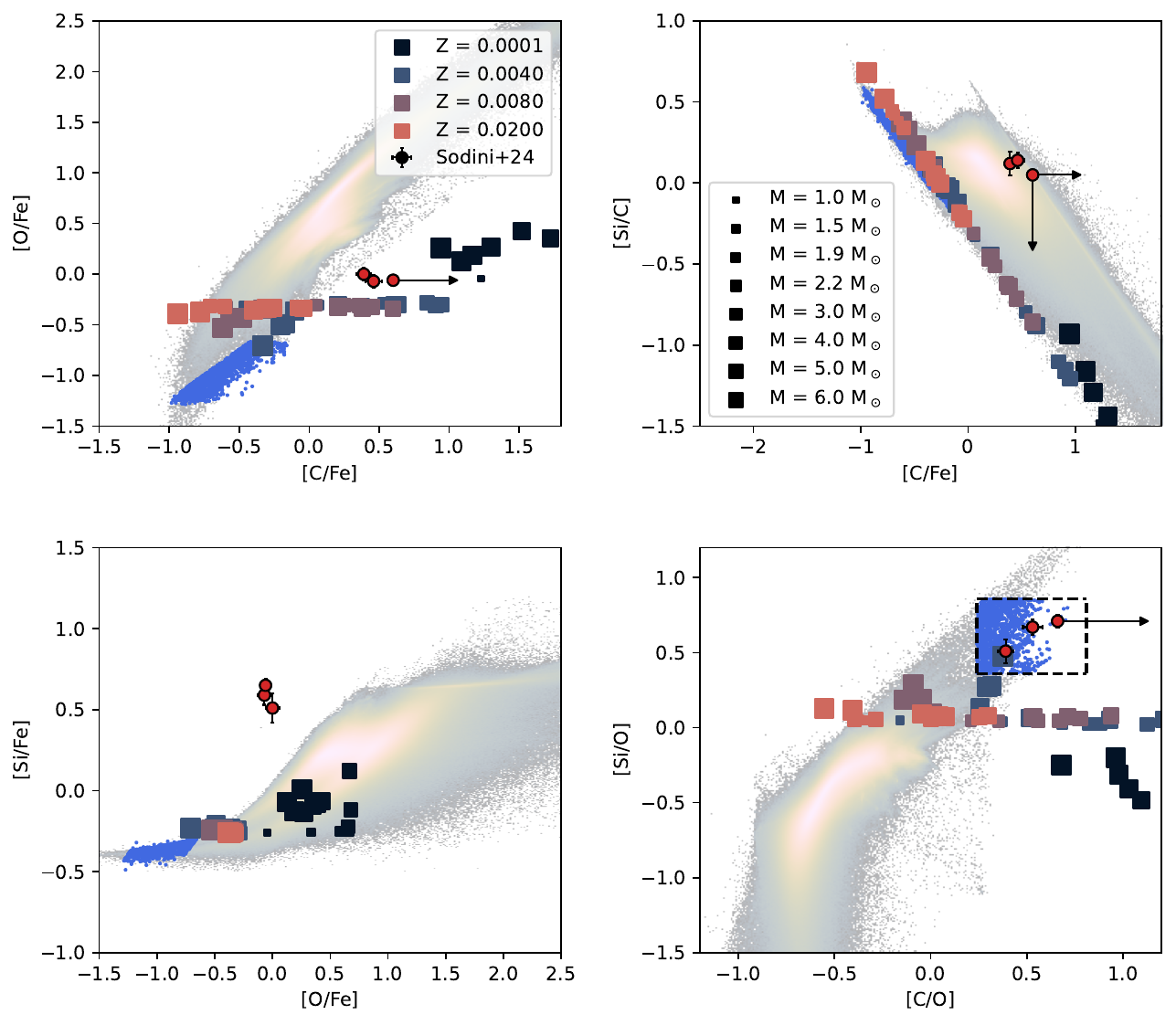}
    \caption{Same as Fig.~\ref{app:fig:snIIyields}, but with coloured squares marking the AGB metal yields from \citet{Karakas2010}. The enrichment from intermediate-mass stars ($\approx 5 - 6\,\Msun$) at $\approx 0.3\,\Zsun$ is similar to the metal abundance pattern displayed by gas cells reproducing the observed [Si/O]--[C/O] abundances.}
    \label{app:fig:AGByields}
\end{figure*}

To understand the origin of the Pop~II--Pop~III confusion on the [Si/O]--[C/O] distribution presented in Section~\ref{sec:possiblepopIII} as well as to interpret observed abundances, we compare our distributions to the theoretical metal yields used in the simulations. 
Specifically in Fig.~\ref{app:fig:snIIyields} we represent the same distribution shown in Fig.~\ref{fig:hist2d_popIII}, adding the SN~II yields obtained as the ratio of ejecta mass between corresponding elements with respect to solar abundances, from the tables in \citet{Portinari1998}. 
Colours mark the initial metallicity of the star, while the circle size identifies the initial mass. 
The figure shows how the DLA-like gas with [Si/O]--[C/O] abundances similar to the potential Pop~III-enriched candidates identified in \citet{Sodini2024} exhibit abundances consistent with those ejected by intermediate mass stars ($\approx 9~\Msun$) SNe at metallicity $\approx 0.6\,\Zsun$.
On the other hand, the observed abundances in \citetalias{Sodini2024} reside closer, especially in the first panel showing the [O/Fe]--[C/Fe] relations, to patterns expected in ejecta from massive stars at higher metallicity ($Z = 0.02$) exploding as SNe.

Similarly, Fig.~\ref{app:fig:AGByields} shows how AGB yields from \citet{Karakas2010} compare to the simulated distributions. Even in this case, AGB metal yields of stars with initial mass $\approx 5-6\,\Msun$ and $Z = 0.004$ are similar to the coloured points marking gas cells with [Si/O]--[C/O] distributions similar to the observed absorbers.
These abundance patterns suggest that the possible Pop~II--Pop~III confusion in the [Si/O]--[C/O] plane might be favoured by enrichment from intermediate-mass stars through AGB winds and SNe explosions. 
At the same time, we cannot rule out the hypothesis that observations of potential Pop~III candidates might arise from gas enriched by massive, metal-rich stars exploding as SNe.

\section{Redshift evolution}
\label{app:redev}

\begin{figure*}
    \includegraphics[width=1.8\columnwidth]{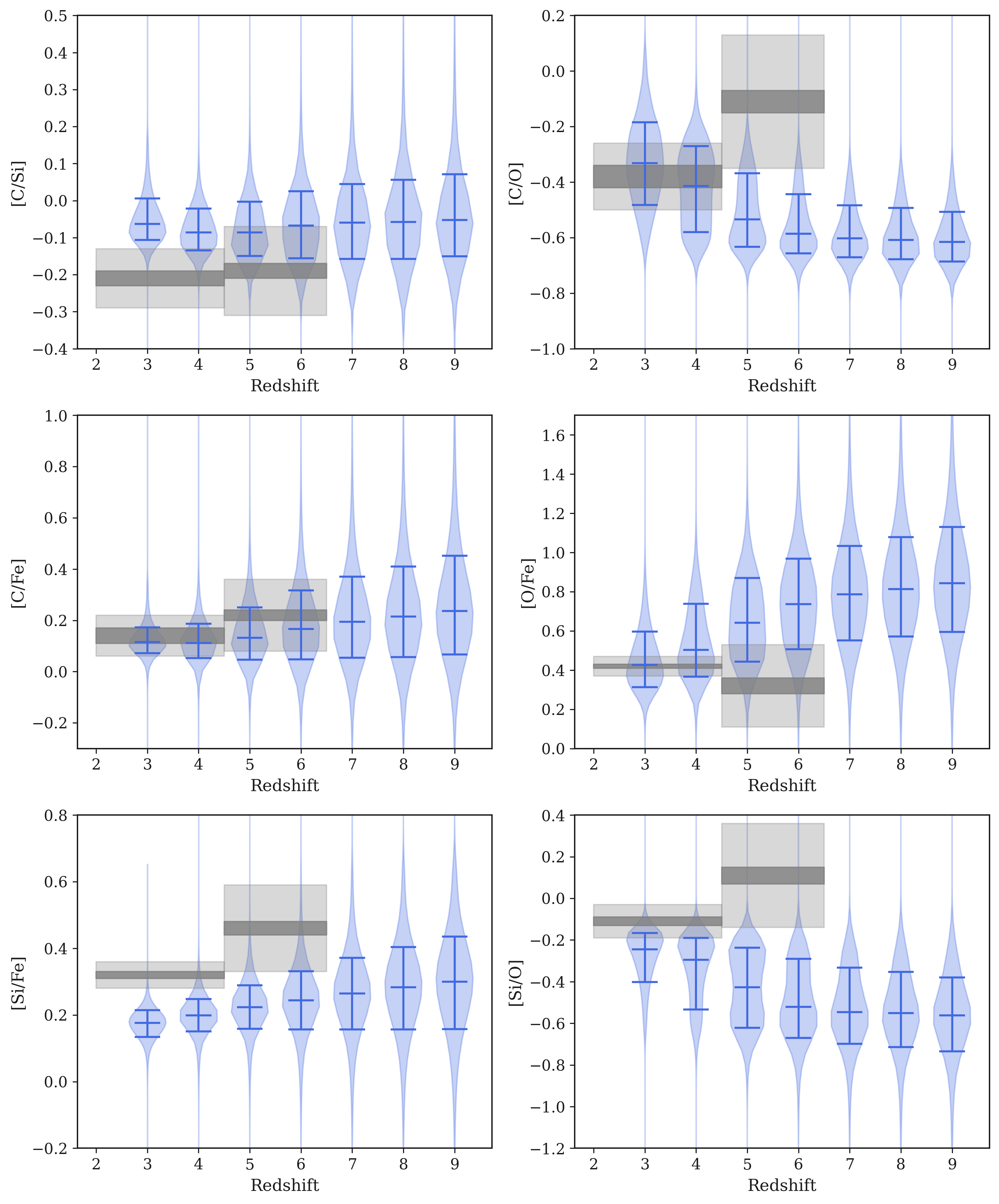}
    \caption{Redshift evolution of metal abundance ratios. The blue violins mark the density distribution of DLA-like gas cells in the zoom-in regions except for the most massive one (m13.0) which is not simulated at $z<6$. The horizontal lines identify the $16^{\rm th}$, $50^{\rm th}$ and $84^{\rm th}$ percentiles of each distribution. The shaded regions from \citet{Sodini2024} are centred on the weighted average of the observed values and mark the standard deviation of the sample (light gray) and the deviation of the mean (dark gray), when dividing the observed systems in multiple works in two redshift bins, $2 \leq z \leq 4.5$ and $4.5 \leq z \leq 6.5$, respectively.}
    \label{fig:violinplot}
\end{figure*}

The redshift evolution of different abundance ratios is represented by the violin plots in Fig.~\ref{fig:violinplot}. 
These violins show the density distributions of abundance ratios when considering DLA-like gas cells across all zoom-in regions except for m13.0, which hosts the most massive target halo. 
This exclusion is due to the fact that in the fiducial \thz simulations this region is reproduced only down to $z=6$, as our model does not include quasar feedback effects, which might become significant in massive galaxies at lower redshift.

While some of the distributions remain overall constant in the analysed redshift range, as for example [C/Si], [C/Fe] and [Si/Fe], a slight increase with cosmic time can be observed for [C/O] and [Si/O], together with a decrease in [O/Fe], indicating a possible slower oxygen enrichment relative to the other elements.
Overall, however, our data do not suggest strong evolutionary trends with redshift, with most distributions showing variations of less than $1\sigma$ from $z=9$ to $z=3$.
Our simulated distributions are consistent within $2\sigma$ with observational results, marked by the gray shaded regions. 
Some discrepancies are nevertheless present, probably arising from limited observational sample, or from differences associated with our selection method.
Observations and semi-analytical models suggest that the larger dispersion in abundance ratio distributions observed at high redshift might be related to Pop~III pollution \citep[][]{Sodini2024}. 
However, in the fiducial \thz simulations a larger dispersion at higher redshift is found in most of the abundance ratios, with [C/O] being the main exception. 
Our analysis, therefore, suggests that increased abundance ratio dispersions at higher redshifts can naturally arise when stellar evolution models are implemented within a hydrodynamical framework, without requiring the contribution of Pop~III stars.
However, we cannot exclude that the introduction of Pop~III stars in the simulations might enhance this dispersion at high redshift.
To test the importance of Pop~III stars in the dispersion we would need to run multiple zoom-in regions with at least one Pop~III stellar model incorporated, which will be addressed in future works.

\section{Metal yields comparison}

\begin{figure*}
    \includegraphics[width=1\linewidth]{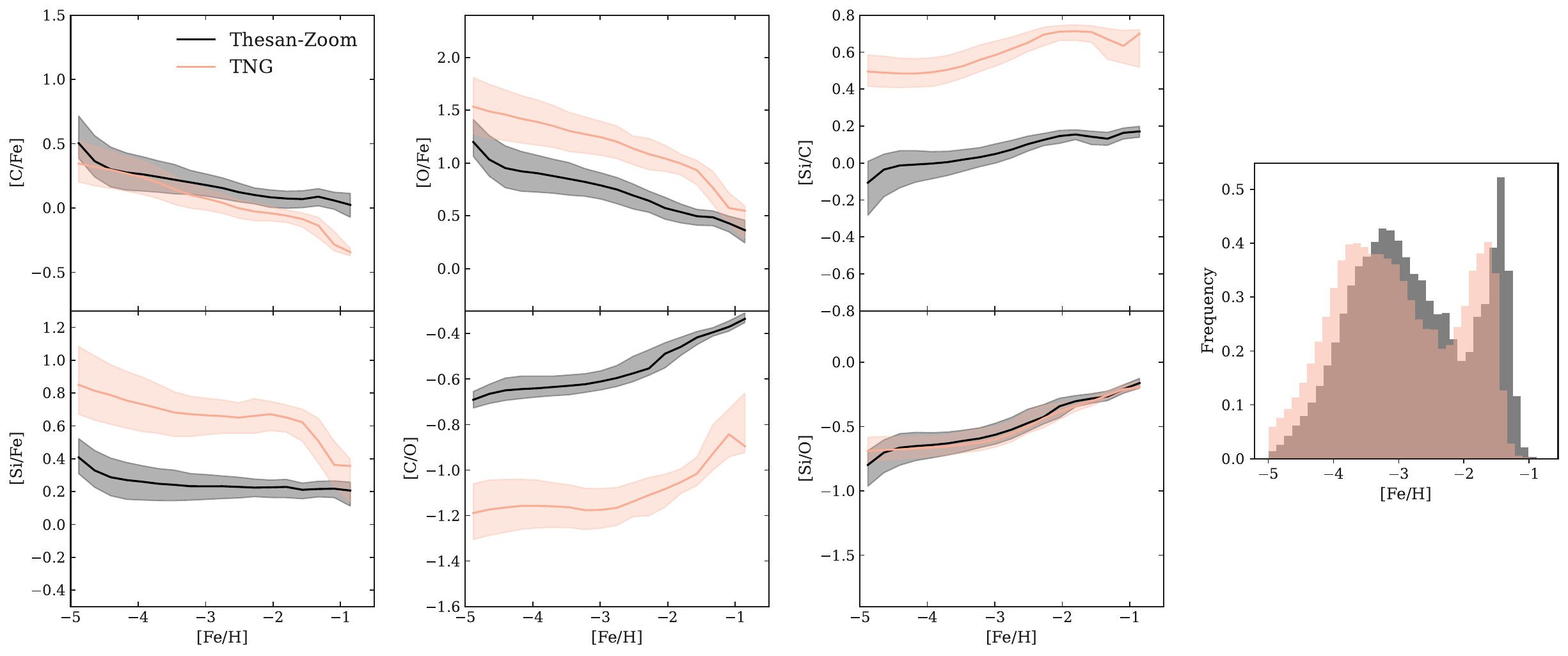}
    \caption{Metal abundance ratios as a function of iron abundance in the gas of the zoom-in region m10.8 at $z=6$ when simulated using different yield tables. The solid lines mark the median abundance ratios in the fiducial \thz model (dark line) and in the simulations with SN~II yields from IllustrisTNG \citep[light-coloured line;][]{Pillepich2018}, as presented in Section~\ref{sec:differentyields}. The shaded regions mark the 16$^{\rm th}$ and 84$^{\rm th}$ percentiles, respectively.
    The last panel on the right shows values of iron enrichment in the simulations, with \thz yields leading to values on average $\approx 0.25$ dex larger than those obtained with IllustrisTNG SN~II yields.
    The latter lead to a lower carbon enrichment and enhanced silicon and oxygen densities.}
    \label{app:fig:metalyieldcomp}
\end{figure*}

In Section~\ref{sec:differentyields}, we show how the abundance ratio distributions differ from the \thz fiducial models when using SN~II yields from the IllustrisTNG simulations.

From Fig.~\ref{fig:Kobayashi}, we notice how the [Si/C]--[C/Fe] and [Si/Fe]--[O/Fe] distributions exhibit the largest differences, with the IllustrisTNG yields predicting higher [Si/C] values and higher [Si/Fe] at high [O/Fe] ratios, similarly to what predicted by the SN~II yields of \citet{WoosleyWeaver1995}.
To understand the origin of this discrepancy, we examine how each metal abundance ratio evolves in the DLA-like gas of the same zoom-in region (m10.8) when using the two different yield models.
These are shown in Fig.~\ref{app:fig:metalyieldcomp} as a function of iron abundance. 
The histograms on the right show the iron enrichment in the two simulations, with the fiducial \thz model displaying values shifted to higher numbers, roughly $0.25$ dex higher with respect to the IllustrisTNG yields.
Despite this slightly higher iron enrichment in our fiducial models, the [C/Fe] abundances of IllustrisTNG are lower than in \thz, suggesting an enhanced carbon enrichment in our models. 
On the other hand, the levels of silicon and oxygen predicted by the IllustrisTNG SN~II yields are too high to be explained by the smaller iron enhancement in \thz, implying a higher enrichment from these two elements in IllustrisTNG.
This also leads to an overall matching [Si/O] abundance ratio in the two simulations, as can be observed in the third panel in the bottom row.
The large differences displayed by [C/O] and [Si/C] abundance ratios result naturally from the previous observations: the IllustrisTNG SN~II yields lead to a higher production of oxygen and silicon, and a lower production of iron and carbon with respect to the fiducial \thz model.

\bsp	
\label{lastpage}
\end{document}